\documentclass[status=final]{article}

\usepackage{biblatex-files}

\usepackage{ifthen}
\usepackage[svgnames,table]{xcolor}
\usepackage{mathptmx}

\usepackage[backend=bibtex,style=authoryear-comp,natbib=true,maxcitenames=2]{biblatex} 
\usepackage[colorlinks,allcolors=Blue]{hyperref} %
\usepackage{multirow}
\usepackage{multicol}
\usepackage{tabularx}
\usepackage[english]{babel}
\usepackage{csquotes} %
\usepackage{enumerate}
\usepackage{ragged2e}
  \newcolumntype{P}[1]{>{\raggedright\arraybackslash}p{#1}}
  \newcolumntype{L}{>{\raggedright\arraybackslash}X}
  \newcolumntype{?}{!{\vrule width 1pt}}
\usepackage{amssymb} 
\usepackage{cleveref} %
  \Crefname{section}{Section}{Sections}
  \Crefname{paragraph}{Section}{Sections}
  \Crefname{table}{Table}{Tables}
  \Crefname{figure}{Figure}{Figures}
\usepackage{stmaryrd}  
\usepackage{paralist}
\usepackage{tikz}
\usetikzlibrary{calc,shapes}
\usepackage{xtab}  %
\usepackage{booktabs} 
\usepackage{pdflscape}

\newenvironment{acknowledgements}{\small\paragraph{Acknowledgements.}}{\normalsize}
\newcommand{\sidecaption}{}

\newcommand{\wrt}{with respect to\ }

\newcommand{\cf}{cf.\ } 		
\newcommand{\ies}{i.e.,\ } 		
\newcommand{\ie}{that is,\ }			
 		
\newcommand{\egs}{e.g.\ } 		
\newcommand{\Eg}{For example,\ }
\newcommand{\eg}{for example,\ }
\newcommand{\Egs}{E.g.\ }

\newcommand{\vs}{vs.\ }

\newcommand{\trackedcontent}[5][arabic]{%
  \refstepcounter{#2}%
  \textbf{#4%
    \ifthenelse{\equal{#1}{arabic}}{\arabic{#2}}{%
      \ifthenelse{\equal{#1}{Alph}}{\Alph{#2}}{\arabic{#2}}%
    }#5%
  }%
  \label{#3}%
}
\newcommand{\RQ}[1]{%
  \trackedcontent{tcRQ}{rq:#1}{RQ}{}}
\newcommand{\Construct}[1]{%
  \trackedcontent{tcConstruct}{construct:#1}{C}{}}

\newcommand{\Question}[1]{%
  \trackedcontent{tcQuestion}{question:#1}{Q}{}}
\newcommand{\Finding}[1]{%
  \trackedcontent{tcFinding}{finding:#1}{(F}{)}}

\newcommand{\RQRef}[1]{RQ\ref{rq:#1}}
\newcommand{\ConstructRef}[1]{C\ref{construct:#1}}

\newcommand{\QuestionRef}[1]{Q\ref{question:#1}}
\newcommand{\FindingRef}[1]{F\ref{finding:#1}}

\newcommand{\threat}[3][min]{\emph{#1 $\lightning$: #2} / #3 \hfill\checkmark}
\newcommand{\threatPMit}[3][min]{\emph{#1 $\lightning$: #2} / #3 \hfill$\circ$}

\usepackage{doi}

\usepackage{acro}
\acsetup{format-include-endings,hyperref=false}
\DeclareAcronym{CMMI}{short = CMMI, long = {Capability Maturity Model Integration}}
\DeclareAcronym{FM}{short = FM, long = {formal method}}
\DeclareAcronym{SE}{short = SE, long = software engineering} 
\DeclareAcronym{TAM}{short = TAM, long = technology acceptance model} 
\DeclareAcronym{GQM}{short = GQM, long = goal-question-metric}
\DeclareAcronym{LTS}{short = LTS, long = labelled transition system}
\DeclareAcronym{RQ}{short = RQ, long = research question}
\DeclareAcronym{SWOT}{short = {SWOT}, long = {strengths, weaknesses, opportunities, and threats}}
\DeclareAcronym{SMT}{short = SMT, long = satisfiability modulo theory}
\DeclareAcronym{AEB}{short = AEB, long = academic educational background}
\DeclareAcronym{NEB}{short = NEB, long = unspecified educational background}
\DeclareAcronym{TLD}{short = TLD, long = top-level domain}
\DeclareAcronym{HQ}{short = HQ, long = head quarter}
\DeclareAcronym{EOU}{short = EOU, long = ease of use} 
\DeclareAcronym{MBE}{short = MbE, long = model-based engineering} 
\DeclareAcronym{PEOU}{short = PEOU, long = perceived ease of use} 
\DeclareAcronym{PU}{short = PU, long = perceived usefulness} 
\DeclareAcronym{UFM}{short = UFM, long = Use of FMs in
  mission-critical SE}
\DeclareAcronym{Usef}{short = U, long = usefulness} 
\DeclareAcronym{MlCh}{short = MC, long = multiple choice}
\DeclareAcronym{CI}{short = CI, long = confidence interval}
\DeclareAcronym{ICT}{short = ICT, long = information and communication
technology}
\DeclareAcronym{IS}{short = IS, long = information system}

\DeclareAcronym{II}{short = II, long = respondents with increased
  usage intent}
\DeclareAcronym{DI}{short = DI, long = respondents with decreased
  usage intent}
\DeclareAcronym{ME}{short = ME, long = more experienced respondents}
\DeclareAcronym{LE}{short = LE, long = less experienced respondents}
\DeclareAcronym{P}{short = P, long = practitioners}
\DeclareAcronym{NP}{short = NP, long = non-practitioners}
\DeclareAcronym{M}{short = M, long = respondents with some motivations to
  use FMs}
\DeclareAcronym{U}{short = NM, long = respondents without any motivations to
  use FMs}
\DeclareAcronym{PV}{short = PV, long = past view of a respondent}
\DeclareAcronym{FV}{short = FV, long = future view of a respondent}

\usepackage{preprint}

\def\bBlind{0}
\newcommand{\blind}[1]{\if\bBlind1{[source blinded]}\else{#1}\fi}

\usepackage{xwatermark}
\newwatermark[firstpage,color=gray!60,angle=90,scale=0.32, xpos=3.9in,ypos=0]{\href{https://arxiv.org/abs/1812.08815}{\color{gray}{1812.08815}}}
\newwatermark[firstpage,color=gray!90,angle=0,scale=0.28, xpos=0in,ypos=-5in]{*correspondence: \texttt{mario.gleirscher@york.ac.uk}}

\usepackage{authblk}

\begin{document}
\newcommand{\shorttitle}{Formal Methods Use}
\title{Formal Methods in Dependable Systems Engineering: A Survey of
  Professionals from Europe and North America\footnotemark} 
\author[1]{Mario Gleirscher}
\author[2]{Diego Marmsoler}
\affil[1]{Department of Computer Science, University of York,\\
  Deramore Lane, Heslington, York YO10 5GH, United Kingdom}
\affil[2]{Institut f\"ur Informatik, Technical University of Munich,\\
  Boltzmannstra{\ss}e 3, 85748 Garching, Germany}

  \maketitle
\begin{abstract}
  \emph{Context:} \Acp{FM} have been around for a while,
  still being unclear how to leverage their benefits, overcome their
  challenges, and set new directions for their improvement towards a
  more successful transfer into practice.
  \emph{Objective:} We study the use of formal methods in
  mission-critical software domains, examining industrial and academic
  views.
  \emph{Method:} We perform a cross-sectional on-line survey.
  \emph{Results:}
  Our results indicate an increased intent to apply \acp{FM} in industry,
  suggesting a positively perceived usefulness. But the results also
  indicate a negatively perceived ease of use.
  Scalability, skills, and education seem to be among the key
  challenges to support this intent.
  \emph{Conclusions:} We present the largest study of this kind so
  far~($N=216$), and our observations provide valuable insights,
  highlighting directions for future theoretical and empirical
  research of formal methods.  Our findings are strongly coherent with
  earlier observations by \citet{Austin1993-Formalmethodssurvey}.
\end{abstract}

   \vspace{1em}

\footnotetext{*
  \emph{Funding:} Partly funded by the Deutsche Forschungsgemeinschaft
  (DFG, German Research Foundation) under the Grant no.~381212925.
  \emph{Conflict of Interest:} The authors declare that they have no
  conflict of interest.
  This is a post-peer-review, pre-copyedit version of an article
  published in \emph{Empirical Software Engineering}. The final authenticated version
  is available online at:
  \doi{10.1007/s10664-020-09836-5}
}

\begin{multicols}{3}
\label{sec:nomenclature}
\footnotesize
\printacronyms
\normalsize
\end{multicols}

\section{Motivation and Challenges}
\label{sec:introduction}
\label{sec:motivation}

Over the past decades, many software errors have been deployed in the
field and some of these errors had a clearly intolerable
impact.\footnote{See anecdotal evidence~(grey literature, press
  articles) on software-related incidents, \eg by \citet{Kaner1998,
    Kaner2018}, \citet{Neumann2018,Charette2018}.}  Cost savings from
reducing such impact have been \emph{the} motivation of \acfp{FM} as a
first-class approach to error prevention, detection, and
removal~\citep{Holloway1997}.

In university courses on software engineering, we learned that
\acp{FM} are among the best we have to design and assure correct
systems.  The question ``Why are \acp{FM} not used more widely?''
\citep{Knight1997} is hence more than justified.  With a Twitter
poll,%
\footnote{See
  \url{https://twitter.com/MarioGleirscher/status/889737625178976256}.}
which emerged from our coffee spot discussions, we solicited opinions
on a timely paraphrase of a statement argued by \citet{Holloway1997}:
``\acp{FM} should be a cornerstone of dependability and security of
highly distributed and adaptive automation.''
What can a tiny opportunity sample of
22 respondents from our social network tell? Not much, well,
\begin{inparaenum}[(i)]
\item 55\% \emph{agree}s, \ies seem to attribute importance to this role of
  \acp{FM},
\item 14\% \emph{disagree}s, \ies oppose that view,
\item 32\% just \emph{don't know}.
\end{inparaenum}
Why should and how could \acp{FM} be a cornerstone?

Since the beginning of \ac{SE} there has been a
debate on the \emph{usefulness of \acp{FM}} to improve \ac{SE}.
In the 1970s and 1980s, several \ac{SE} and \ac{FM} researchers had started to
examine this usefulness and to identify error possibilities despite
the rigour in
\acp{FM}~\citep{Gerhart1976-ObservationsFallibilityApplications}, with the
aim of responding to critical observations of
practitioners~\citep{Jackson1987-PowerLimitationsFormal}.

\citet{Hall1990} and \citet{Bowen1995} illuminate 14 myths~(\egs
``formal methods are unnecessary''), providing their insights on when
\acp{FM} are best used and highlighting that \acp{FM} can be overkill in some
cases but are highly recommended in others.
The transfer of \acp{FM} into \ac{SE} practice is by far not straightforward.
\citet{Knight1997} examine reasons for the low adoption of \acp{FM} in
practice.  \citet{Barroca1992} ask: ``To what extent should \acp{FM} form
part of the [safety-critical \ac{SE}] method?''

\citet[pp.~148--149, 165--166]{Glass2002} and \citet{Parnas2010}
observe that ``many [\ac{SE}] researchers advocate rather than
investigate'' by assuming the need for more methodologies.
\citeauthor{Glass2002} summarises that \acp{FM} were supposed to help
represent firm requirements concisely and support rigorous
inspections\footnote{\Eg walking through development artefacts in a
  structured and moderated discussion group and with bug pattern
  checklists~\citep{Fagan1976-DesignCodeInspections}.}  and testing.
He observes that \emph{changing requirements} has become an
established practice even in critical domains, and inspections, even
if based on \acp{FM}, are insufficient for complete error removal.  In line
with \citet[p.~591]{Barroca1992}, %
he notes that \acp{FM} have occasionally been sold as to make error removal
complete, but there is no silver bullet~\citep[pp.~108--109]{Glass2002}.
Bad communication between theorists and practitioners sustains the
issue that \acp{FM} are taught but rarely
applied~\citep[pp.~68--70]{Glass2002,Holloway1996-Impedimentsindustrialuse}.
\citet{Parnas2010} compares alternative paradigms in \ac{FM}
research~(\egs axiomatic \vs relational calculi) and points to
challenges of \ac{FM} adoption~(\egs valid simple abstractions).

In contrast, \citet{Miller2010} draw positive conclusions from recent
applications of \emph{model checking} and highlight lessons learned.
In his keynote, \citet{OHearn2018} conveys positive experiences
in scaling \acp{FM} through adequate tool support for \emph{continuous
  reasoning} in agile projects~(see, \egs\cite{Chudnov2018}).
Many researchers~(see, \egs\cite{Aichernig2003}) have been working on
the improvement of \acp{FM} towards their successful transfer.
\citet{Boulanger2012} and \citet{Gnesi2013} summarise promising
industry-ready \acp{FM} and present larger case studies.

Have software errors been overlooked because of hidden inconsistencies
that can be detected when properly formalised?  Are such errors 
compelling arguments for the wider use of \acp{FM}?  Strong evidence for
\emph{the ease of use of \acp{FM} and their efficacy and usefulness} is
scarce and largely anecdotal, rarely drawn from \emph{comparative
  studies}~(\egs\cite{Pfleeger1997-Investigatinginfluenceformal,Sobel2002}),
often primarily conducted in research
labs~(\egs\cite{Galloway1998,Chudnov2018} and many others).
In late response to
\citeauthor{Holloway1996-Impedimentsindustrialuse}'s request for
empirical data~\citep{Holloway1996-Impedimentsindustrialuse},
\citet{Graydon2015} still observes a lack of evidence for the
effectiveness of \acp{FM} in assurance argumentation for safety-critical
systems, suggesting empirical studies to examine hypotheses and
collect evidence.

\acp{FM} have many potentials but \ac{SE} research has reached a stage
of maturity where strong empirical evidence is crucial for
\emph{research progress and transfer}.  \citet{Jeffery2015} identify
questions and metrics %
for \emph{\ac{FM} productivity assessment}, supporting \ac{FM}
research transfer.

\paragraph{Contributions.}
\label{sec:contributions}

We contribute to \ac{SE} and \ac{FM} research 
\begin{inparaenum}[(1)]
\item by presenting results of the largest cross-sectional survey of \ac{FM}
  use among \ac{SE} researchers and practitioners to this date,
\item by answering research questions about the past and intended use of
  \acp{FM} and the perception of 
  systematically mapped \ac{FM} challenges,
\item by relating our findings to the perceived ease of use and
  usefulness of \acp{FM} using a simplified variant of the technology
  acceptance model for evaluating engineering methods and techniques,
  and
\item by providing a research design for repetitive~(\egs longitudinal)
  \ac{FM} studies.
\end{inparaenum}

\paragraph{Overview.} 
\label{sec:overview}

The next section introduces important terms.  \Cref{sec:relwork}
relates our work to existing research.  In \Cref{sec:research-design},
we explain our research design.  We describe our data and answer our
research questions in \Cref{sec:results}.  In \Cref{sec:discussion},
we summarise and interpret our findings in the light of existing
evidence and \wrt threats to validity.  \Cref{sec:conclusions}
highlights our conclusions and potential follow-up work.

\section{Background and Terminology} 
\label{sec:background}

By \emph{formal methods}, we refer to \emph{explicit} mathematical
models and \emph{sound} logical reasoning about \emph{critical
  properties}~\citep{Rushby1994}---such as reliability, safety,
security, more generally, dependability and performance---of
electrical, electronic, and programmable electronic or software
systems in mission- or property-critical application domains.
Model checking, theorem proving, abstract interpretation, assertion
checking, and formal contracts are examples of \acp{FM}.  
By \emph{use of \acp{FM}}, we refer to their application in the
development and analysis of critical systems and to substantially
integrating \acp{FM} with the used programming methodologies~(\egs
structured development, \ac{MBE}, assertion-based programming,
test-driven development), notations~(\egs UML, SysML), and tools.

\paragraph{Tool and Method Evaluation.}

In the following, we give an overview of several evaluation approaches
and explain in \Cref{sec:construct} which approach we take.

The widely used \emph{\acl{TAM}}~(\acs{TAM};
\cite{Davis1989-PerceivedUsefulnessPerceived}) is a psychological test
that allows the assessment of end-user IT based on the two constructs
\emph{\acl{PEOU}}~(\acs{PEOU}, \ies positive and negative experiences
while using an IT system) and \emph{\acl{PU}}~(\acs{PU}, \ies positive
experiences of accomplishing a task using an IT system compared to not
using this system for accomplishing the same task).

Complementary to \ac{TAM},
\citet{Basili1985-Quantitativeevaluationsoftware} proposes the
\ac{GQM} approach to method and tool evaluation.
While \ac{GQM} serves as a good basis for quantitative follow-up studies,
we follow the user-focused \ac{TAM}.
Maturity models according to the
\acl{CMMI}~\citep{SEI2010-CMMIDevelopment} do not fit our purposes
because they focus on engineering process improvement beyond
particular development techniques.
\citet{Poston1992-Evaluatingselectingtesting} present tool survey
guidelines based on technology-focused classification and selection
criteria with a very limited view on tool usefulness and usability.
\citet{Miyoshi1993-empiricalstudyevaluating} evaluate \emph{ease of
  use} of development environments~(\ies specification and modelling
tools) using metrics from the ISO/IEC~9126 quality model.

From comparing two models 
of predicting an individual's intention to use a tool, 
\citet{Mathieson1991-PredictingUserIntentions} supports \ac{TAM}'s validity
and convenience but indicates its limits in providing enough
information on users' opinions.
For software methods and programming techniques,
\citet{Murphy1999-Evaluatingemergingsoftware} show how surveys, case
studies, and experiments can be used to compensate for this lack of
information about usefulness and usability.

Because \acp{FM} are by definition based on a formal language and usually
supported by tools, it is reasonable to adopt the \ac{TAM} for the
assessment of \acp{FM}.
Unfortunately, the body of literature on the evaluation of \acp{FM} in \ac{TAM}
style is very small.
However, \citet{Riemenschneider2002-Explainingsoftwaredeveloper} apply
\ac{TAM} %
to methods~(\egs UML-based architecture design), concluding that ``if
a methodology is not regarded as useful by developers, its prospects
for successful deployment may be severely undermined.''
According to their approach, \ac{FM} usage intentions would be driven by
\begin{inparaenum}[(1)]
\item an organisational mandate to use \acp{FM}, %
\item the compatibility of \acp{FM} with how engineers perform their work,
  and %
\item %
  the opinions of developers' coworkers and supervisors toward
  using \acp{FM}. %
\end{inparaenum}
Overall, the application of \ac{TAM} to \acp{FM} allows causal
reasoning from \ac{FM} user acceptance towards intention of \ac{FM} use.

Specialising the approach in
\citet{Riemenschneider2002-Explainingsoftwaredeveloper},
\emph{\ac{EOU}} of a \ac{FM} characterises the type and amount of
effort a user is likely to spend to learn, adopt, and apply this \ac{FM}.
\emph{Usefulness}~(U) determines how fit a \ac{FM} is for its purpose, \ie
how well it supports the engineer to accomplish an appropriate task.
If \ac{EOU} and U are measured by a survey whose data points are user
perceptions then we talk of \emph{\acf{PEOU}} and \emph{\acf{PU}}.
Together, \ac{PEOU} and \ac{PU} form the \emph{user acceptance of a
  \ac{FM}} %
and, by support of
\citet{Mathieson1991-PredictingUserIntentions,Riemenschneider2002-Explainingsoftwaredeveloper},
can predict the intention to use this \ac{FM}.

Whereas \ac{TAM} is a model based on the two user-focused constructs
\ac{PEOU} and \ac{PU},
\citet{Kitchenham1997-DESMETmethodologyevaluating} propose a
meta-evaluation approach called DESMET for tools and methods based on
multiple performance indicators~(\egs with \ac{TAM} as one of the
indicators).

\section{Related Work}
\label{sec:relwork}

\Cref{tab:rw} shows a systematic map~\citep{Petersen2008} of 36
studies on \ac{FM} research evaluation and transfer.  For each study,
we estimate\footnote{This estimate is based on opinions and attitudes
  expressed by the original authors and, where unavailable, on our own
  interpretation when reading the studies.} the authors' attitude
against or in favour of \acp{FM}, the motivation of the study, the
approach followed, and the type of result obtained.
Most of these works present personal experiences, opinions, case
studies, or literature summaries.  In contrast, the work presented in
this paper focuses on the analysis of experience from a wide range of
practitioners and experts.  However, we found four similar studies.

\citet{Austin1993-Formalmethodssurvey} sought to explain the low
acceptance of \acp{FM} in industry around 1992.  Using a questionnaire
similar to ours with only open questions, they evaluated 111 responses
from a sample of size 444, using a sampling method similar to
ours~(then using different channels).  Responses from \ac{FM} users
are distinguished from general responses.  Their questions examine
benefits, limitations, barriers, suggestions to overcome those
barriers, personal reasons for or against the use of \acp{FM}, and
ways of assessing \acp{FM}.

In a second study in 2001, \citet{Snook2001} conduct interviews with
representatives of five companies to discover the main issues involved
in \ac{FM} use, in particular, issues of understandability and the
difficulty of creating and utilising formal specifications.

A similar, though more comprehensive interview study was performed by
\citet{Woodcock2009} in 2009.  They assess the state of the art of the
application of \acp{FM}, using questionnaires to collect data on~62
industrial projects.

\citet[pp.~102--103]{Liebel2016-Modelbasedengineering} assess effects
on and shortcomings of the adoption of \ac{MBE} in embedded \ac{SE}
including a discussion of \ac{FM} adoption.  The authors observe a
lack of tool support, bad reputation, and rigid development processes
as obstacles to \ac{FM} adoption.  Their data suggests a need of
\ac{FM} adoption.  30\% of the responses from industry declare the
need for \acp{FM} as a reason to adopt \ac{MBE}.  Moreover, responses
indicate that \ac{MBE} adoption has a positive effect on \ac{FM} adoption.
One limitation of their study is the small number of responses from
\ac{FM} users.

While these studies focus on the elicitation of the state of the art
and the state of practice, the main focus of our study is to compare
the current \ac{FM} adoption or use with the intention to adopt and
use \acp{FM} in the future.  To the best of our knowledge, our study
offers the largest set of data points investigating the use of
\acp{FM} in \ac{SE}, so far.  In \Cref{sec:relevi}, we provide a
further discussion of how our findings relate to the findings of these
studies, particularly to the works of
\citet{Austin1993-Formalmethodssurvey,Woodcock2009}.

\begin{table}[t]
  \caption{Overview of related work on \ac{FM} use and adoption, grouped by
    primary focus and motivation
    \label{tab:rw}}
  \footnotesize
  \begin{tabularx}{\columnwidth}
    {Xlllccc}
    \toprule
    \textbf{Study}
    & \textbf{A}
    & \textbf{Motivation}
    & \textbf{Support}
    & \textbf{E}
    & \textbf{C}
    & \textbf{R}
    \\\midrule
    \multicolumn{7}{l}{\textbf{Surveys}}
    \\
    \citet{Austin1993-Formalmethodssurvey}
    & = & LoEv & Interviews & &\textbullet&\textbullet \\
    \citet{Snook2001}
    & = & LoEv & Interviews &\textbullet \\
    \citet{Oliveira2004}
    & = & Edu./Train. & Course websites &\textbullet \\
    \citet{Woodcock2009}$^a$
    & = & LoEv & Interviews & &\textbullet \\
    \citet{Davis2013-StudyBarriersIndustrial}
    & + & TechTx & Interviews & &\textbullet &\textbullet\\
    \citet{Liebel2016-Modelbasedengineering}
    & + & LoEv & Online questionnaire &\textbullet \\ 
    \citet{Ferrari2019-SurveyFormalMethods}
    & + & TechTx & Literature study &\textbullet \\
    \midrule
    \multicolumn{7}{l}{\textbf{Literature Studies and Summaries}} \\
    \citet{Wing1990-specifiersintroductionformal}
    & + & SotA & O/E &\textbullet&\textbullet \\
    \citet{Bloomfield1991}
    & = & SotA & &\textbullet \\
    \citet{Fraser1994-Strategiesincorporatingformal}
    & = & TechTx & & & &\textbullet \\
    \citet{Heitmeyer1998}
    & = & TechTx & & & &\textbullet \\
    \citet{Gleirscher2019-NewOpportunitiesIntegrated}
    & + & TechTx & \acs{SWOT} analysis & &\textbullet &\textbullet \\
    \midrule
    \multicolumn{7}{l}{\textbf{Expert Opinions and Experience Reports}} \\
    \citet{Jackson1987-PowerLimitationsFormal}
    & = & TechTx & & &\textbullet \\
    \citet{Bjorner1987}
    & = & TechTx & & &\textbullet \\
    \citet{Barroca1992}
    & = & SotA & Multiple cases & &\textbullet \\
    \citet{Bowen1995}
    & + & Hyp. Testing & & & &\textbullet \\
    \citet{Bowen1995a}
    & + & TechTx & & & &\textbullet \\
    \citet{Hinchey1996}
    & -- & TechTx & & &\textbullet \\
    \citet{Heisel1996}
    & + & TechTx & & & &\textbullet \\
    \citet{Holloway1996-Impedimentsindustrialuse}
    & + & LoEv & & &\textbullet \\
    \citet{Lai1996}
    & + & TechTx & & &\textbullet \\
    \citet{Bowen2005}
    & + & Hyp. Testing & Literature study & & &\textbullet \\
    \citet{Parnas2010}
    & = & TechTx & & &\textbullet&\textbullet \\
    \midrule
    \multicolumn{7}{l}{\textbf{Case Studies and Experiments}} \\
    \citet{Gerhart1976-ObservationsFallibilityApplications}
    & = & LoEv & Multiple cases, O/E &\textbullet &\textbullet &\textbullet \\
    \citet{Hall1990}
    & + & Hyp. Testing & O/E & & &\textbullet \\
    \citet{Craigen1995a}$^b$
    & + & SotA & Multiple cases, O/E &\textbullet \\
    \citet{Knight1997}
    & = & TechTx & Field experiment &\textbullet \\
    \citet{Pfleeger1997-Investigatinginfluenceformal}
    & = & Hyp. Testing & Effect analysis &\textbullet \\
    \citet{Galloway1998}
    & + & TechTx & Single case in lab &\textbullet&\textbullet \\
    \citet{Sobel2002}
    & = & Hyp. Testing & Lab experiment &\textbullet \\  %
    \citet{Miller2010}
    & = & TechTx & Multiple cases, O/E &\textbullet \\
    \citet{Klein2018-Formallyverifiedsoftware}
    & + & TechTx & &\textbullet \\
    \citet{Chudnov2018}
    & = & TechTx & &\textbullet \\  %
    \bottomrule	
  \end{tabularx}
  {\footnotesize
    $^a$See also \citet{Bicarregui2009},
    $^b$see also \citet{Craigen1993,Craigen1995};
    (A)ttitude,
    (E)valuation/analysis,
    (C)hallenges,
    (R)ecommendations,
    +/=/-- \dots positive/neutral/negative,
    LoEv \dots lack of empirical evidence,
    Hyp. Testing \dots hypotheses testing,
    Edu./Train. \dots education/training,
    O/E \dots opinion/experience report,
    SotA \dots state of the art,
    \acs{SWOT} \dots \acl{SWOT},
    TechTx \dots technology transfer
  }
\end{table}

\section{Research Method}
\label{sec:research-design}

In this section, we describe our research design, our survey
instrument, and our procedure for data collection and analysis.  For
this research, we follow the guidelines of \citet{Kitchenham2008} for
self-administered surveys and use our experience from a previous more
general survey~\citep{Gleirscher2018-safetysurvey}.

\subsection{Research Goal and Questions}
\label{sec:hypoth-rese-quest}

The questions in \Cref{sec:motivation} have led to this survey on the
\emph{use, usage intent, and challenges of \acp{FM}}.  Our interest is
devoted to the following \emph{\acp{RQ}}:

\begin{enumerate}
\item[\RQ{1}] In which typical domains, for which purposes, in which roles,
  and to what extent have \emph{\acp{FM} been used}?
\item[\RQ{2}] Which \emph{relationships} can we observe between \emph{past
    experience in using \acp{FM}} and the \emph{intent to use
    \acp{FM}}?
\item[\RQ{3}] How difficult do study participants perceive widely known \ac{FM}
  \emph{challenges} to be?
\item[\RQ{4}] What can we say about the \emph{perceived ease of use} and the
  \emph{perceived usefulness} of \acp{FM}?
\end{enumerate}

\subsection{Construct and Link to Research Questions} 
\label{sec:construct}

\begin{table}
  \footnotesize
  \caption{Concepts and scales for the construct ``Use
    of \acp{FM} in mission-critical \ac{SE}'' (\acs{UFM})
  \label{tab:constructs}}
  \begin{tabularx}{\linewidth}[b]{lc>{\hsize=.7\hsize}X>{\hsize=.3\hsize}X}
    \toprule
    \textbf{Concept}
    & \textbf{Id.}
    & \textbf{Description [Scale]}
    & \textbf{Point [Question]}
    \\\midrule
    \multicolumn{4}{l}{\emph{Measured twice \dots}}
    \\
    Domain
    & \Construct{AppDom}
    & Application domains of \acp{FM} [\acs{MlCh} among domains] %
    & Past [\QuestionRef{D1_appdom_past}], Intent [\QuestionRef{F1_appdom_future}]
    \\
    Role
    & \Construct{Role}
    & Role in using \acp{FM} [\acs{MlCh} among roles] %
    & Past [\QuestionRef{P1_role_past}], Intent [\QuestionRef{F2_role_future}]
    \\
    Use
    & \Construct{Use}
    & Use of \acp{FM} [experience level/relative frequency per \ac{FM} class] %
    & Past [\QuestionRef{P2_use_past}/\QuestionRef{P3_use_past}],
      Intent [\QuestionRef{F3_use_future}/\QuestionRef{F4_use_future}]
    \\
    Purpose
    & \Construct{Purpose}
    & Purpose of using \acp{FM} [absolute/relative frequency per purpose] %
    & Past [\QuestionRef{P4_purpose_past}], Intent
      [\QuestionRef{F5_purpose_future}]
    \\
    \multicolumn{4}{l}{\emph{Measured once \dots}}
    \\
    Experience
    & \Construct{ExpLev}
    & Level of \ac{FM} experience [duration ranges in years] %
    & Single [\QuestionRef{D2_explev}]
    \\
    Motivation
    & \Construct{Motiv}
    & Motivation to use \acp{FM} [degree per motivational factor] %
    & Single [\QuestionRef{D3_motiv}]
    \\
    Obstacles
    & \Construct{Obst}
    & Difficulty of obstacles to using \acp{FM} [degree per challenge] %
    & Single [\QuestionRef{O1_obst}]
    \\\bottomrule
  \end{tabularx}
  {\footnotesize MC\dots multiple-choice}
\end{table}

\Cref{tab:constructs} lists the (C)oncepts that constitute the construct
\emph{Use of \acp{FM} in mission-critical \ac{SE}}~(\acs{UFM}), the corresponding
\emph{scales}, the points of measurement, and references to (Q)ues\-tions
from the questionnaire.

\paragraph{Measuring Past and Intended Use.}

For \RQRef{1}~($\mathit{UFM}$), we examine potential application \emph{domains} for
\acp{FM}~(\ConstructRef{AppDom}), \emph{roles} when using
\acp{FM}~(\ConstructRef{Role}), \emph{motivations} and \emph{purposes} of
using \acp{FM}~(\ConstructRef{Motiv}, \ConstructRef{Purpose}), and the
extent of $\mathit{UFM}$ at the general~(\ConstructRef{ExpLev}) and
specific~(\ConstructRef{Use}) experience level of our study
participants when using \acp{FM}.

For \RQRef{2}, we compare the \emph{past}~($\mathit{UFM}_p$) and \emph{intended
  use}~($\mathit{UFM}_i$) of \acp{FM} regarding the
domain~(\ConstructRef{AppDom}), role~(\ConstructRef{Role}), \ac{FM}
class~(\ConstructRef{Use}), and purpose~(\ConstructRef{Purpose}).  We
measure $\mathit{UFM}_i$ by relative frequency~(\Cref{tab:scales}) \wrt a
participants' current situation, \ac{FM} class, and purpose of use.
Using a \emph{relative} instead of an \emph{absolute} frequency scale
slightly reduces the burden on respondents to make detailed and, hence,
uncertain predictions of $\mathit{UFM}_i$.

For \RQRef{3}, we measure the perception of difficulty of several
obstacles~(\ConstructRef{Obst}) known from the literature and from our
experience.

\paragraph{Method Evaluation and \ac{TAM}-style Interpretation.}
\label{sec:tam-style-ufm}

We follow DESMET~\citep{Kitchenham1997-DESMETmethodologyevaluating}
and \citet{Murphy1999-Evaluatingemergingsoftware} insofar as we combine
a \emph{qualitative survey}~(\ies \ac{FM} evaluation by \ac{SE} practitioners
and researchers)
and a \emph{qualitative effects analysis} based on the past and intent
measurements for \ConstructRef{Purpose}~(\ies subjective assessment of
effects of \acp{FM} by asking \ac{SE} practitioners and
researchers). %

We assume $\mathit{UFM}$ is, nowadays, to a large extent implying the use of
the tools automating the corresponding \acp{FM}.  This assumption is
justified inasmuch as for all \acp{FM} referred to in this survey,
tools are available.  In fact, in the past two decades~(the period
most survey respondents could have possibly used \acp{FM}), the
development of a \ac{FM} has mostly gone hand in hand with the
development of its supporting tools.

For \RQRef{4}, we associate our findings from \RQRef{2} and \RQRef{3} with \ac{PEOU} and
\ac{PU}.
Whereas \ac{TAM} predicts $\mathit{UFM}_i$ %
of a specific tool by measuring \ac{PEOU} and \ac{PU}, %
we directly interrogate past~(like in
\cite[Figure~2]{Mohagheghi2012-empiricalstudystate}) and intended use of
classes of \acp{FM}. %
We measure $\mathit{UFM}_i$~(\ConstructRef{AppDom},
\ConstructRef{Role}, \ConstructRef{Use}, \ConstructRef{Purpose}) in
more detail than \ac{TAM}. %
Our approach relates to \ac{TAM} for
methods~\citep[Table~2]{Riemenschneider2002-Explainingsoftwaredeveloper}
inasmuch as we collect data for \ac{PEOU} through asking about potential
obstacles to the further use of \acp{FM}~(\ConstructRef{Obst}) %
based on experience with past \ac{FM} use~($\mathit{UFM}_p$).  For this, respondents
are asked to rate the difficulty of several known challenges to be
tackled in typical \ac{FM} applications.  Furthermore, $\mathit{UFM}_i$ is known to
be correlated with \ac{PU}. %
We then interpret the answers to \RQRef{3} to examine the \ac{PEOU}
and, furthermore, interpret the answers to \RQRef{2} to reason about \ac{PU}.
In \Cref{sec:questionnaire}, we discuss our questionnaire including
the questions for measuring the sub-constructs.

\subsection{Study Participants and Population}
\label{sec:study-subj-popul}

Our target group for this survey includes persons with
\begin{inparaenum}[(1)]
\item an educational background in engineering and the sciences
  related to critical computer-based or software-intensive
  systems, preferably having gained their first degree,
  \emph{or}
\item a practical engineering background in a reasonably
  critical system or product domain involving software
  practice.
\end{inparaenum}
We use \emph{(study or survey) participant} and \emph{respondent} as
synonyms.  We talk of \emph{\ac{FM} users} to refer to the part of the
population that has already used \acp{FM} in one or another way.  See
\Cref{sec:rq1-extval} and \Cref{tab:respcat} for a more fine-grained
analysis of the population.

\subsection{Survey Instrument: On-line Questionnaire}
\label{sec:questionnaire}
\label{sec:legend}

\Cref{tab:questions} summarises the questionnaire we use to measure
$\mathit{UFM}$~(\Cref{tab:constructs}).  The scales used for encoding the answers
are described in~\Cref{tab:scales}.

Although we do not collect personal data, respondents could leave us
their email address if they want to receive our results.  We expect
participants to spend about 8 to 15 minutes to complete the
questionnaire.  However, we thought it to be unnecessary in our case
to instrument the questionnaire or our tooling to allow us to
determine the time spent for submitting complete data points.

\begin{table*}
  \caption{Summary of questions from the questionnaire
  \label{tab:questions}}
  \footnotesize
  \begin{tabularx}{\textwidth}{l>{\hsize=.7\hsize}X>{\hsize=.3\hsize}Xll}
    \toprule
    \textbf{Id.}
    & \textbf{Question or Question Template}
    & \textbf{Scale} (see \Cref{tab:scales})
    & \textbf{Sec.}
    & \textbf{Fig.}
    \\\midrule
    \Question{D1_appdom_past}
    & In which \emph{application domains}~(\ConstructRef{AppDom}) in industry or academia have you mainly used \acp{FM}?
    & \acs{MlCh} among domains
    & \ref{sec:D1_appdom}
    & \ref{fig:D1_appdom} 
    \\
    \Question{D2_explev} 
    & How many years of \emph{\ac{FM} experience}~(including the study of
      \acp{FM}, \ConstructRef{ExpLev}) have you gained?
    & Duration range in years
    & \ref{sec:D2_explev}
    & \ref{fig:D2_explev} 
    \\
    \Question{D3_motiv}
    & Which have been your \emph{motivations}~(\ConstructRef{Motiv}) to use \acp{FM}?
    & Degree per motivational factor
    & \ref{sec:D3_motiv}
    & \ref{fig:D3_motiv} 
    \\
    \Question{P1_role_past}
    & In which roles~(\ConstructRef{Role}) have you used \acp{FM}? 
    & \acs{MlCh} among roles
    & \ref{sec:sum-P1_role_past}
    & \ref{fig:P1_role_past} 
    \\
    \Question{P2_use_past}
    & Describe your \emph{level of experience}~(\ConstructRef{Use})
      for \emph{$\langle$class of formal description techniques$\rangle$}.
    & Level of experience per class
    & \ref{sec:sum-P2_use_past}
    & \ref{fig:P2_use_past} 
    \\
    \Question{P3_use_past}
    & Describe your \emph{level of experience}~(\ConstructRef{Use})
      for \emph{$\langle$class of formal reasoning techniques$\rangle$}.
    & Level of experience per class
    & \ref{sec:sum-P3_use_past}
    & \ref{fig:P3_use_past} 
    \\
    \Question{P4_purpose_past}
    & I have mainly \emph{used \acp{FM} for}~(\ConstructRef{Purpose}) ...
    & Absolute frequency per purpose
    & \ref{sec:sum-P4_purpose_past}
    & \ref{fig:P4_purpose_past} 
    \\
    \Question{F1_appdom_future}
    & In which \emph{domains}~(\ConstructRef{AppDom}) in
      industry or academia do you intend to use \acp{FM}? 
    & \acs{MlCh} among domains
    & \ref{sec:sum-F1_appdom_future}
    & \ref{fig:rq2_comparison_D1_F1} %
    \\
    \Question{F2_role_future}
    &In which \emph{roles}~(\ConstructRef{Role}) would~(or do) you intend to use \acp{FM}?
    & \acs{MlCh} among roles
    & \ref{sec:sum-F2_role_future}
    & \ref{fig:rq2_comparison_P1_F2} %
    \\
    \Question{F3_use_future}
    & I~(would) \emph{intend to use}~(\ConstructRef{Use}) \emph{$\langle$class of
      formal description techniques$\rangle$} \emph{$\langle$this$\rangle$} often.
    & Relative frequency per class
    & \ref{sec:sum-F3_use_future}
    & \ref{fig:F3_use_future} 
    \\
    \Question{F4_use_future}
    & I~(would) \emph{intend to use}~(\ConstructRef{Use}) \emph{$\langle$class of
      formal reasoning techniques$\rangle$} \emph{$\langle$this$\rangle$} often.
    & Relative frequency per class
    & \ref{sec:sum-F4_use_future}
    & \ref{fig:F4_use_future} 
    \\
    \Question{F5_purpose_future}
    & I~(would) intend to \emph{use \acp{FM} for}~(\ConstructRef{Purpose}) \emph{$\langle$purpose$\rangle$}.
    & Relative frequency per purpose
    & \ref{sec:sum-F5_purpose_future}
    & \ref{fig:F5_purpose_future} 
    \\
    \Question{O1_obst}
    & For any use of \acp{FM} in my future activities, I consider
      \emph{$\langle$obstacle$\rangle$}~(\ConstructRef{Obst}) as \emph{$\langle$that$\rangle$} difficult.
    & Degree of difficulty per obstacle
    & \ref{sec:sum-O1_obst}
    & \ref{fig:O1_obst} 
    \\\bottomrule	
  \end{tabularx}
  {\footnotesize MC\dots multiple-choice}
\end{table*}

\begin{table}
  \caption{Scales used in the questionnaire
    \label{tab:scales}}
  \footnotesize
  \begin{tabularx}{1.0\linewidth}{>{\hsize=.15\hsize}X>{\hsize=.7\hsize}Xl}
    \toprule
    \textbf{Name}
    & \textbf{Values}
    & \textbf{Type} \\
    \midrule
    \emph{degree of\newline motivation}
    & \textbf{``no motivation''},
      ``moderate motivation'',
      ``strong motivation (or requirement)''
    & L3
    \\
    \emph{degree of\newline difficulty}
    & ``not as an issue.'',
      ``as a moderate challenge.'',
      ``as a tough challenge.'',
      \textbf{``I don't know.''}
    & L3
    \\\midrule
    \emph{experience level\newline (duration-based)}
    & \textbf{``I do not have any knowledge of or experience in \acp{FM}.''},
      ``less than 3 years'',
      ``3 to 7 years'',
      ``8 to 15 years'',
      ``16 to 25 years'',
      ``more than 25 years''
    & O
    \\
    \emph{experience level\newline (task-based)}
    & \textbf{``no experience or no knowledge''},
      ``studied in (university) course'',
      ``applied in lab, experiments, case studies'',
      ``applied once in engineering practice'',
      ``applied several times in engineering practice''
    & O
    \\\midrule
    \emph{frequency\newline (absolute)}
    & \textbf{``not at all.''},
      ``once.'',
      ``in 2 to 5 separate tasks.'',
      ``in more than 5 separate tasks.''
    & O
    \\
    \emph{frequency\newline (relative)}
    & \textbf{``no more or not at all.''},
      ``less often than in the past.'',
      ``as often as in the past.'',
      ``more often than in the past.'',
      \textbf{``I don't know.''}
    & L3
    \\\midrule
    \emph{choice}
    & single/multiple: (\emph{ch})ecked, (\emph{un})checked
    & N
    \\\bottomrule
  \end{tabularx}
  
  {\footnotesize 
    bold\dots express lack of knowledge or indecision;
    (N)ominal, (O)rdinal, Ln \dots Likert-type
    scale with $n$ values}
\end{table}

\paragraph{Face and Content Validity.}

We derived answer options from the literature, our own experience with
\acp{FM}, \ac{SE} research training, discussions with other \ac{SE}
researchers and colleagues from industry, pilot responses, and coding
of open answers. %
Particularly, the classification of \ac{FM}
methods~(\ConstructRef{Use}; \QuestionRef{P2_use_past},
\QuestionRef{P3_use_past}, \QuestionRef{F3_use_future},
\QuestionRef{F4_use_future}) and the list of obstacles or
challenges~(\ConstructRef{Obst}; \QuestionRef{O1_obst}) were derived
from our own training, literature knowledge prior to this study, and
experience as well as from occasional personal discussions with
\ac{SE} experts from academia and industry.
Most questions are half-open, allowing respondents to go beyond given
answer options.
We treat \emph{degree} and \emph{relative frequency} as 3-level
Likert-type scales.

For each question, we provide \emph{``do not know''~(dnk)}-options to
include participants without previous knowledge of \acp{FM} in any academic
or practical context.  If participants are not able to provide an
answer they can choose, \egs ``do not know'', ``not yet used'', ``no
experience'', or ``not at all'', and proceed.  This way, we reduce
bias by forced response.
We indicate \emph{dnk}-answers whenever we \emph{exclude}
them.  Our questionnaire tool~(\Cref{sec:tooling}) supports us
\emph{with getting complete data points}, reducing the effort to
deal with missing answers.

\subsection{Data Collection: Sampling Procedure}
\label{sec:daten-collection}

We could not find an open, non-commercial panel of engineers.
Large-scale \emph{panel services} are either
commercial~(\egs\cite{DecisionAnalyst2018}) or they do not allow the
sampling of software engineers~(\egs\cite{Leiner2014}).  Hence, we opt
for a mixture of opportunity, volunteer, and cluster-based sampling.
To draw a diverse sample of potential \ac{FM} users, we
\begin{enumerate}
\item advertise our survey on various on-line discussion channels,
\item invite software practitioners and researchers from our social
  networks, and
\item ask these people to disseminate our survey.
\end{enumerate}
\label{sec:representation}
We examine \ConstructRef{ExpLev}, \ConstructRef{AppDom},
\ConstructRef{Role}, and \ConstructRef{Use} from \Cref{tab:constructs}
to check how well our \emph{sample covers the given concept
  categories}.  The better the coverage of these categories the wider
is the range of analyses possible from our data set.  Less covered
categories might indicate inappropriate concepts as well as the case
that our sample just does not touch this fraction of the target
population.  Under the assumption that the sample is drawn from the
target population in a uniformly random fashion, we would be able to
draw conclusions about the constitution of the target population.
However, as noted, this assumption was in our case not
controllable.

\begin{table}
  \sidecaption
  \footnotesize
  \begin{tabularx}{.7\columnwidth}[b]{Xl}
    \toprule
    \textbf{Channel Type} & \textbf{Examples \& References}
    \\\midrule
    General panels & SurveyCircle, \url{www.surveycircle.com} \\ 
    LinkedIn groups & \Egs on ARP 4754, DO-178, FME, ISO 26262  \\
    Mailing lists & \Egs system safety (U Bielefeld, formerly U York) \\
    Newsletters & BCS FACS; GI RE, SWT, TAV \\
    Personal pages & \Egs Facebook, Twitter, LinkedIn, Xing \\
    ResearchGate & Q\&A forums on \url{www.researchgate.net} \\
    Xing groups & \Egs Safety Engineering, RE \\
    \bottomrule
  \end{tabularx}
  \caption{Channels used for sampling 
    \label{tab:adplatforms}}
\end{table}

\subsection{Data Evaluation and Analysis}
\label{sec:data-analysis}
\label{sec:proc:rqs}

For \RQRef{1}, we summarise the data and apply descriptive statistics for
categorical and ordinal variables in \Cref{sec:summ-answ-quest}.  We
answer \RQRef{2} by comparison of the data for the past and future
views regarding the domain~(\ConstructRef{AppDom}),
role~(\ConstructRef{Role}), \ac{FM} class~(\ConstructRef{Use}), and
purpose~(\ConstructRef{Purpose}) in \Cref{sec:analysis-rq2}.
Then, in \Cref{sec:analysis-rq3}, we answer \RQRef{3} by
\begin{itemize}
\item describing the \emph{challenge difficulty ratings} after
  associating one of
  \begin{inparaenum}[(1)]
  \item domain,
  \item motivational factor,
  \item role, 
  \item purpose, and
  \item \ac{FM} class
  \end{inparaenum}
  with challenge~(\ConstructRef{Obst}) and
\item distinguishing
  \begin{inparaenum}[(1)]
  \item more experienced~(\acs{ME}, $> 3$ years) from less experienced
    respondents~(\acs{LE}, $\leq 3$ years),
  \item practitioners~(\acs{P}, practised at least once) from
    non-practitioners (\acs{NP}, not used or only in course or lab),
  \item motivated~(\acs{M}, moderately or strongly motivated by at least one
    specified factor) from unmotivated respondents~(\acs{U}, no motivating
    factor specified),
  \item respondents' past and future views, and
  \item respondents with increased usage
    intent~(\acs{II}) from ones with decreased usage
    intent~(\acs{DI}).
  \end{inparaenum}
\end{itemize}
We apply association analysis between these categorical and ordinal
variables, using \emph{pairs of
  matrices}~(\egs\Cref{fig:heatmapP2P3O1}).
We answer \RQRef{4} by arguing from results for \RQRef{1}, 2, and 3.

\paragraph{Half-Open and Open Questions.}
\label{sec:openquestions}

We code open answers in additional text fields as follows: If we can
subsume an open answer into one of the given options, we add a
corresponding rating~(if necessary).  If we cannot do this then we
introduce a new category ``Other'' and estimate the rating.  Finally,
we cluster the added answers and split the ``Other'' category~(if
necessary).  For \QuestionRef{O1_obst}, we performed the latter step
combined with independent coding~\citep{Neuendorf2016} to confirm that
the understanding of the challenge categories is consistent among the
authors of the present study.
For \acs{MlCh} questions, we eliminate the choice of ``I do/have
not\dots'' options from the data if ordinary answer options where also
checked.

\paragraph{Tooling.}
\label{sec:tooling}

We use Google Forms~\citep{Google2018} for implementing our
questionnaire~(\Cref{sec:questionnaire-complete}) and for data
collection~(\Cref{sec:daten-collection}) and storage.  For statistical
analysis and data visualisation~(\Cref{sec:data-analysis}), we use GNU
R~\citep{RProject2018}~(with the packages \texttt{likert},
\texttt{gplots}, and \texttt{ggplot2} and some helpers from the
``Cookbook for R'' and the ``Stack Exchange Stats''
community\footnote{See \url{http://www.cookbook-r.com} and
  \url{https://stats.stackexchange.com}.}).
Content analysis and coding takes place in a spreadsheet application.
A draft of \Cref{sec:appendix} has been archived in
\citet{Gleirscher2018-fmsurvey-supl}.

\section{Execution, Results, and Analysis}
\label{sec:results}

In this section, we summarise the responses to the questions in
\Cref{tab:questions} and answer the \acp{RQ}~1, 2, and 3 as explained
in \Cref{sec:hypoth-rese-quest}.  To answer \RQRef{1}, we describe the
sample in \Cref{sec:descr-data-points} and discuss some facets of
\ac{FM} use in \Cref{sec:summ-answ-quest}.  For \RQRef{2}, we summarise data
about past use and usage intent in \Cref{sec:sensitivity-analysis}.
For \RQRef{3}, we analyse further data in \Cref{sec:analysis-rq3}.

\subsection{Survey Execution}
\label{sec:survey-execution}

\begin{figure}
  \includegraphics[width=\linewidth]{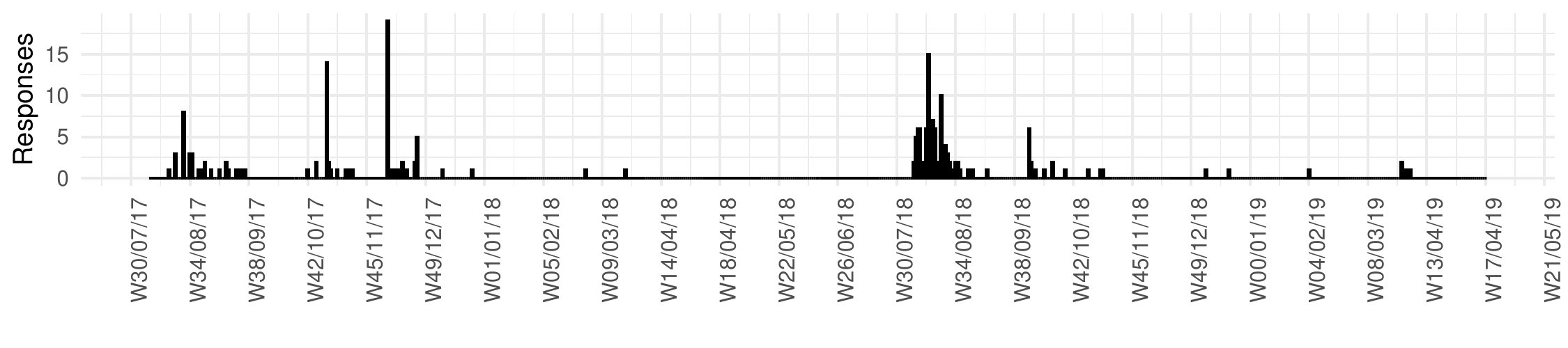}
  \caption{Distribution of responses over time
  \label{fig:responsehist}}
\end{figure}

For data collection, we
\begin{inparaenum}[(1)]
\item advertised our survey on the channels in
  \Cref{tab:adplatforms} and
\item personally invited $>$ 30 persons.
\end{inparaenum}
The sampling period lasted \emph{from August 2017 til March 2019}.
In this period, we \emph{repeated step 1 up to three times} to
increase the number of participants.  \Cref{fig:responsehist}
summarises the distribution of responses.  The channels in
\Cref{tab:adplatforms} particularly cover the European and North
American areas.

\subsection{Description of the Sample (answering RQ~1)}
\label{sec:descr-data-points}

A size estimation of the channels in \Cref{tab:adplatforms} yields
around 65K \emph{channel memberships}~(for some channels we make a
best guess but, \egs for LinkedIn the counts are given).  Assuming
participants are, on average, member of at least three of the
channels, we could have \emph{reached} up
to~20K %
\emph{real persons}.
Given a recent estimate of worldwide 23 million \ac{SE}
practitioners~\citep{EvansData2018}
and assuming that at least 1\% are mission-critical \ac{SE}
practitioners, our \emph{population} might comprise at least 230K
persons, possibly around 38K in the US and 61K in Europe.%
\footnote{An estimation in
  \citet{Gleirscher2019-NewOpportunitiesIntegrated} suggests that
  about 5\% of the overall \acs{ICT}/\acs{IS} developer population are
  embedded systems practitioners in critical and non-critical domains.
  Moreover, \citet{EvansData2018} and \citet{wiki:swedemographics}
  describe data from 2016 and 2017, suggesting that 3.87
  million~(19\%) \ac{SE} practitioners live in the US and about 13.3
  million~(39\%) in Europe, the Middle East, and Africa.  According to
  an analysis of data from Stack Overflow by
  \citet{ATOMICO2019-StateEuropeanTech}, there is a ``software
  engineering talent pool'' of about 6.1 million in Europe.}
We received $N=216$ responses resulting in an estimated \emph{response
  rate} between 1 and 2\% and a \emph{population coverage} of at
most 0.1\% globally and 0.2\% in the US and in
Europe. %
About 40\% of our respondents %
provided their email addresses, the majority from the
US, UK, Germany, France, and a sixth from other EU and non-EU
countries.

In the following, we summarise the responses to the questions about
the application domain~(\QuestionRef{D1_appdom_past}), the level of
experience~(\QuestionRef{D2_explev}), and the
motivations~(\QuestionRef{D3_motiv}) of a \ac{FM} user.

\paragraph{Guide to the Figures.}
\label{sec:guide2fig}

For Likert-type ordered \emph{scales}, we use centred
diverging stacked bar charts~(see, \egs \Cref{fig:D3_motiv}) as
recommended by~\citet{Robbins2011-PlottingLikertOther}.
The \emph{horizontal bars} in each line show the answer fractions
according to the legend at the bottom and are annotated with the
percentages of the left-most, middle, and right-most answer options.
These bars are aligned by the midpoint of the middle group (for 3- and
5-level scales) or by the boundary between the two central groups (for
4-level scales).
\emph{Bar labels} often abbreviate the corresponding answer options in
the questionnaire.  The questionnaire copy in
\Cref{sec:questionnaire-complete} contains short definitions,
explanations, and examples to clarify the answer options.  For sake of
brevity, we do not repeat this information here.
``M'' denotes the median, ``CI'' the 95\% confidence interval for the
median calculated according to
\citet{Campbell1988-Calculatingconfidenceintervals}, ``X'' the number
of excluded data points per answer option, and ``NA'' the number of
invalid data points.

\paragraph{\QuestionRef{D1_appdom_past}: Application Domain.}
\label{sec:D1_appdom}

\begin{figure}
  \sidecaption
  \includegraphics[width=.7\columnwidth]{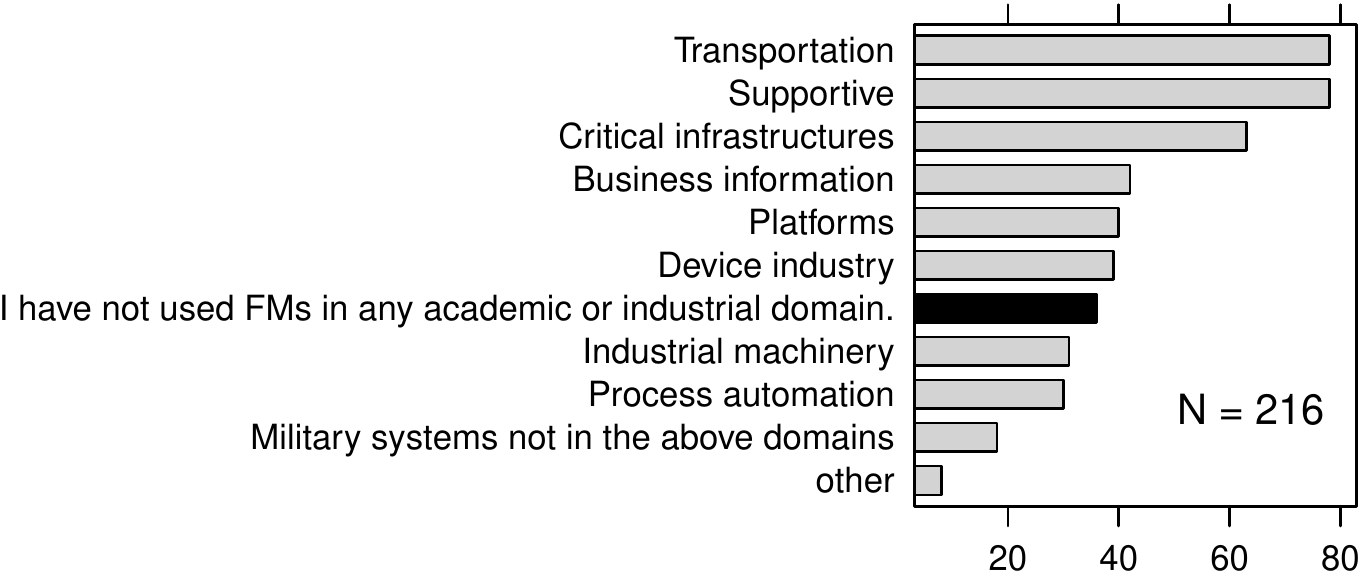}
  \caption{(\QuestionRef{D1_appdom_past})
    In which application domains in industry or academia
    have you mainly used \acp{FM}?~(\acs{MlCh})
    \label{fig:D1_appdom}}
\end{figure}

For each domain, \Cref{fig:D1_appdom} shows the number of participants
having experience in that domain.\footnote{\ac{MlCh} entails that the sum of
  answers can exceed $N$.}  Note that 180 %
of the respondents do have experience with applying \ac{FM} in
different industrial contexts, while only 36 have not applied \acp{FM}
to any application domain.  Medical healthcare is an example where
participants could have checked more than one answer category because
medical devices would belong to ``device industry'' and emergency
management IT would belong to ``critical infrastructures''.  See
\Cref{sec:questionnaire-complete} for more information about the
answer categories.

\paragraph{\QuestionRef{D2_explev}: FM Experience.}
\label{sec:D2_explev}

\begin{figure}
  \sidecaption
  \includegraphics[width=.5\columnwidth]{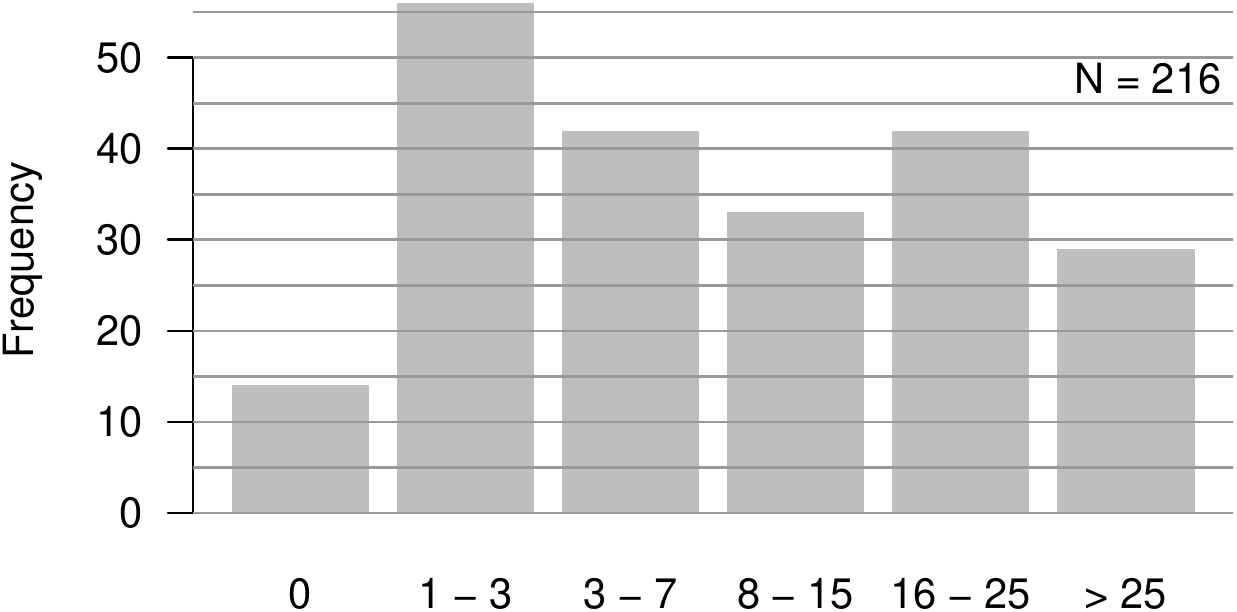}
  \caption{(\QuestionRef{D2_explev})
    How many years of \ac{FM} experience~(including the study of \acp{FM}) have you gained? 
    \label{fig:D2_explev}}
\end{figure}

\begin{figure}
  \sidecaption
  \includegraphics[width=.7\columnwidth]{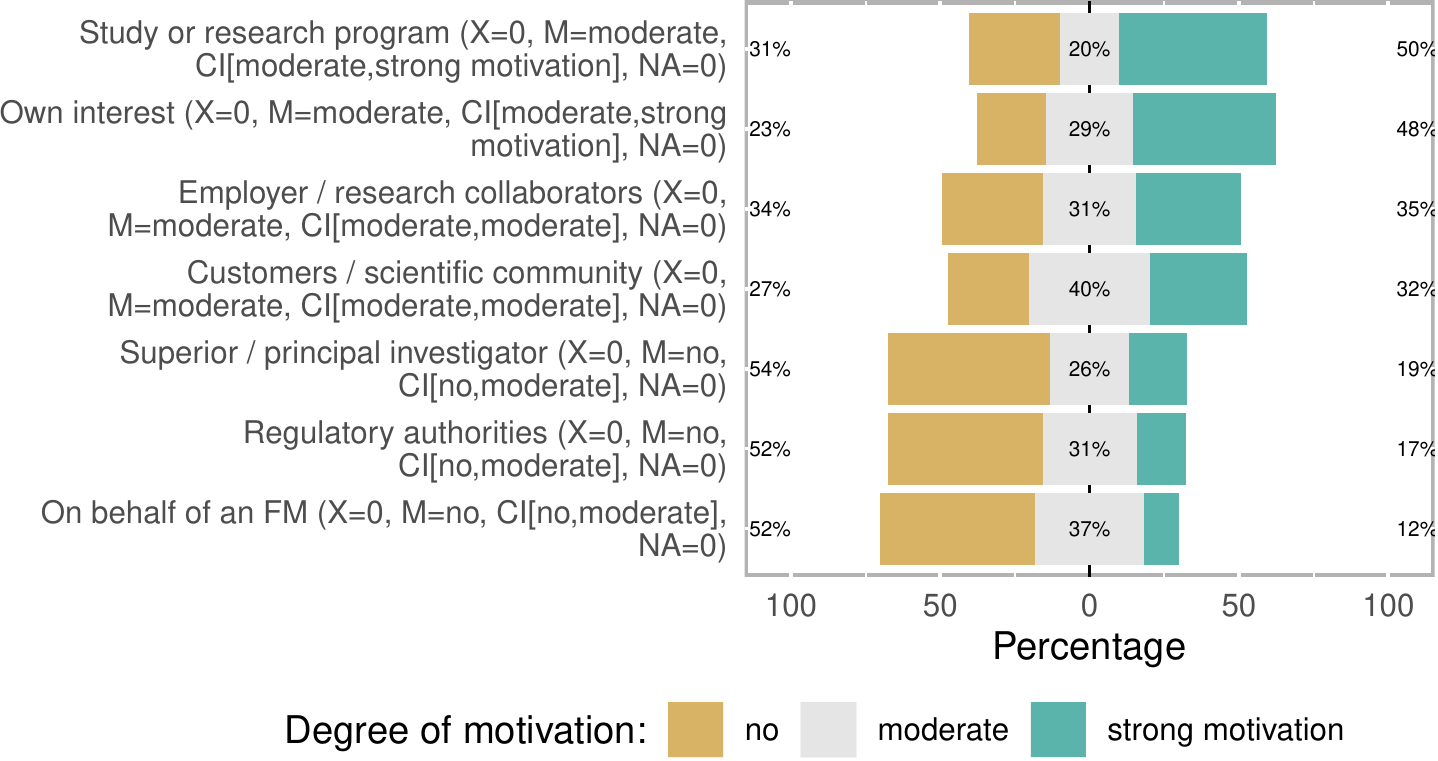}
  \caption{(\QuestionRef{D3_motiv})
    Which have been your motivations to use \acp{FM}?
    \label{fig:D3_motiv}}
\end{figure}

\Cref{fig:D2_explev} depicts participants' years of experience in
using \acp{FM}, showing that the sample covers all experience levels.
However, the fraction of respondents with no experience~(\ies category
``0'') is comparatively low.  According to \Cref{sec:data-analysis},
one third of the participants can be considered \acp{LE} with up to three
years of experience, and two thirds can be considered \acp{ME} with at
least three years of experience~($29$ of those with even more than
$25$ years).
A further analysis of the study participants' experience profile is
available
from \Cref{tab:respcat} in \Cref{sec:rq1-extval} on \cpageref{tab:respcat}.

\paragraph{\QuestionRef{D3_motiv}: Motivation.}
\label{sec:D3_motiv}

\Cref{fig:D3_motiv} suggests that \emph{regulatory authorities} play
a subordinate role in triggering the use of \acp{FM}.  In contrast,
\emph{intrinsic motivation}~(in terms of private interest) seems to be
the major factor for using \acp{FM}.  For 9 respondents, none of the given
factors was motivating at all.  The 88 open responses for this
question %
could either be subsumed in at least one of the given categories~(65
in ``Own~(private) interest'', 11 in other categories) or be declared
as a comment~(3) or not a further motivation~(9).  Hence, coding did
not require an additional answer category to \QuestionRef{D3_motiv}.

\subsection{Facets of Formal Methods Use  (answering RQ~1)}
\label{sec:summ-answ-quest}

In the following, we summarise the responses to the questions about
the role of a user~(\QuestionRef{P1_role_past}), use in
specification~(\QuestionRef{P2_use_past}), use in
analysis~(\QuestionRef{P3_use_past}), and the underlying
purpose~(\QuestionRef{P4_purpose_past}) of such use.

\paragraph{\QuestionRef{P1_role_past}: Role.}
\label{sec:sum-P1_role_past}

\begin{figure}
  \sidecaption
  \includegraphics[width=.7\columnwidth]{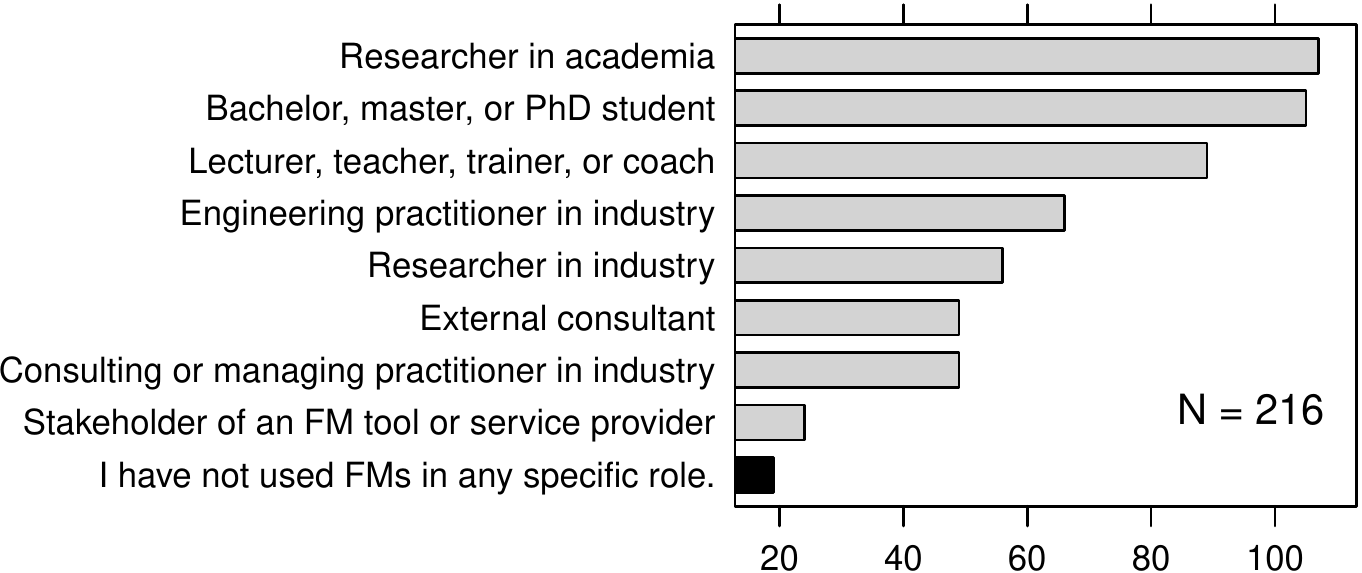}
  \caption{(\QuestionRef{P1_role_past})
    In which roles have you used \acp{FM}?~(\acs{MlCh})
    \label{fig:P1_role_past}}
\end{figure}

\Cref{fig:P1_role_past} shows in which roles the respondents applied
\acp{FM}.  An analysis of the \ac{MlCh} answers shows that 72\% of the
participants used \acp{FM} in an \emph{academic environment}, as a
researcher, lecturer, or student.  50\% of the participants applied
\acp{FM} in \emph{practice}, as an engineer or consultant~(see
also %
\cite{Gleirscher2018-fmsurvey-supl}).

\paragraph{\QuestionRef{P2_use_past}: Use in Specification.}
\label{sec:sum-P2_use_past}

\begin{figure}
  \sidecaption
  \includegraphics[width=.7\columnwidth]{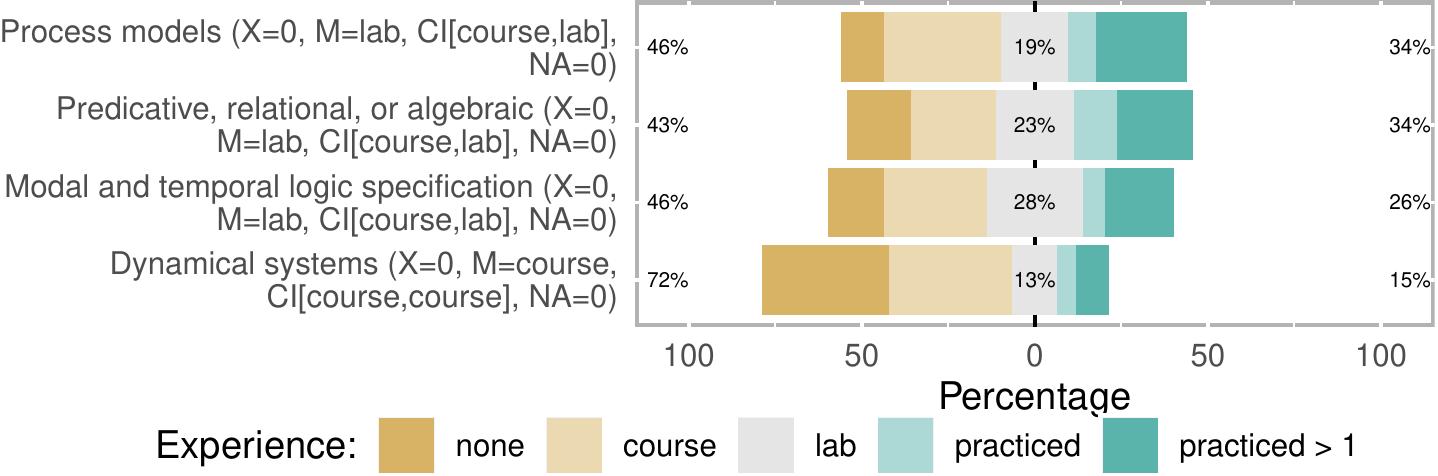}
  \caption{(\QuestionRef{P2_use_past})
    Describe your level of experience with each of the following classes of formal description techniques.
    \label{fig:P2_use_past}}
\end{figure}

The degree of usage of \acp{FM} for specification is depicted in
\Cref{fig:P2_use_past}.  There is an \emph{almost balanced} proportion
between theoretical and practical experience with the use of various
specification techniques.  Only the use of \acp{FM} for the description of
dynamical systems seems to be remarkably low.

\paragraph{\QuestionRef{P3_use_past}: Use in Analysis.}
\label{sec:sum-P3_use_past}

\begin{figure}
  \sidecaption
  \includegraphics[width=.7\columnwidth]{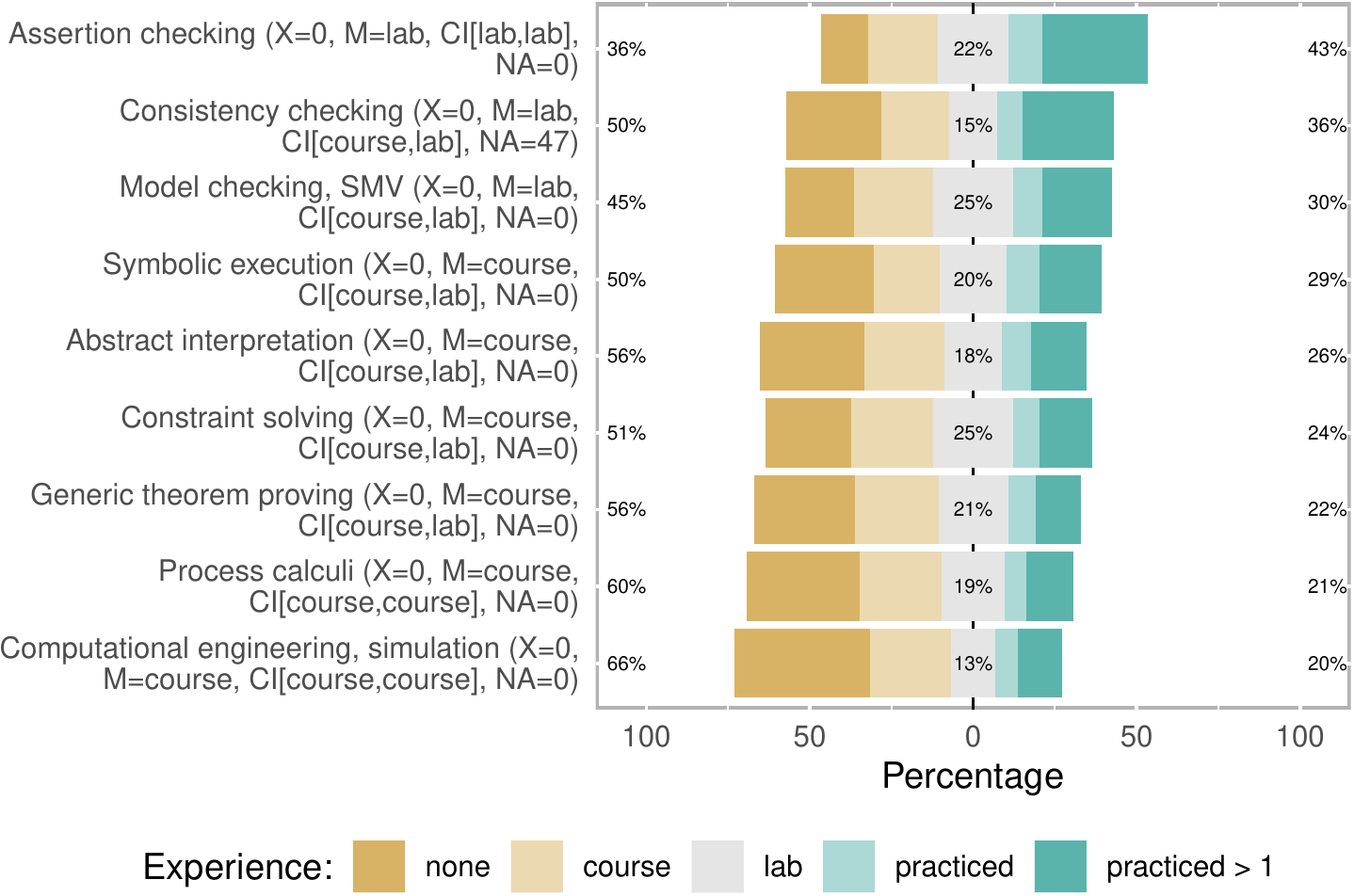}
  \caption{(\QuestionRef{P3_use_past})
    Describe your level of experience with each of the following
    classes of formal reasoning techniques. 
    \label{fig:P3_use_past}}
\end{figure}

The use of \acp{FM} for analysis is depicted in \Cref{fig:P3_use_past}.
Similar to specification techniques, we observe an \emph{almost
  balanced} proportion between theoretical and practical experience
with the usage of various analysis techniques.  Outstanding is the use
of assertion checking techniques, such as contracts.
As expected from the observations for \QuestionRef{P2_use_past}, the
use of \acp{FM} in computational engineering, such as algebraic reasoning
about differential equations, is again exceptionally low.

\paragraph{\QuestionRef{P4_purpose_past}: Purpose.}
\label{sec:sum-P4_purpose_past}

\begin{figure}
  \sidecaption
  \includegraphics[width=.7\columnwidth]{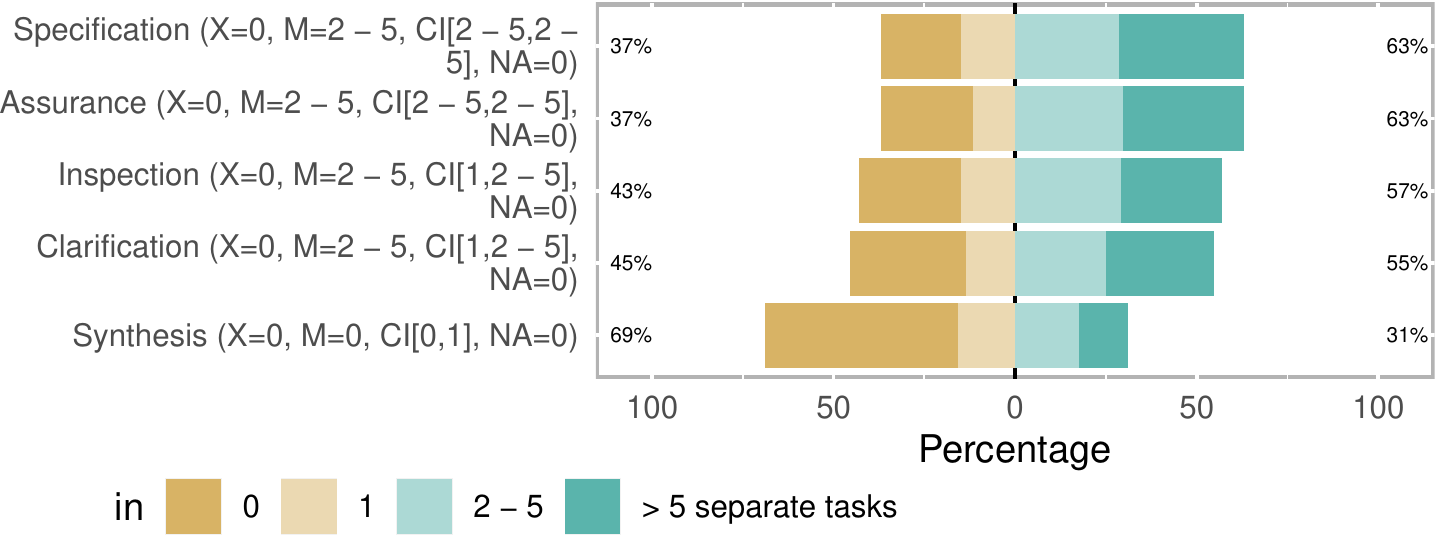}
  \caption{(\QuestionRef{P4_purpose_past})
    I have mainly used \acp{FM} for ... 
    \label{fig:P4_purpose_past}}
\end{figure}

\Cref{fig:P4_purpose_past} depicts the participants' purposes to apply
\acp{FM}.  It seems that the respondents employ \acp{FM} mainly for assurance, %
specification, and inspection.  Synthesis, on the other hand, to them
seems to be only a subordinate purpose in the use of \acp{FM}.

\subsection{Past Use versus Usage Intent  (answering RQ~2)}
\label{sec:sensitivity-analysis}
\label{sec:analysis-rq2}

We investigate the usage intent of \acp{FM} across various domains and
roles as well as the participants' intent to use various \acp{FM} and their
intended purpose to use \acp{FM}.

\paragraph{Application Domain.}
\label{sec:sum-F1_appdom_future}

\begin{figure}
  \includegraphics[width=\columnwidth]{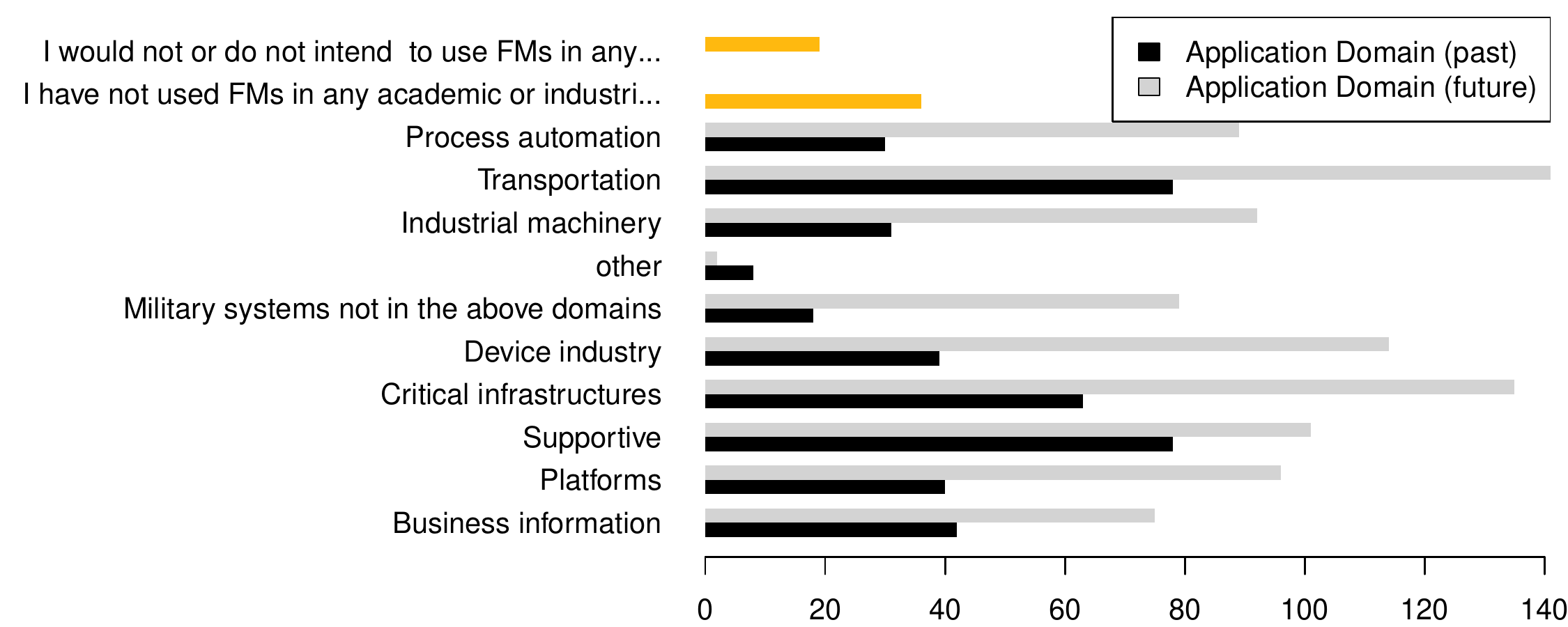}
  \caption{Number of respondents using \acp{FM} by domain~(past \vs intent)
    \label{fig:rq2_comparison_D1_F1}}
\end{figure}

\Cref{fig:rq2_comparison_D1_F1} compares the respondents' past 
domains of \ac{FM} application with their intended
domains~(see~\QuestionRef{F1_appdom_future}).  This figure reveals two
insights into the participants' intentions to use \acp{FM}:
\begin{inparaenum}[(i)]
\item Fewer participants do not want to apply \acp{FM} in the
  future~(19) than participants that have not used \acp{FM}~(36, see
  yellow bars).  Ten participants fall into both categories, they have
  not used \acp{FM} and do not intend to use \acp{FM}.
\item The intended application of \acp{FM} outperforms the current
  application of \acp{FM} across \emph{all} domains.  Hence, there is a
  tendency to increase the use of \acp{FM} across all application domains.
\end{inparaenum}

\paragraph{Role.}
\label{sec:sum-F2_role_future}

\begin{figure}
  \includegraphics[width=\columnwidth]{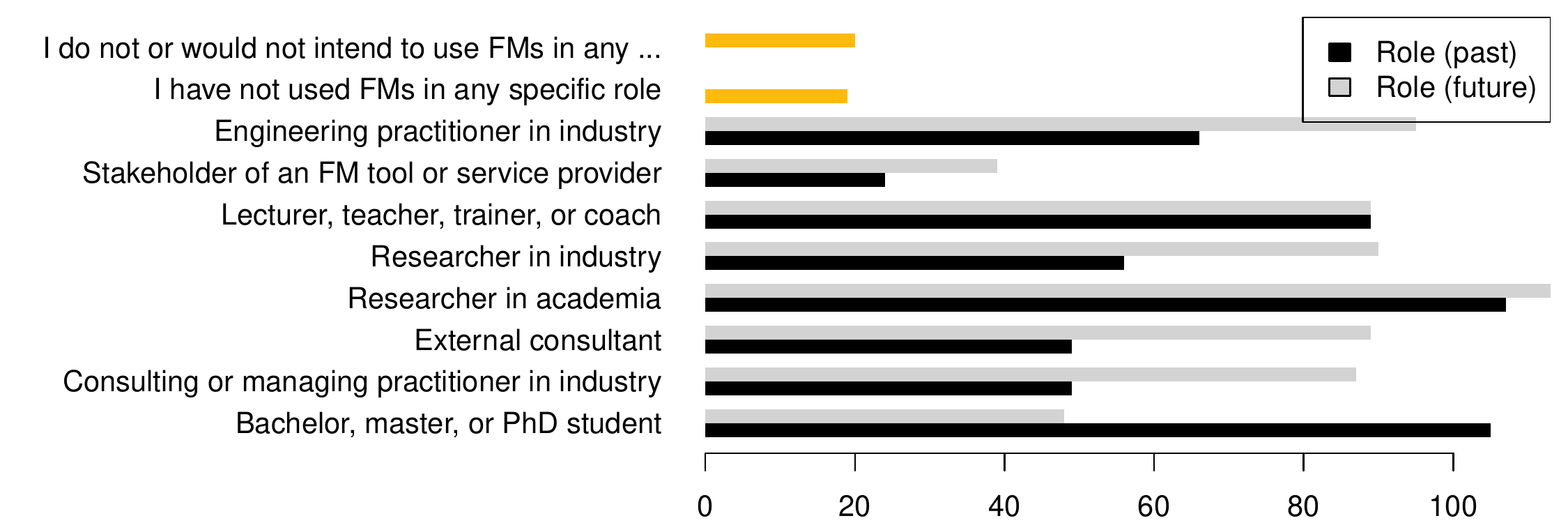}
  \caption{Number of respondents applying \acp{FM} by role~(past \vs intent)
    \label{fig:rq2_comparison_P1_F2}}
\end{figure}

\Cref{fig:rq2_comparison_P1_F2} compares the participants' roles in
which they applied \acp{FM} in the past with their intended role to apply
\acp{FM} in the future~(see~\QuestionRef{F2_role_future}).  Similar to the
results for the application domain, we observe that some participants,
who have not applied \acp{FM} in any role so far, intend to apply such
methods in the future.  However, the comparison reveals that
\emph{academic} disciplines~(\ies researcher and lecturer) seem to
be \emph{stable}.  There is only a small difference between the number
of participants who applied \acp{FM} in academic domains in the past and
the number of participants who want to apply such methods to these
domains in the future.

In contrast, there is a \emph{significant} increase in the number of
participants aiming to apply \acp{FM}, across all \emph{industrial} roles.

Furthermore, the diagram shows a strong contrast between past and
indented use in the category ``Bachelor, master, or PhD student.'' We
can see several reasons for this difference.  From the respondents who
``used \acp{FM} as a student,'' many (i) might not be able to ``use \acp{FM} as
a student'' anymore because of having graduated, (ii) did not find \acp{FM}
or the way \acp{FM} were taught as helpful, or (iii) moved into a business
domain with no foreseeable demand for the application of \acp{FM}.
   
\paragraph{\QuestionRef{F3_use_future}: Intended use for Specification.}
\label{sec:sum-F3_use_future}

\begin{figure}
  \sidecaption
  \includegraphics[width=.7\columnwidth]{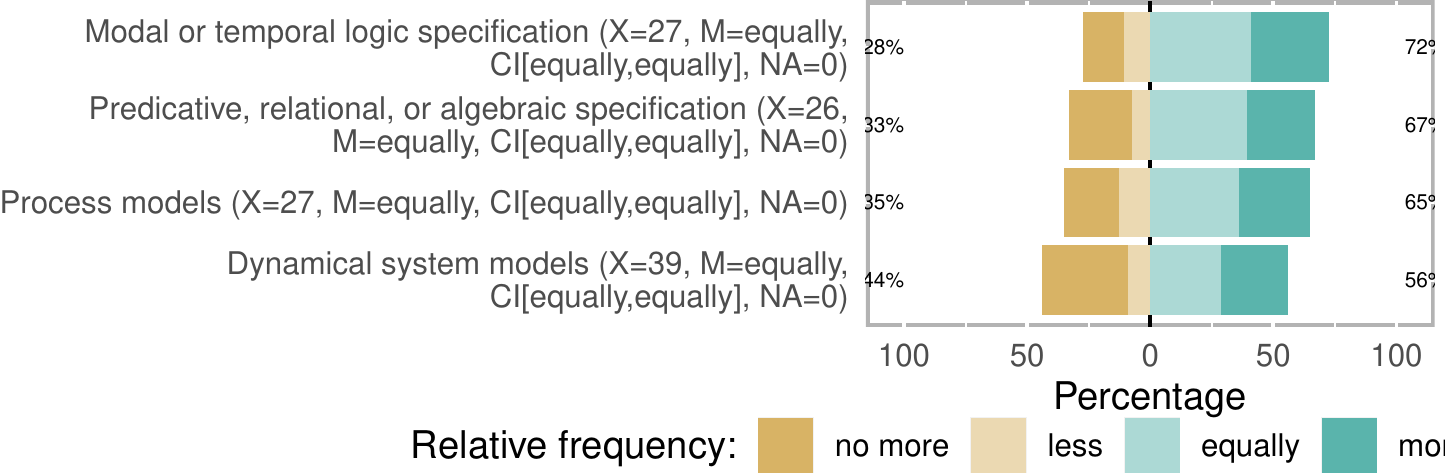}
  \caption{(\QuestionRef{F3_use_future})
    I~(would) intend to use ... 
    \label{fig:F3_use_future}}
\end{figure}

\Cref{fig:F3_use_future} depicts the respondents' intended
\emph{future} use of various \acp{FM} for system specification~(\ies formal
description techniques).  The figure shows an \emph{almost equal}
amount of participants aiming to decrease ~(\ies ``no more'' and
``less'') and increase~(\ies ``more often'') their use of \acp{FM} for
specification.  Only \emph{dynamical} system models again seem to be
an exception: more participants want to decrease their use of this
technology, compared to participants who want to increase it.

\paragraph{\QuestionRef{F4_use_future}: Intended use for Analysis.}
\label{sec:sum-F4_use_future}

\begin{figure}
  \sidecaption
  \includegraphics[width=.7\columnwidth]{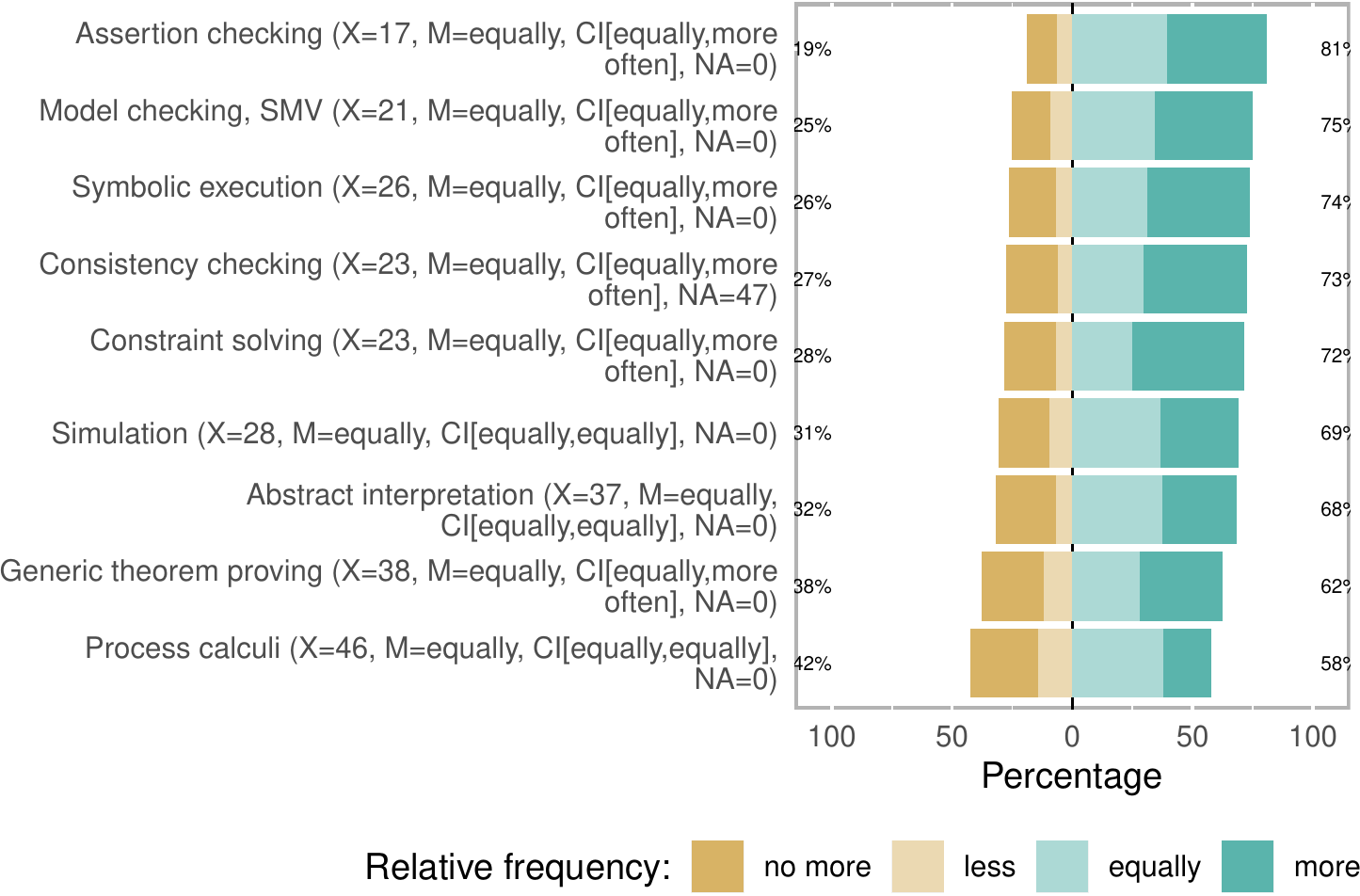}
  \caption{(\QuestionRef{F4_use_future})
    I~(would) intend to use ... 
    \label{fig:F4_use_future}}
\end{figure}

The respondents' intended use of \acp{FM} for the analysis of
specifications~(\ies formal reasoning techniques) is depicted in
\Cref{fig:F4_use_future}.  Except for process calculi, we observe a
general tendency of the participants to \emph{increase} their future
\ac{FM} use.

\paragraph{\QuestionRef{F5_purpose_future}: Intended Purpose.}
\label{sec:sum-F5_purpose_future}

\begin{figure}
  \sidecaption
  \includegraphics[width=.7\columnwidth]{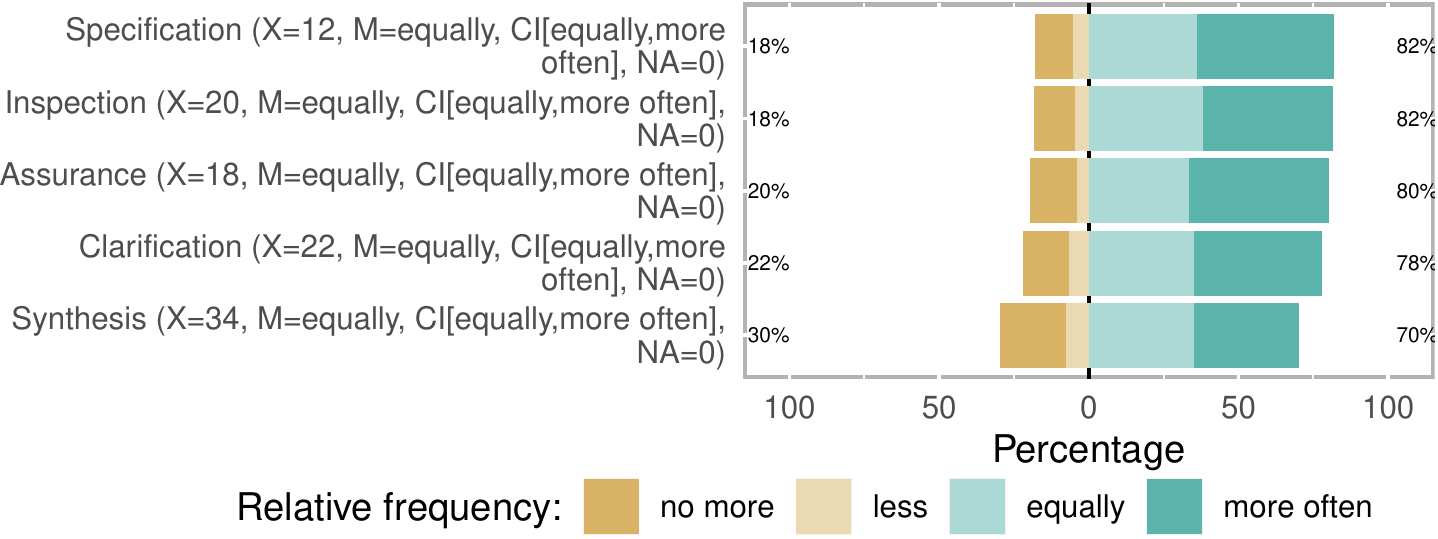}
  \caption{(\QuestionRef{F5_purpose_future})
    I~(would) intend to use \acp{FM} for ... 
    \label{fig:F5_purpose_future}}
\end{figure}

\Cref{fig:F5_purpose_future} indicates why respondents intend to apply
\acp{FM}.  Again, there is a tendency of the participants to
\emph{increase} \ac{FM} use across all listed purposes.

\paragraph{\QuestionRef{P4_purpose_past} and
  \QuestionRef{F5_purpose_future}: Comparison of Code- and Model-based
  \acp{FM}.}
\label{sec:comparison_P4_F5}

In the following, we regard \emph{practitioners} with experience level
``applied several times in engineering practice'' or ``applied once in
engineering practice'' and frequency ``applied in 2 to 5 separate
tasks'' or ``applied in more than 5 separate tasks'' (see
\Cref{tab:scales}).  We compare \emph{users of code-based \acp{FM}}
(CBs; including ``abstract interpretation'', ``assertion checking'',
``symbolic execution'', ``consistency checking''; with N=128) with
\emph{users of model-based \acp{FM}} (MBs; including ``process
calculi'', ``model checking'', ``theorem proving'', and
``simulation''; with N=114).  While some of the \ac{FM} classes can be
seen as both, code- and model-based, we made a choice based on our
experience but left out ``constraint solving'' because it is a
fundamental technique intensively applied in both.

The comparison of past and future use for code-based (top half of
\Cref{fig:rq2_comparison_cbfm_mbfm} in
\Cref{sec:data-intent-cbfm-mbfm}) and model-based \acp{FM} (bottom
half of \Cref{fig:rq2_comparison_cbfm_mbfm}), \eg in
\emph{inspection}~(\egs error detection, bug finding) shows the
following:
\begin{itemize}
\item CBs show slightly more frequently an increased intent (the
  ``more often'' group) than MBs; for both sub-groups,
  respondents with 2 to 5 and with more than 5 past uses.
\item MBs show slightly more frequently a decreased intent (the
  ``no more'' group) than CBs.
\end{itemize}
Looking at \emph{assurance}~(\egs proof, error removal) shows the
following:
\begin{itemize}
\item MBs show slightly more frequently an increased intent than CBs
  when looking at respondents who have used \acp{FM} more than 5
  times.  However, MBs indicate slightly less frequently an increased
  intent than CBs when looking at respondents with 2 to 5 uses.
\item CBs indicate more \emph{dnk}s after 2 to 5 uses and slightly
  more frequently a decreased intent after 5 uses in comparison with
  MBs.
\end{itemize}

\paragraph{\QuestionRef{D1_appdom_past}, \QuestionRef{P2_use_past},
  and \QuestionRef{P3_use_past}: Practised \ac{FM} Classes by Application
  Domain.} 

\begin{figure}
  \includegraphics[width=\columnwidth]{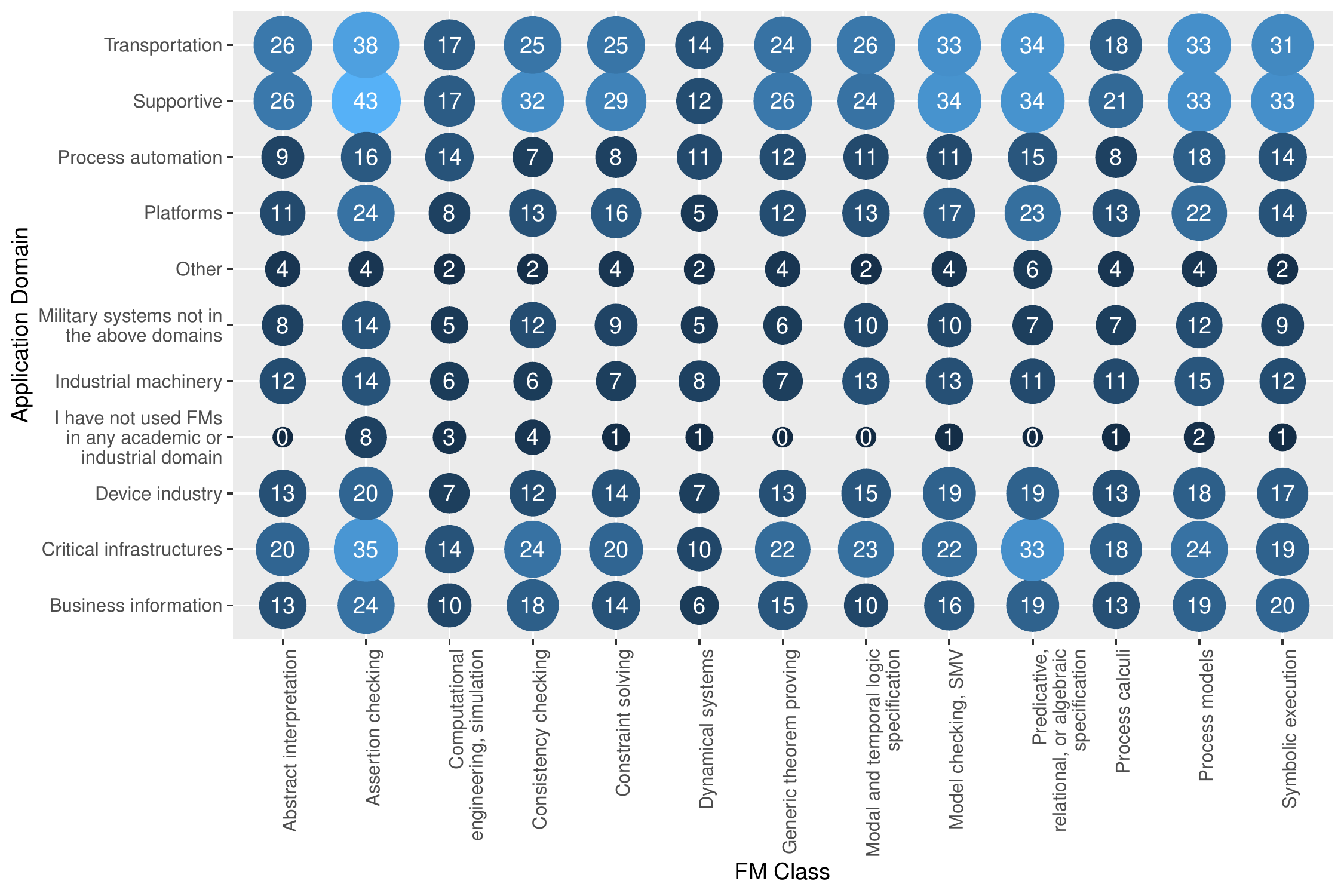}
  \caption{Approximation (likelihood) of practised use ($\mathit{UFM}_p$) by \ac{FM}
    class and application domain 
    \label{fig:rq2_techbydom_past}}
\end{figure}

\begin{figure}
  \includegraphics[width=\columnwidth]{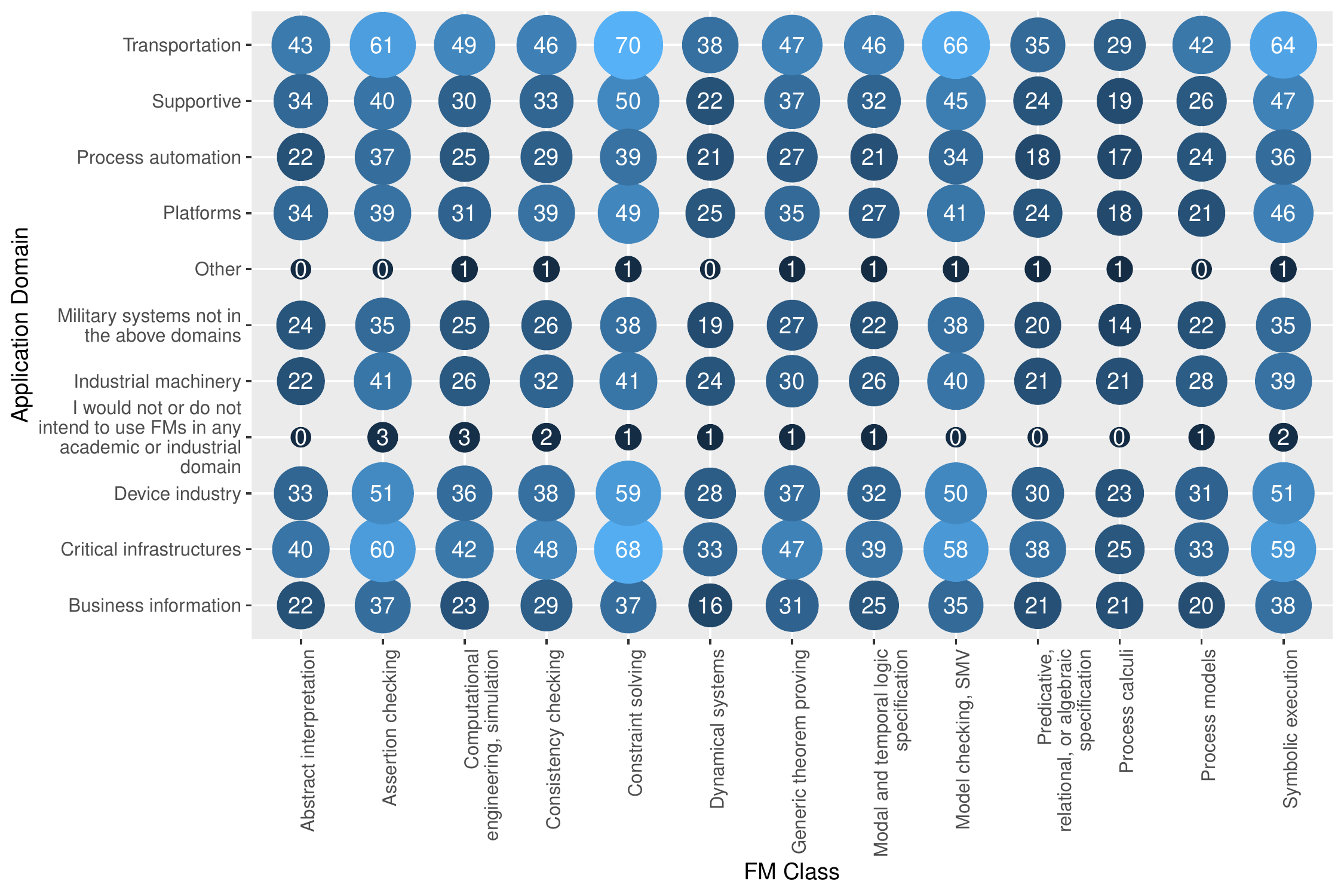}
  \caption{Approximation (likelihood) of increased usage intent
    ($\mathit{UFM}_i$) by \ac{FM} class and application domain
    \label{fig:rq2_techbydom_future}}
\end{figure}

We asked respondents about their use of each \ac{FM} class
\emph{independent} of the application domain and about their general
use of \acp{FM} in each such domain.  Hence, we can only approximate
past usage per \ac{FM} class and application domain assuming that the
overall usage per respondent is uniformly distributed among the
specified \ac{FM} classes and domains.
For that, we interpret~(and count) each respondent who specifies a
domain in combination with ``applied once in engineering practice'' or
``applied several times in engineering practice'' for an \ac{FM} class
as a practitioner who \emph{has used}~($\mathit{UFM}_p$) or, respectively,
\emph{wants to use}~($\mathit{UFM}_i$) \acp{FM} of that class in that domain.
More generally, we count a respondent who specifies $n$ domains, say
$d_1$ to $d_n$, in combination with ``applied once in engineering
practice'' or ``applied several times in engineering practice'' for
$m$ \ac{FM} classes, say $c_1$ to $c_m$, as a practitioner who
\emph{has used}~($\mathit{UFM}_p$) or, respectively, \emph{wants to
  use}~($\mathit{UFM}_i$) \acp{FM} of the classes $c_1$ to $c_m$ in the domains
$d_1$ to $d_n$.
\Cref{fig:rq2_techbydom_past,fig:rq2_techbydom_future} show these
approximations for $\mathit{UFM}_p$ and $\mathit{UFM}_i$.

\subsection{Perception of Challenges (answering RQ~3)}
\label{sec:analysis-rq3}

\Cref{tab:challenges} lists the \ac{FM} challenges subject to discussion,
their background, and literature referring to them.  We apply the
procedure described in \Cref{sec:proc:rqs}.

\begin{table}
  \caption{Feedback on given and additional challenges
    (see \Cref{sec:map-lit-chall} for a full list of references)
    \label{tab:challenges}}
  \footnotesize
  \begin{tabularx}{\textwidth}
    {>{\hsize=.3\hsize}L
    c
    >{\hsize=.2\hsize}L
    >{\hsize=.35\hsize}L}
    \toprule
    \textbf{Challenge Name \& Description}
    & \textbf{Src.}
    & \textbf{Supported by (oldest, newest)}
    & \textbf{Findings for \RQRef{3} (\Cref{sec:analysis-rq3})}
    \\\midrule
    \textbf{Scalability:} Useful in handling large and
      technologically heterogeneous systems
    & Q
    & 7 studies, \egs\citet{Hall1990,Miller2010}
      \nocite{Bowen1995,Lai1995,Lai1996,Craigen1993,Craigen1995a}
    &
      toughest in \Cref{fig:O1_obst}; %
      by \acp{P} more than by \acp{NP};
      when using \acp{FM} for assurance and clarification;
      independent of \ac{FM} class
    \\\midrule
    \textbf{Skills \& Education:} Methods known (little
    misconception); trained and experienced users available
    & Q
    & 13 studies, \egs\citet{Bjorner1987}, \citet{Bicarregui2009}
      \nocite{Hall1990,Barroca1992,Hinchey1996,Bowen1995a,Lai1995,Lai1996,Heisel1996,Galloway1998,Craigen1993,Craigen1995a,Snook2001}
    & 2nd toughest; %
      agreed by \acp{LE} and \acp{ME}; 
      largely independent of \ac{FM} class;
      comparatively small tough-proportions by \acp{M}
    \\\midrule
    \textbf{Transfer of Proofs:} 
    Relation %
    between models and reality (\egs code),
    handling incomplete specifications %
    & Q
    & 8 studies, \egs\citet{Jackson1987-PowerLimitationsFormal,Parnas2010}
      \nocite{Hall1990,Craigen1993,Craigen1995a,Snook2001,Bloomfield1991,Barroca1992}
    & Agreed by \acp{LE} and \acp{ME};
      top-rated by \acp{DI} %
      and \acp{U}; largely independent of \ac{FM} class
    \\\midrule
    \textbf{Reusability:}
    Parametric proofs, reusable specifications and verification results
    & Q
    & \citet{Barroca1992,Bowen1995a}
    & Top-rated by tool provider stakeholders and lectures
    \\\midrule
    \textbf{Abstraction:} Useful and correct (automated) abstractions
      from irrelevant detail (for comprehension and validation)
    & Q
    & 12 studies, \egs\citet{Jackson1987-PowerLimitationsFormal,Parnas2010,Miller2010}
      \nocite{Hall1990,Barroca1992,Bowen1995a,Lai1996,Galloway1998,Heitmeyer1998,Heisel1996,Knight1997,Snook2001}
    & Varies notably across \ac{FM} classes
    \\\midrule
    \textbf{Tools \& Automation:}
    Useful notations %
    and trustworthy tools (for manipulation, checking, collaboration, documentation)
    & Q
    & 16 studies, \egs\citet{Bjorner1987,OHearn2018}
      \nocite{Hall1990,Bloomfield1991,Bowen1995,Hinchey1996,Bowen1995a,Bowen2005,Bicarregui2009,Woodcock2009,Parnas2010,Lai1996,Heitmeyer1998,Craigen1993,Craigen1995a,Knight1997}
    & Top-rated by \acp{DI}; %
      but comparatively small tough-proportions from practitioners
    \\\midrule
    \textbf{Maintainability:} %
    Stable proofs, easily modifiable specifications, and adaptable verification results
    & Q
    & \citet{Barroca1992,Knight1997,Parnas2010}
    & Comparatively small tough-proportions from practitioners
    \\\midrule
    \cellcolor{lightgray}
    \textbf{Resources:}
    Sufficient resources, %
    good cost-benefit ratio (despite adoption, training, licenses)
    & 4R
    & 11 studies, \egs\citet{Hall1990,Woodcock2009}
      \nocite{Craigen1993,Craigen1995a,Bloomfield1991,Bowen1995,Bowen1995a,Lai1995,Heisel1996,Knight1997,Bicarregui2009}
    & \multirow{3}{4cm}[-1.5em]{No detailed data was collected: Because these
      challenges were mentioned
      several times each, we classify them to be at least of moderate difficulty.}
    \\\cmidrule{1-3}
    \cellcolor{lightgray}
    \textbf{Process Compatibility:}
    Integration into existing process, method 
    culture, standards, and regulations
    & 6R
    & 12 studies, \egs\citet{Bjorner1987,OHearn2018}
      \nocite{Bloomfield1991,,Bowen1995,Bowen1995a,Lai1995,Hinchey1996,Lai1996,Heitmeyer1998,Heisel1996,Knight1997,Craigen1995a}
    \\\cmidrule{1-3}
    \cellcolor{lightgray}
    \textbf{Practicality \& Reputation:}
    Benefit awareness and sufficient empirical evidence for benefits
    & 7R
    & 6 studies, \egs\citet{Lai1995,Parnas2010}
      \nocite{Lai1996,Galloway1998,Glass2002,Bicarregui2009}
    \\\bottomrule	
  \end{tabularx}
  {\footnotesize Src.\dots source, Q \dots in questionnaire, $n$R
    \dots additionally raised by $n$ Respondents}
\end{table}

\paragraph{General Ranking (\QuestionRef{O1_obst}).}
\label{sec:sum-O1_obst}

\begin{figure}
  \sidecaption
  \includegraphics[width=.7\columnwidth]{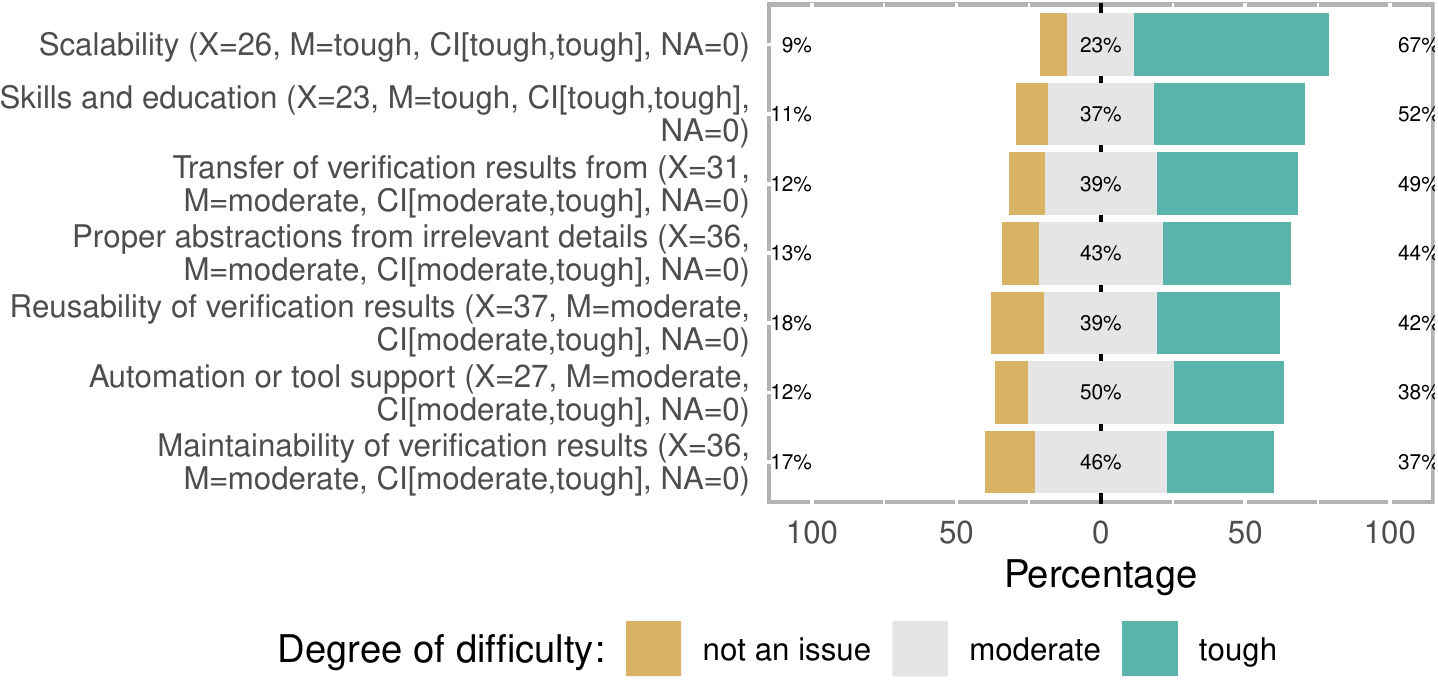}
  \caption{(\QuestionRef{O1_obst})
    For any use of \acp{FM} in my future activities, I consider
    $\langle$\emph{obstacle}$\rangle$ as [not an$\mid$a
    moderate$\mid$a tough] issue.
    \label{fig:O1_obst}}
\end{figure}

\Cref{fig:O1_obst} shows the respondents' ratings of all challenges.
Most of them believe that \emph{scalability} will be the
toughest challenge and \emph{maintainability} is considered the least
difficult of all rated obstacles.  For \emph{reuse of proof results},
\emph{proper abstractions}, and \emph{tool support}, the participants
distribute more uniformly across moderate and high difficulty.

In the following, we compare specific groups of respondents by how they
perceive the difficulty of the various challenges.  We group
respondents according to the criteria in \Cref{sec:data-analysis} and
according to the role, motivating factor, \ac{FM} class, and purpose they
specified.  \Cref{sec:data-rq3} provides some background material for the
following association analyses.

\paragraph{Less Experienced (\acs{LE}) versus More Experienced (\acs{ME})
  Respondents (\QuestionRef{D2_explev}).}

The comparison of the difficulty ratings of \acp{LE} with the ratings
of \acp{ME} shows that
\begin{inparaenum}[(i)]
\item \acp{LE} less often perceive the given challenges as tough,
\item \acp{ME} significantly more often rate \emph{scalability} as tough,
\item both groups show the closest agreement on \emph{transfer of
  verification results} and \emph{skills and education}.
\end{inparaenum}

\paragraph{Non-Practitioners (\acs{NP}) versus Practitioners (\acs{P}) by Past
  Purpose (\QuestionRef{P4_purpose_past}).}

The perception of \emph{skills and education} and \emph{scalability}
as the most difficult challenges is largely independent of the
purpose, again \acp{P} attributing more significance to \emph{scalability}.
Scalability, the forerunner in \Cref{fig:O1_obst}, exhibits the most
tough-ratings from \acp{NP} in \emph{synthesis} and from \acp{P} in
\emph{assurance} and \emph{clarification}~(see the top half of
\Cref{fig:rq3_heatmapP4_O1} in \Cref{sec:data-rq3}).

\paragraph{Decreased Intent (\acs{DI}) versus Increased Intent (\acs{II}) by
  Purpose (\QuestionRef{F5_purpose_future}).}

The comparison of the difficulty ratings of respondents with no or
decreased intent to use \acp{FM} \emph{for a specific purpose} and of
respondents with equal or increased intent shows:
\begin{inparaenum}[(i)]
\item \emph{Scalability} and \emph{skills and education}, both
  forerunners in \Cref{fig:O1_obst}, show the most tough-ratings from
  \acp{II} for \emph{assurance}~(67\%) and \emph{inspection}~(66\%) and
  from \acp{DI} for \emph{synthesis}~(53\%).
\item The trend in \Cref{fig:O1_obst} is more clearly observable from
  \acp{II} than from \acp{DI}, where \emph{transfer of verification results} and
  \emph{automation and tool support} seem to be tougher than
  \emph{skills and education}.
\end{inparaenum}

\paragraph{Non-Practitioners (\acs{NP}) versus Practitioners (\acs{P})
  by \acs{FM} Class (\QuestionRef{P2_use_past},
  \QuestionRef{P3_use_past}).}

\begin{figure}
  \sidecaption
  \includegraphics[width=\columnwidth]{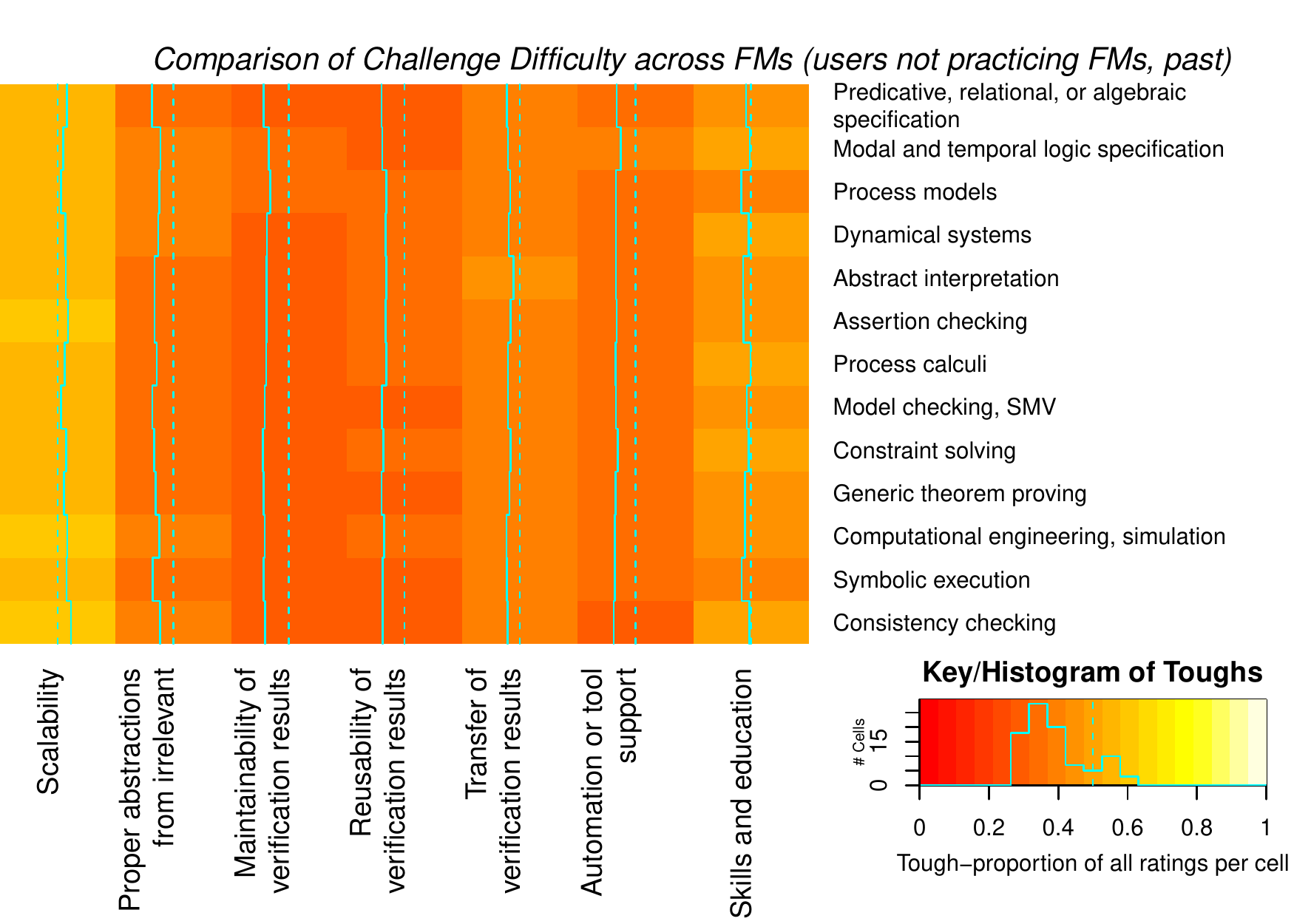}
  \includegraphics[width=\columnwidth]{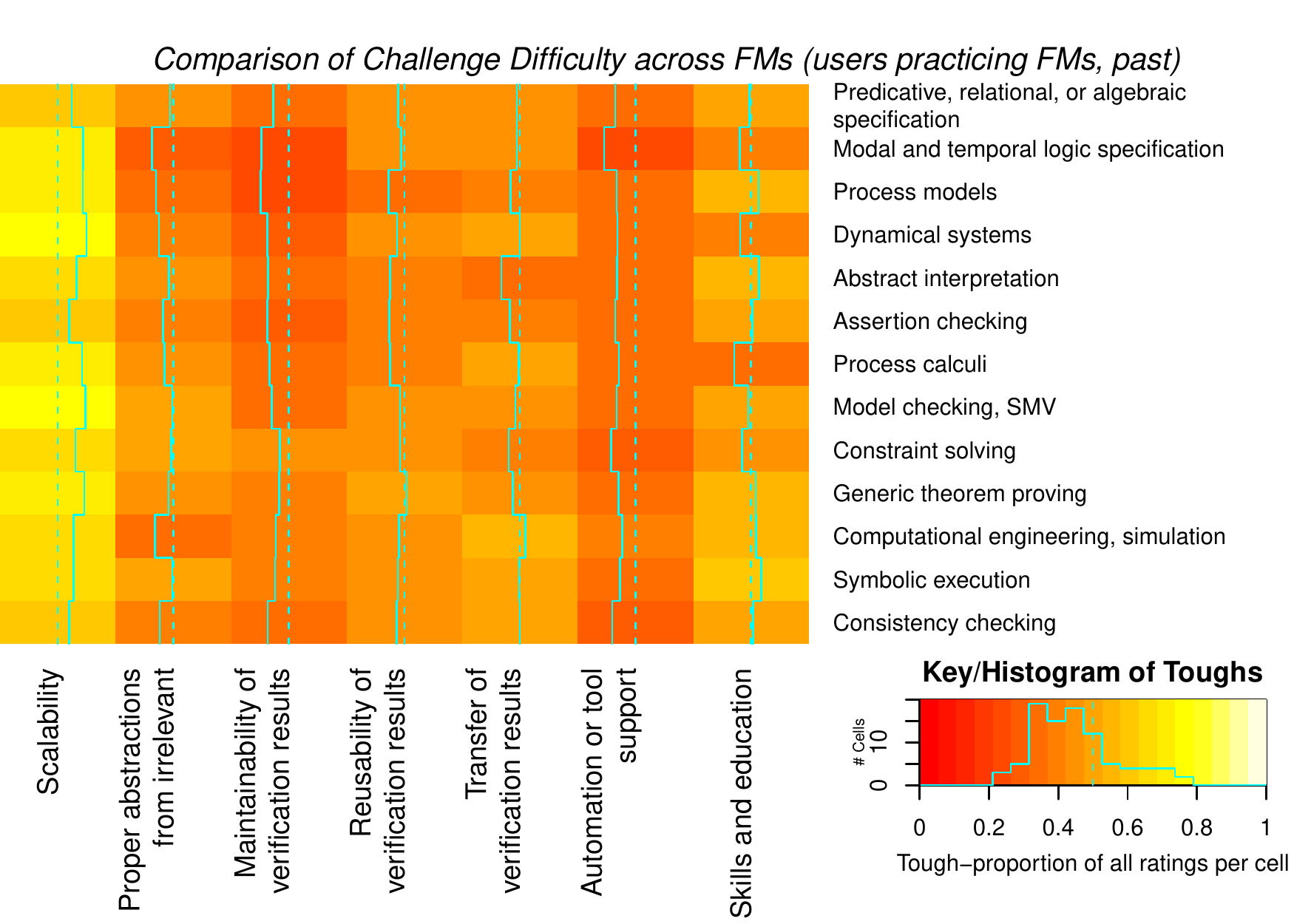}
  \caption{Difficulty of challenges (cols):
    \acp{NP} (top) compared to
    \acp{P} (bottom) by class of used \ac{FM} (rows).
    \emph{Legend:} In each cell of an association matrix, both the
    solid vertical line and the colour~(gradient from red to white)
    represent the tough proportions~(from 0 to 100\%), with the dotted
    vertical line marking the 50\% margin.  The
    histogram~(to the lower right corner of each matrix) counts the
    combinations~(cells) in each 5\%-band of tough ratings.  \Egs $\sim$70\% of
    ``process calculi'' users perceive ``scalability'' as a tough challenge.
    \label{fig:heatmapP2P3O1}}
\end{figure}

The top half of \Cref{fig:heatmapP2P3O1} shows for \acp{NP}, the trend in
\Cref{fig:O1_obst} is \textbf{largely independent of the \ac{FM}
  class}, except for \emph{consistency checking} and \emph{logic}
leading with \emph{tough} proportions of 49\%.  

The bottom half of \Cref{fig:heatmapP2P3O1} shows for \acp{P}, difficulty
ratings across \ac{FM} classes vary more: The foremost challenges in
\Cref{fig:O1_obst} received the most \emph{tough}-ratings from users
of \emph{process models}, \emph{dynamical systems}, \emph{process
  calculi}, \emph{model checking}, and \emph{theorem proving}.
Difficulty ratings of users are often centred on moderate or tough,
\emph{proper abstraction} and \emph{skills and education} show a
comparatively wide variety across \ac{FM} classes.

The histograms in the lower right corners in \Cref{fig:heatmapP2P3O1}
indicate that
\begin{inparaenum}[(i)]
\item \acp{NP}' difficulty ratings vary less than \acp{P}' ratings,
\item \acp{NP}' ratings are more independent from the \ac{FM} classes, and
\item \acp{NP}' difficulty ratings are lower on average than \acp{P}' ratings.
\end{inparaenum}
\Cref{sec:data-rq3} contains several such association matrices with
more detailed data in the matrix cells.

\paragraph{Decreased Intent (\acs{DI}) versus Increased Intent
  (\acs{II}) by \ac{FM} Class (\QuestionRef{F3_use_future},
  \QuestionRef{F4_use_future}).}

The trend in \Cref{fig:O1_obst} is supported by many tough
ratings~(48\%) for \emph{transfer of verification results} from \acp{DI} in
\emph{consistency checking}.
However, \acp{DI} in \emph{process calculi} provide comparatively many
tough-ratings~(39\%) for the generally low-ranked \emph{automation and
  tool support}.
\emph{Assertion checking} exhibits comparatively low tough-proportions
across all challenges whereas \emph{process calculi} exhibit
comparatively high tough-ratings.
Mirroring the trend in \Cref{fig:O1_obst}, \acp{II} show less variance than
\acp{DI} across all \ac{FM} classes.

\paragraph{Unmotivated (\acs{U}) versus Motivated (\acs{M})
  respondents by Motivating Factor (\QuestionRef{D3_motiv}).}

Respondents with moderate to strong motivation to use \acp{FM} more likely
identify the given challenges as moderate to tough, \textbf{regardless of the
  motivating factor.} %
The trend in \Cref{fig:O1_obst} seems explainable by many
tough ratings from respondents motivated by \emph{regulatory
  authorities}~(69\%), not motivated by \emph{tool providers}~(56\%),
and not motivated by \emph{superiors/principal investigators}~(56\%,
see \Cref{fig:rq3_heatmapD3_O1} in \Cref{sec:data-rq3}).
\acp{U}' tough-ratings are \textbf{notably lower than \acp{M}'}
tough-ratings.

\paragraph{Past and Future Views by Role (\QuestionRef{P1_role_past},
  \QuestionRef{F2_role_future}).}

Although participants show role-based discrepancies between their past
and intended use of \acp{FM}~(\Cref{fig:rq2_comparison_P1_F2}), the
\textbf{perception of difficulty} of the rated challenges seems to be
\textbf{largely similar}, following the trend in \Cref{fig:O1_obst}.
The high ranking of \emph{scalability}~(and \emph{reusability of
  verification results}) is supported by many tough-ratings from
\emph{tool provider stakeholders} for the past view and many from
\emph{lecturers} for the future view.
Respondents not having used \acp{FM} or not planning to use \acp{FM}
exhibit the lowest tough-ratings but also the highest fractions of
$dnk$-answers.

\paragraph{Past and Future Views by Domain
  (\QuestionRef{D1_appdom_past}, \QuestionRef{F1_appdom_future}).}

The trend in \Cref{fig:O1_obst} is underpinned by highest
tough-proportions for respondents from the \emph{transportation},
\emph{military systems}, \emph{industrial machinery}, and
\emph{supportive} domains.

\section{Discussion}
\label{sec:discussion}

In this section, we discuss and interpret our findings, relate them to
existing evidence, outline general feedback on the questionnaire, and
critically assess the validity of our study.

\subsection{Findings and their Interpretation}
\label{sec:interpr-results}

The following~(F)indings are based on the data summarised and analysed
in the
\Cref{sec:sensitivity-analysis,sec:summ-answ-quest,sec:descr-data-points}.
All findings are then collected in \Cref{tab:findings} on
\cpageref{tab:findings}.

\paragraph{Findings for RQ~1.}

\Finding{RQ1_noreg} %
\emph{Regulatory authorities} with their norms, codes, or policies
represent only a minor motivating factor to use \acp{FM}.
\emph{Intrinsic motivation}~(maybe market-triggered) seems to be
stronger.  This finding is consistent with what we know from the
literature survey in
\citet{Gleirscher2019-NewOpportunitiesIntegrated}: \acp{FM} are not
formally required by corresponding standards today, not even for the
highest safety integrity levels.  If regulatory authorities change
their recommendations to requirements, then this might spike as a
motivating factor.

\Finding{RQ1_lownonusers} %
The low fraction of respondents with no experience in
\Cref{fig:D2_explev} may have been caused
\begin{inparaenum}[(1)]
\item by our choice of expert channels in \Cref{tab:adplatforms} where
  the likelihood of encountering \ac{FM} users %
  is probably higher than in more generic \ac{SE} channels (\egs Stack
  Overflow) and
\item by the fact that \ac{SE} students will usually have an \ac{FM}
  course or some lectures about \acp{FM} such that they would choose
  ``1--3 years'' in \QuestionRef{D2_explev} and ``studied in course''
  in \QuestionRef{P2_use_past}.
\end{inparaenum}

\Finding{RQ1_dynsys} %
We observe the least use of \acp{FM} in computational engineering and for
reasoning about dynamical systems, \eg reasoning about the correctness
of algorithms, and their implementation in embedded software,
controlling such systems.  One explanation for this is that our sample
mainly comprises software and systems engineers who will work less
intensively with such \acp{FM} than, \eg mechanical or control engineers.
Another explanation is that such \acp{FM} are still less widely known, less
well developed, or less well supported by tools than \acp{FM} focusing on
the reasoning about pure software.

\paragraph{Findings for RQ~2.}

\Finding{RQ2_increase} %
It seems that in \emph{all given
  domains}~(\Cref{fig:rq2_comparison_D1_F1}, except for \emph{other})
respondents intend to \emph{increase} their future use of \acp{FM}.
Moreover, we observe that this tendency is \emph{independent} of the
particular \emph{\ac{FM} class}~(except process calculi) or
\emph{purpose}.  The data also suggest that the use of \acp{FM} by
teachers and researchers is saturated.  This saturation indicates a
stable intent to teach \acp{FM}, to perform research in \acp{FM}, or
to otherwise use \acp{FM} in teaching or research.  However, there is an
increased intent to apply \acp{FM} in \emph{industrial contexts} in
the future.  One explanation could be that engineers have already
wanted to use \acp{FM} but have not had the opportunity or were not
told or permitted to do so.  Another explanation for an increased
intent of \ac{FM} non-users could be due to some bias when answering
questions about whether someone would do~(\egs try out) something.

\Finding{RQ2_expsupuse} %
Our data suggest that experience in using a certain \ac{FM} class is
positively associated with the intent to use this \ac{FM} class in the
future.  To investigate this suspicion, we analysed the intended use
of a \ac{FM} class based on the experience of participants in using
this class~(also by association analysis as described in
\Cref{sec:data-analysis}).  We observe that the \emph{more experience}
one has with using a specific \ac{FM} class, the \emph{more likely
  they will apply it in the future}~(see the two charts in the
\Cref{sec:data-intent-by-purpose,sec:data-intent-by-tech}).  %
No experience with a specific \ac{FM} class correlates with a
\emph{low intent} to use that class.  Participants not having used
\acp{FM} and, hence, unfamiliar with them might not have had the need
in the first place.  Only little experience with a certain \ac{FM}
class \emph{significantly increases the intent to apply it again in
  the future}.  Similar observations can be made for the use of
\acp{FM} in general for a specific purpose.

\Finding{RQ2_cbfm_mbfm} %
The differences in past and intended use between code- and model-based
\acp{FM}~(\Cref{sec:comparison_P4_F5}), \eg when looking at inspection
and assurance, are marginal.  Moreover, we cannot find a significant
difference or a trend between these two categories of \acp{FM} when
considering different purposes, experience levels, and usage
frequencies.

\Finding{RQ2_techbydom} %
The approximation in the
\Cref{fig:rq2_techbydom_past,fig:rq2_techbydom_future} allows the,
albeit vague, interpretation of the numbers as the likelihood that
respondents \emph{have used}~(\Cref{fig:rq2_techbydom_past}) or
\emph{want to use}~(\Cref{fig:rq2_techbydom_future}) a particular
\ac{FM} in a particular domain.  Assuming this model,
\Cref{fig:rq2_techbydom_future} indicates the highest likelihoods of an
increased $\mathit{UFM}_i$ for methods such as ``assertion checking'',
``constraint solving'', ``model checking'', and ``symbolic execution''
in domains such as ``transportation'', ``critical infrastructures'',
and the ``device industry''.

\paragraph{Findings for RQ~3.}

\begin{figure}
  \sidecaption
  \includegraphics[width=.6\columnwidth]{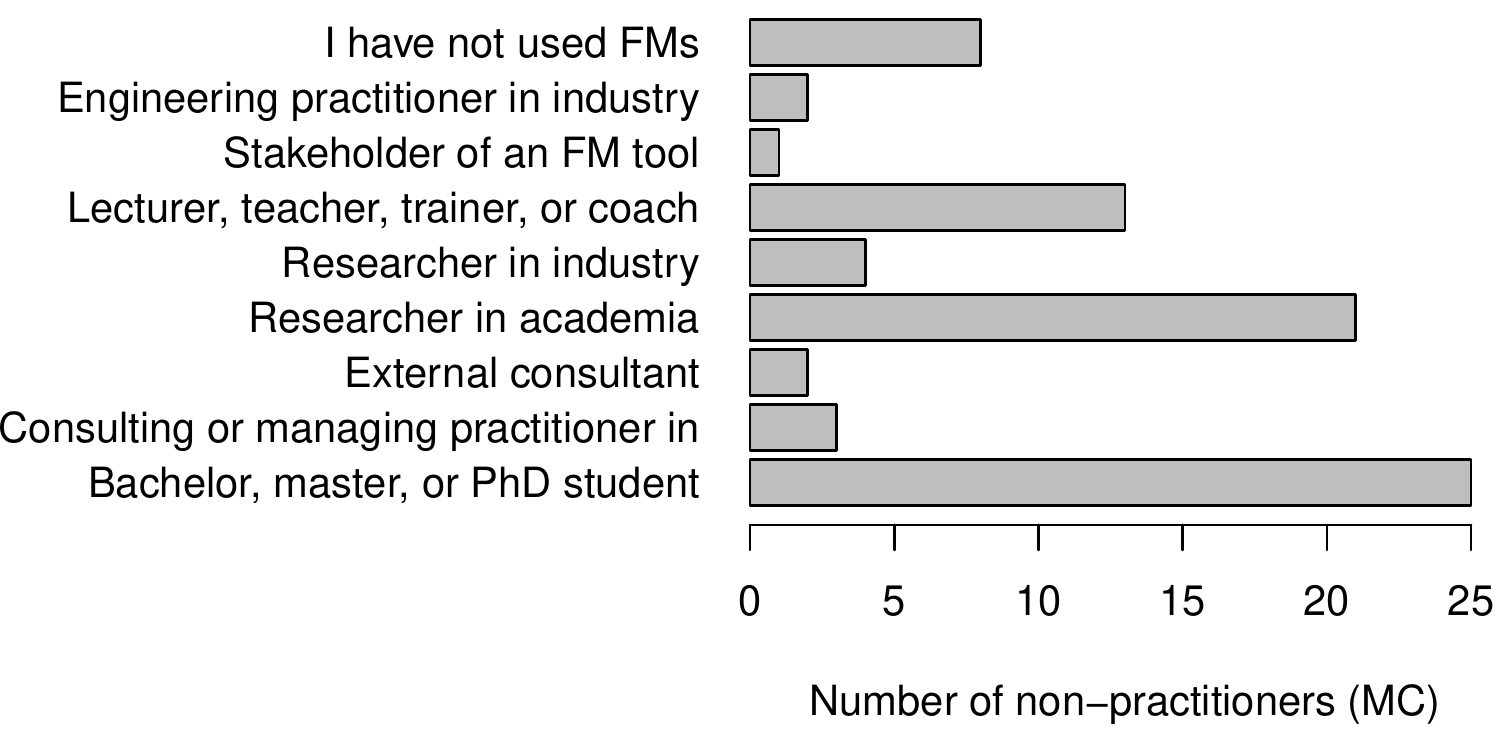}
  \caption{The past role profile of the 46 non-practitioners (out of
    216 respondents) helps to explain
    finding~\FindingRef{RQ3_scalableads} 
    \label{fig:rq3_np_profile}}
\end{figure}

\Finding{RQ3_scalableads} %
\emph{Scalability} and \emph{skills and education} lead the challenge
ranking, independent of the domain, \ac{FM} class, motivating factor, and
purpose.  Practitioners see scalability as more problematic than
non-practitioners, whereas non-practitioners perceive \emph{skills and
  education} as more problematic than practitioners.
\Cref{fig:rq3_np_profile} may explain the latter by showing a high fraction of
students among the 46 non-practitioners.

\Finding{RQ3_maintlow} %
\emph{Maintainability of proof results} or other verification
artefacts was found to be the least difficult challenge.  However, in
the lower half of \Cref{fig:heatmapP2P3O1}, the challenge column
``maintainability'' shows relatively low frequencies for ``modal and
temporal logic'' and ``model checking'' (possibly because of the high
level of automation) whereas ``theorem proving'' (possibly because of
a low level of automation) and ``constraint solving'' (possibly
because of being too versatile or generic for the present purpose)
show the highest frequencies of tough ratings.  See
\Cref{fig:rq3_heatmapF3_F4_O1} in the \Cref{sec:data-rq3} for more
details.
\Finding{RQ3_reusenew} %
\emph{Reusability of proof results} was rated as tough by several
practitioner groups.

\Finding{RQ3_DIsVsTools} %
\ac{FM} users with decreased usage intent rate \emph{tool deficiencies} as
their top obstacle to \ac{FM} adoption.
\Finding{RQ3_newchall} %
Furthermore, our respondents raised three additional challenges~(\ies
resources, process compatibility, and practicability \& reputation)
which we cross-validated with the literature~(see highlighted rows in
\Cref{tab:challenges}).  The fact that these obstacles were mentioned
several times in addition to the given obstacles justifies them to be
highly relevant and at least moderate.  However, our data does not
allow to rank them more precisely.

\Finding{RQ3_simperc} %
Challenges are perceived \emph{as moderate or tough}, largely
similarly between the pairs of groups we distinguish in
\Cref{sec:data-analysis}.

\Finding{RQ3_toughest} %
With 72\% of tough ratings for \emph{scalability}, process
calculi~(\egs ACP, CCS, CSP) perform in the midfield despite their
high reputation as compositional methods.  \emph{Scalability} of
process models~(\egs Petri nets, Mealy machines, labelled transition
systems, Markov models) is also ranked in the middle field of tough
challenges.  The ranking of these models, however, is unsurprising in
the light of the difficult scalability of model checking, a frequently
used verification technique for process models and the leader in this
ranking~(\cf\Cref{fig:heatmapP2P3O1}).
One explanation for the high number of tough-ratings from \acp{NP} in
\emph{synthesis} could be that \acp{NP} might either not associate
\acp{FM} with synthesis in general, or because automated synthesis of
sophisticated artefacts is known to be an unsolved problem in many
cases, independent of the use of \acp{FM}.

\subsection{Relationship to TAM for Methods (answering RQ~4)}
\label{sec:rel-peou-pu}

In analogy to the reasoning in
\citet{Davis1989-PerceivedUsefulnessPerceived}, an increased positive
experience with practically applying \acp{FM} forms a high degree of
\ac{PU}~(\Cref{sec:background}).
\citet[pp.~329, 331]{Davis1989-PerceivedUsefulnessPerceived} observed that
current and intended usage %
are significantly correlated with \ac{PU}, less
with \ac{PEOU}.
In fact, \FindingRef{RQ2_increase} suggests an increased intent to use
\acp{FM} in the future.  Moreover, \FindingRef{RQ2_expsupuse} 
suggests a positive association of the degree of experience with
$\mathit{UFM}_i$, \ie more experience increases the intent.
\Finding{RQ4_PUpos} %
Because the use of \acp{FM} is not mandatory for most respondents, a
likely explanation for an increased intent~($\mathit{UFM}_i$) is that our
respondents perceive the \emph{usefulness of \acp{FM}} to be more
positive than negative.

Inspired by \citet{Riemenschneider2002-Explainingsoftwaredeveloper},
in the last paragraph of \Cref{sec:tam-style-ufm}, we justify the use
of challenge scales to collect data for \acs{PEOU} and \acs{PU}.
We justify the validity of the \ac{FM}-specific challenge scale
using the studies in \Cref{tab:rw}.  The column
``supported by'' in \Cref{tab:challenges} indicates studies discussing
the corresponding challenges.  From these discussions, we infer that
tackling these challenges contributes to an increased \acs{EOU}
and~\acs{Usef}.
First, the studies suggest that \acp{FM} are \emph{easier to use} if
users have sufficient skills and education, if the methods scale to
large systems, if mature tools and automation are available, and if
proofs are easily maintainable and reusable.
Second, the studies suggest that \acp{FM} are \emph{more useful} if
they are compatible with the process, if their cost-benefit ratio is
low, if their abstractions are correct and expressive, and if proofs
can be correctly transferred to reality.
Hence, these challenges represent \ac{FM}-specific
substrata~\citep[p.~325]{Davis1989-PerceivedUsefulnessPerceived} of
\acs{EOU} and \acs{Usef} for \acp{FM}.
Moreover, a high degree of \ac{PEOU} corresponds to an increased positive user
experience with \acp{FM} which translates to a low proportion of tough
ratings for the obstacles measured in \QuestionRef{O1_obst}.
However, from \FindingRef{RQ3_simperc}, we observe that respondents
rate most challenges as moderate to tough, largely independent of
other variables~(\FindingRef{RQ3_scalableads}).

\Finding{RQ4_PEOUneg} %
Overall, it thus seems that our respondents perceive the \emph{ease of
  use of \acp{FM}} to be more negative than positive.  According to
\Cref{tab:challenges}, many of the surveyed studies discuss
\emph{skills \& education}~(12 studies) and \emph{tools \&
  automation}~(16) as important challenges.  Moreover,
\Cref{fig:O1_obst} suggests that conceptual difficulties~(possibly,
from a lack of education and training, from difficulties in \ac{FM}
teaching, from a lack of \ac{FM} students) seem to be at least as
responsible for the negative ease of use as the lower ranked tool
deficiencies.  Indeed, in a recent discussion of ``push-button
verifiers'', \citet{OHearn2018} highlights that both conceptual
expertise and tool deficiencies are still significant bottlenecks.
However, an investigation of respondents' experiences with \ac{FM}
tools in comparison to their experiences with \ac{FM} concepts goes
beyond the possibilities of the data collected for this study.

\subsection{Relationship to Existing Evidence}
\label{sec:relevi}

Our systematic map shows that our list of challenges is completely
backed by substantial literature~(see \Cref{tab:challenges}) raising
and discussing these challenges.
\Finding{RelEv_lowdiffobst} %
However, the fact that maintainability and reusability were least
covered by our literature is, on the one hand, in line with
\FindingRef{RQ3_maintlow} but, on the other hand, not with
\FindingRef{RQ3_reusenew} and typical cultures of reuse in practice.

Beyond the general findings about \ac{FM} benefits in
\citet{Austin1993-Formalmethodssurvey}, we steered our half-open
questionnaire towards a refined classification of responses, comparing
past with intended use, and interrogating recently perceived obstacles
among a methodologically and geographically more diverse sample.
Their sample mainly covers Z and VDM users in the UK.  Our
questionnaire has less focus on representation and methodology and
excludes both questions on benefits and on suggestions to overcome
obstacles.  Regarding the latter,
\citet{Austin1993-Formalmethodssurvey} mention the improvement of
education and standardisation, the preparation of case studies, and
the definition of \ac{FM} effectiveness metrics.

\Finding{RelEv_nearrep} %
The report of \citet{Austin1993-Formalmethodssurvey} from the National
Physical Laboratory archive was unfortunately no more available
to us.  We finally managed to get access to a paper copy provided by a
friendly colleague.  This, however, only happened after conducting
this survey.  Anyway, we found that our conclusions are nearly
identical to \citeauthor{Austin1993-Formalmethodssurvey}'s.  The data
from \Cref{fig:O1_obst} and \Cref{tab:challenges} confirms that many
of the obstacles~(\ies limitations and barriers) they identified back
in 1991/2
remain~(\egs understanding the notation and the underlying
mathematics, resistance to process changes),
some have been lightly addressed~(\egs lack of cost/benefit evidence)
and
some have been more strongly addressed~(\egs lack of expressiveness,
lack of appropriate tools).
Not mentioned in \citet{Austin1993-Formalmethodssurvey} is scalability,
rated by our respondents to be the toughest obstacle.

\FindingRef{RQ2_expsupuse} is in line with other observations in
\citet{Woodcock2009,Bicarregui2009} that the repeated use of a \ac{FM}
results in lower overheads~(\ies an experienced effort or cost
reduction and improved error removal), up to an order of magnitude
less than its first use~\citep{Miller2010}.
Finally, our study generalises the main findings about barriers in
\citet{Davis2013-StudyBarriersIndustrial} to several geographies and
application domains, however, using an on-line questionnaire instead
of interviews and not asking for barrier mitigations.

\begin{table}
  \caption{Summary of findings per research question}
  \label{tab:findings}
  \footnotesize
  \begin{tabularx}{1.0\linewidth}{lX}
    \toprule
    \multicolumn{2}{X}{\textbf{\RQRef{1}:} In which typical domains, for which purposes, in which roles, and to
    what extent have \emph{\acp{FM} been used}?}\\
    \FindingRef{RQ1_noreg}
    & Intrinsic motivation to use \acp{FM} is stronger than norms or codes
      of regulatory authorities. \\
    \FindingRef{RQ1_lownonusers}
    & The fraction of respondents with no experience at all is
      comparatively low. \\
    \FindingRef{RQ1_dynsys}
    & Respondents use \acp{FM} the least in computational engineering and
      for dynamical systems.\\
    \midrule
    \multicolumn{2}{l}{\textbf{\RQRef{2}:} Which \emph{relationships} can we
    observe between \emph{past experience in using \acp{FM}} and 
    \emph{intent to use \acp{FM}}?}\\ 
    \FindingRef{RQ2_increase}
    & Increased intent to use \acp{FM} observable across all application domains. \\
    \FindingRef{RQ2_expsupuse}
    & Amount of experience is positively associated with the strength
      of usage intent. \\
    \FindingRef{RQ2_cbfm_mbfm}
    & The responses do not show any significant differences between
      code- and model-based \acp{FM}.
    \\
    \FindingRef{RQ2_techbydom}
    & Respondents show high likelihoods of an increased intent to use
      \acp{FM} such as ``model checking'' or ``assertion checking'' in
      areas such as ``transportation'' or ``critical infrastructures''.
    \\
    \midrule
    \multicolumn{2}{l}{\textbf{\RQRef{3}:} How difficult do study
    participants perceive widely known \ac{FM}
      \emph{challenges}?}\\
    \FindingRef{RQ3_scalableads}
    & Scalability and skills \& education lead the challenge difficulty ranking. \\
    \FindingRef{RQ3_maintlow}
    & Maintainability of proof results is found to be the least worrying challenge. \\
    \FindingRef{RQ3_reusenew}
    & Reusability of proof results is rated as tough by several practitioner groups. \\
    \FindingRef{RQ3_DIsVsTools}
    & \ac{FM} users with decreased usage intent rate \emph{tool deficiencies} as
    their top obstacle.\\
    \FindingRef{RQ3_newchall}
    & Respondents identified resources, process compatibility, and
      reputation as further obstacles. \\
    \FindingRef{RQ3_simperc}
    & All considered challenges are generally perceived as moderate or tough. \\
    \FindingRef{RQ3_toughest}
    & Among the \ac{FM} classes, process models are most 
      positively associated with tough scalability. \\
    \midrule
    \multicolumn{2}{l}{\textbf{\RQRef{4}:} What can we say about the \emph{perceived ease of use} and the
      \emph{perceived usefulness} of \acp{FM}?}\\
    \FindingRef{RQ4_PUpos}
    & Respondents perceive the usefulness of \acp{FM} as mainly
      positive and intend to increase their use. \\
    \FindingRef{RQ4_PEOUneg}
    & Respondents perceive the ease of use of \acp{FM} as mainly
      negative. \\
    \midrule
    \multicolumn{2}{l}{\textbf{Relationship to Existing Evidence} (from the literature):}
    \\
    \FindingRef{RelEv_lowdiffobst}
    & Proof maintainability and reusability are least covered by the literature. \\
    \FindingRef{RelEv_nearrep}
    & We repeat \citet{Austin1993-Formalmethodssurvey}, excluding
      benefit analysis but with a
      broader sample and more detailed questions. \\
    \bottomrule
  \end{tabularx}
\end{table}

\subsection{Threats to Validity}
\label{sec:threats-validity}

We assess our research design with regard to four common
criteria~\citep{Shull2008,Wohlin2012}.  Per threat~($\lightning$), we
estimate its criticality~(minor or major), describe it, and discuss
its partial~($\circ$) or full~(\checkmark) mitigation.

\subsubsection{Construct Validity} 
\label{sec:validity-construct}

Why would the construct~(\Cref{sec:construct}) appropriately
represent the phenomenon?

\threatPMit[maj]{Inappropriate questions and conceptual
  misalignment}{To support \emph{face validity}, we applied our own
  experience from \ac{FM} use to develop a core set of questions.  For
  the design of our questionnaire, we use feedback from colleagues,
  from respondents we personally know, and from the general feedback
  on the survey to improve and support \emph{content validity}.  A
  positive comparison with the questionnaire in
  \citet{Austin1993-Formalmethodssurvey} finally confirms the
  appropriateness of our questions.  However, we might have needed
  additional questions to check for conceptual alignment, \eg to more
  precisely determine whether the respondents' understanding of
  \emph{\acp{FM}} and of the \emph{use or application of \acp{FM}}
  closely matches ours.  However, from 18 respondents giving feedback
  on our questionnaire, only one commented on the definition and one
  on the classification of \acp{FM}.  That suggests that many
  respondents did not have or were not aware of misunderstandings
  worth mentioning.}

\threat{Questionnaire limited for measurement of \ac{PEOU}~(\egs per
  \ac{FM} class) and \ac{PU}}{We avoid deriving conclusions specific
  to a \ac{FM} or a corresponding tool from our data.}

\threat{Bias by omitted scale values~(\egs \ac{FM} class, domain,
  purpose)}{Respondents are encouraged to provide open answers to all
  questions, helping us to check scale completeness.  Between 8\% and
  40\% of the respondents made use of the text field ``Other.''
  Our systematic map confirms that we have not listed unknown
  challenges in \QuestionRef{O1_obst}.  We identified three additional
  challenges via open answers and the literature.
  We believe to have achieved good \emph{criterion validity} through
  questions and scales for distinguishing important sub-groups~(see
  \Cref{sec:proc:rqs}) of our population.}

\threat{Educational background asked indirectly}{We approximate what
  we need to know by using data from \QuestionRef{D1_appdom_past},
  \QuestionRef{D3_motiv}, \QuestionRef{P1_role_past}, and
  \QuestionRef{P2_use_past}.}

\subsubsection{Internal Validity} 
\label{sec:validity-internal}

Why would the procedure in \Cref{sec:research-design} lead to
reasonable and justified results?

\threat{Incomplete data points}{After the 47th response, feedback
  from colleagues and respondents resulted in an extension of
  \QuestionRef{D3_motiv} with the option ``on behalf of \ac{FM} tool
  provider''~(\Cref{fig:D3_motiv}) and of \QuestionRef{P3_use_past}
  and \QuestionRef{F4_use_future} with the option ``consistency
  checking''~(\Cref{fig:P3_use_past}).
  The enhancement of 169 complete data points to
  216 maintained all trends.
}

\threat{Duplicate \& invalid answers}{To identify intentional
  misconduct, we checked for timestamp anomalies and %
  for duplicate or meaningless phrases in open answers.  Voluntarily provided
  email addresses~(90/220) indicate only 4 double
  participants. %
  We remove these 4 data points from our data set.
  
  Google Forms includes data points only if all mandatory questions
  are answered and the submit button is pressed.  We also performed a
  consistency check of \ac{MlCh} questions and corrected 5 data points where
  ``I do/have not\dots'' was combined with other checked options.}

\threatPMit{Inter- vs. intra-$\mathit{UFM}$ inference}{Our study design is not
  suitable for ``inter-$\mathit{UFM}$ predictions'', \eg to predict that
  (dis)satisfied model checking practitioners have an increased (a
  decreased) intent to use theorem proving.  However, the
  argumentation in the \Cref{sec:tam-style-ufm,sec:rel-peou-pu} aims
  at ``intra-$\mathit{UFM}$ predictions'', that is, inferring an increased or
  decreased intent to use model checking from the quantity and quality
  of past experience in using model checking.  Such predictions may
  inherit possible limitations of \ac{TAM} studies.}

\subsubsection{External Validity} 
\label{sec:validity-external}

Why would the procedure in \Cref{sec:research-design} lead to similar
results with more general populations?

\threat[maj]{Low response rate}{
  We believe our estimates in \Cref{sec:descr-data-points} to be
  sensible.  
  We tried to
  \begin{inparaenum}[(i)]
  \item improve targeting by repetitively advertising on multiple
    appropriate channels,
  \item spot unreliable contact information, 
  \item provide incentive~(study results via email),
  \item keep the questionnaire short and comprehensible, 
  \item avoid forced answers, and
  \item allow lack of topic knowledge.
  \end{inparaenum}
  Some uncertainties remain, \eg
  lack of sympathy, personal motivation, and interest, or strong
  loyalty, and high expectations in the outcome, or intentional bias.
  However, from an estimated population of around 100K~(\ies the
  rounded sum of 38K and 61K), the minimum sample size for 95\%
  confidence intervals with continuous scale error margins of less
  than 7\% is 196, consistent with the ballpark figure in
  \citet[p.~117:29]{Gleirscher2019-NewOpportunitiesIntegrated}.  Our
  sample~(N=216) exceeds this number.  The 95\% confidence intervals
  for the Likert items show that the margin of error for the median
  sometimes deviates by one category~(\egs \Cref{fig:D3_motiv}).

  In this first study, we aim at understanding common perceptions,
  such as \emph{``\acp{FM} are not practically useful''} or
  \emph{``\acp{FM} are difficult to apply''}.  Because these
  statements address \acp{FM} as a whole, we believe such local
  errors %
  do not affect our general conclusions.  However, the response
  rate~(1 to 2\%) and population coverage~(0.1\%,
  \cf\Cref{sec:survey-execution}) were too low to avoid such errors
  and refute specific null-hypotheses, such as \emph{``\ac{FM} $m$ is
    effective for role $r$ and purpose $p$ in domain $d$''}~(by the
  \ac{FM} community) or \emph{``\ac{FM} $m$ is difficult to apply for
    role $r$ and purpose $p$ in domain $d$''}~(by \ac{SE}
  practitioners), with satisfactory statistical power.}

\threatPMit[maj]{Bias towards specific groups
  \citep[p.~181]{Shull2008}}{We distributed our questionnaire over
  general \ac{SE} channels.  We mix opportunity~(only 5 to 10\% chain
  referral), volunteer, and cluster-based sampling.  Selection bias, a
  problem in snowball
  sampling~\citep{Biernacki1981-SnowballSamplingProblems}, is limited
  by good visibility and accessibility of the target population in
  these channels~(\Cref{sec:descr-data-points}) as well as little use
  and control of referral chains among respondents.
  Our sample includes 50\% practitioners according to
  \Cref{sec:data-analysis}, %
  $\approx$ 21\% \acp{NP}~(incl.~laypersons), %
  and $\approx$ 31\% %
  pure academics.  A bias towards \ac{FM} experts~(\Cref{fig:D2_explev})
  does not harm our \ac{PEOU} discussion led by practitioners but shapes
  our \ac{PU} discussion.  Regarding application domains, our conclusions
  cannot be generalised to, \egs critical IT systems in the finance
  and e-voting sectors.
}

\threatPMit{Non-response}{We decided not to enforce responses or
  provide incentives.  Still, our data suggests that our advertisement
  stimulated responses from \ac{FM}-critical minds.}
   
\threat{Lack of \ac{FM} knowledge}{11 to 18\% of our respondents did
  not know specific challenges~(\Cref{fig:O1_obst}).  For
  \RQRef{1}~(\Crefrange{fig:D1_appdom}{fig:O1_obst}), $dnk$-data points have
  no influence because the findings of \RQRef{1} directly describe and
  interpret the status quo of $\mathit{UFM}_p$.  For test purposes, we included
  $dnk$-data points in the analyses of \RQRef{2} and
  \RQRef{3}~(\Crefrange{fig:F3_use_future}{fig:O1_obst}), with no relevant
  influence.}

\threatPMit{Geographical background missing}{Respondents were not
  required to own a Google account to avoid tracking and to increase
  anonymity and the response rate.  The limited geographical knowledge
  about our sample constrains the generalisability of our conclusions,
  \egs to geographies such as China, India, or Brazil.}

\subsubsection{Reliability}
\label{sec:validity-reliability}

Why would a repetition of the procedure in \Cref{sec:research-design}
with different samples from the same population lead to the same
results?

\threatPMit[maj]{Internal consistency}{%
  All 7 items for the concept ``obstacle to \ac{FM}
  effectiveness''~(\ConstructRef{Obst}) show good internal consistency
  for our sample with a Cronbach $\alpha = 0.84$, the
  \ac{PEOU}-part of \ConstructRef{Obst} consisting of 5 items shows an
  $\alpha = 0.79$~\citep{Shull2008}.  The other concepts are not
  measured with multiple items.}

\threatPMit[maj]{Change of proportions}{The limited sample and the low
  response rate make it hard to mitigate this risk.
  However, we compared the first~(til 4.8.2018, $N_1=114$) and
  second~(from 5.8.2018, $N_2=102$) half of our sample to simulate a
  repetition of our survey with the same questionnaire.  A two-sided
  Mann-Whitney U test for difference does not show a significant
  difference %
  between these two groups~(\egs for \QuestionRef{O1_obst} and
  \QuestionRef{P1_role_past}).  Only for the \QuestionRef{D3_motiv}
  item ``On behalf of \ac{FM} tool provider,'' a $p=0.07$ indicates a
  potential difference.  The addition of that item only after
  the 47th respondent might explain this difference.}

\section{Conclusions}
\label{sec:conclusions}

We conducted an on-line survey of mission-critical software
engineering practitioners and researchers to examine how formal
methods have been used, how these professionals intend to use them,
and how they perceive challenges in using them.  This study
aims to contribute to the body of knowledge of the software
engineering and formal methods communities.

\paragraph{Overall Findings.}

From the evidence we gathered for the use of formal methods, we make
the following observations:
\begin{itemize}
\item %
  \emph{Intrinsic motivation} is stronger than the regulatory one.
\item %
  Despite the challenges, our respondents show an \emph{increased
    intent} to use \acp{FM} in industrial contexts.
\item %
  Past experience is \emph{correlated} with usage intent. 
\item %
  All challenges were rated \emph{either moderately or highly}
  difficult, with
  scalability, skills, and education leading.
  Experienced respondents rate challenges as highly difficult more
  often than less experienced respondents.
\item %
  From the literature and the responses, we identified three
  additional challenges: \emph{sufficient resources}, \emph{process
    compatibility}, \emph{good practicality/reputation}.
\item %
  The negative responses to the questions about obstacles to FM
  effectiveness suggest that the \emph{ease of use of \acp{FM}} is
  perceived more negative than positive.
\item %
  Gaining experience and confidence in the application of a \ac{FM} seems
  to play a role in developing a \emph{positive perception of usefulness
    of that \ac{FM}}.
\end{itemize}
\citet{Barroca1992} present evidence to show that \acp{FM} can be used
in industry effectively and more widely.  Their observation from 1992
is that \ac{FM} use had been limited, benefits were clear but
limitations were subtle.  In response to \citeauthor{Barroca1992}'s
finding ``\acp{FM} are both oversold and under-used'', our insights
from the analysis of RQ~2 and 3 lead us to conclude that today
\acp{FM} are probably more underused than oversold.  However, our data
also suggests that these methods still need substantial improvement
and support in several areas in order for their benefits to be better
utilised.

\paragraph{General Feedback on the Survey.}
\label{sec:gener-feedb-surv}

The questionnaire seems to be well-received by the
participants.  One of them found it an ``interesting set of
questions.''  This impression is confirmed by another
participant:
\begin{quote}
  \small
  ``Well chosen questions which do not leave me guessing.  Relevant to
  future \ac{FM} research and practice.''
\end{quote}
Another respondent noted:
\begin{quote}
  \small
  ``Thank you very much for this survey. It is very constructive and
  important. It handles most of the issues encountered by any
  practitioner and user of \acp{FM}.''
\end{quote}
Only one participant found the questionnaire difficult for \ac{FM} beginners.

\paragraph{Implications Towards a Research Agenda.}

In the spirit of \citet{Jeffery2015} and complementing the suggestions
from the \ac{SWOT} analysis in
\citet{Gleirscher2019-NewOpportunitiesIntegrated}, we want to make
another step in setting out an agenda for future \ac{FM} research.

To address \emph{scalability}, we need more research on how
compositional methods~(\egs automated assume-guarantee reasoning,
\cite{Cofer2012-CompositionalVerificationArchitectural}; %
automated assertion checking,
\cite{Leino2017-AccessibleSoftwareVerification}) can be better
leveraged in practical settings.
To address \emph{skills and education}, we need an enhanced and
up-to-date \emph{\ac{FM} body of knowledge}~(FMBoK; \cite{Oliveira2018}).
From his survey of ``\acp{FM} courses in European higher education'',
\citet{Oliveira2004} observes that
\begin{inparaenum}[(i)]
\item ``model-oriented specification'', ``formalising distribution,
  concurrency and mobility'', and ``logical foundations of formal
  methods'' showed to be the topic areas most frequently taught by \ac{FM}
  lecturers, and
\item Z, B, SML, CSP, and Haskell showed to be the most popular formal
  notations and languages taught in these courses.
\end{inparaenum}
A comparison of the current state with \citeauthor{Oliveira2004}'s
observations can help to evaluate and revise current \ac{FM}
curricula~(\egs for undergraduate \ac{SE} as suggested in
\cite{Davis2013-StudyBarriersIndustrial}) and to derive
recommendations for improved \ac{FM} courses fostering good modelling,
composition, and refinement skills in \ac{SE} practice.
To address \emph{controllable abstractions}, 
we need semantics workbenches for underpinning domain-specific
languages %
with formal semantics.
We believe that further steps in \emph{theory integration and
  unification}~\citep{Gleirscher2019-NewOpportunitiesIntegrated} can
help establish proof hierarchies and, hence, \emph{reusability} and
\emph{proof transfer}.

To address \emph{process compatibility}, we need more research in
\emph{continuous reasoning}~(\egs\cite{OHearn2018,Chudnov2018}), a
revival of activities, possibly even regulations, in tool integration
and model data interchange, and guidance on how to update engineering
development processes.  
To address \emph{reputation}, we need to provide more incentives for
practitioners to use \acp{FM} and take recent progress in \ac{FM}
research into account when changing current software processes,
policies, regulations, and standards.
This includes convincing practitioners to invest in the support of
large-scale studies for monitoring \ac{FM} use in industry.
Cost-savings analyses %
of \ac{FM} applications~(\egs\cite{Jeffery2015}) supported by strong
empirical designs~(\ies controlled field experiments) can help to
collect the necessary evidence for decision making, successful
knowledge transfer, and for implementing this vision.

This survey underpins and enhances the analysis of strengths and
weaknesses of \acp{FM} in
\citet{Gleirscher2019-NewOpportunitiesIntegrated} and can be a guide
\begin{inparaenum}[(1)]
\item for consulting and managing practitioners when considering the
  introduction of \acp{FM} into a engineering organisation, 
\item for research managers when shaping a grant programme for \ac{FM}
  experimentation and transfer, and
\item for associate editors when organising a journal special section
  on applied \ac{FM} research.
\end{inparaenum}

\paragraph{Future Work.}
\label{sec:futurework}

Our survey is another important step in the research of effectively
applying FM-based technologies in practice.  To put it with the words
of one of our participants: ``[A] closed questionnaire is just a
start.''

Hence, we aim at a follow-up study
\begin{inparaenum}[(i)]
\item to find out which particular \ac{FM} (and tool) is used in which
  domain for which particular purpose and role~(\egs was \acs{SMT}
  solving used for model checking in certification or for task
  scheduling at run-time?),
\item to measure where particular techniques work well~(\egs which
  types of formal contracts work well in control software requirements
  management in a DO-178C context?),
\item to measure key indicators for successful use of \acp{FM},
\item to identify management techniques needed to accommodate the
  changes in working practices, and, finally,
\item to provide guidance to future projects wishing to adopt \acp{FM}.
\end{inparaenum}

In a next survey, we like to ask about typical \ac{FM} benefits,
about suggestions for barrier
mitigation~\citep{Davis2013-StudyBarriersIndustrial}, pose more
specific questions on scalability and useful abstraction, the
geographical\footnote{According to
  \url{https://en.wikipedia.org/wiki/United_Nations_geoscheme}.}  and
educational background, and for conceptual alignment.  Further
analysis of obstacles, benefits, and usage intent could also benefit
from a more fine-grained distinction between \acp{FM} directly applied
to program code and \acp{FM} focusing on more abstract models.  We
would also like to change from 3-level to 5-level Likert-type
scales to receive fine-granular responses.  Our research design
accounts for repeatability, hence, allowing us to go for a
longitudinal study.

The research design, and even our current data set, allows the
derivation of the usage intent~($\mathit{UFM}_i$) for each \ac{FM} class,
application domain, and obstacle.  These $\mathit{UFM}_i$ values could be used
to analyse whether a particular \ac{FM} might be
\begin{inparaenum}[(1)]
\item underused~(\ies domains with an increased usage intent indicate
  a potential for more applicability) or
\item oversold~(\ies domains with a lower usage intent and were
  obstacles are perceived as being particularly tough and, hence,
  \acp{FM} as being less effective).
\end{inparaenum}

\begin{acknowledgements}
\label{sec:ack}
It is our pleasure to thank all survey participants for their time
spent and their valuable responses, and all channel moderators for
forwarding our postings.
We are much obliged to Jim Woodcock, who has led previous studies in
our direction, and supported us to critically reflect our work and
relate it to existing evidence.  He connected us with John Fitzgerald,
who made his paper copy of \citet{Austin1993-Formalmethodssurvey}
available to us such that we were able to complete our investigation.
We are grateful to John Fitzgerald and also to John McDermid for
helpful feedback and for encouraging us to do further research in this
direction.  
We would like to spend sincere gratitude to Krzysztof Brzezinski,
Louis Brabant, and Emmanuel Eze for pointing us to several related
works.
\end{acknowledgements}

\footnotesize
\printbibliography

\newpage
\appendix 
\normalsize

\section{Supplementary Material for 
  ``Formal Methods in Dependable Software Engineering: A Survey''}
\label{sec:appendix}

In the following, we provide additional material to the survey,
including
\begin{enumerate}
\item a more detailed analysis of responses to certain questions~(\Cref{sec:rq1-extval}),
\item further visualizations of the collected data~(\Crefrange{sec:data-geosample}{sec:data-rq3}),
\item more details on our analysis of related
  work~(\Cref{tab:relwork:details} in \Cref{sec:relat-work-deta}), 
\item more details on the mapping from studies to
  challenges~(\Cref{sec:map-lit-chall}),
\item a comprehensive table of open answers~(\Cref{sec:answ-open-quest}),
\item a copy of the advertisement flyer~(\Cref{sec:questionnaire-flyer}), 
\item a screenshot of the Twitter poll~(\Cref{sec:questionnaire-twitter}), and
\item a copy of the whole questionnaire~(\Cref{sec:questionnaire-complete}).
\end{enumerate}

\subsection{Data for Analysis of \RQRef{1} and Estimation of External
  Validity}
\label{sec:rq1-extval}

Based on the responses for question~\QuestionRef{D1_appdom_past}, the
\Cref{tab:respcat} provides an overview of categories of respondents
referred to in our analysis (particularly, in \Cref{sec:D1_appdom} and
\Cref{fig:rq2_comparison_P1_F2}) along with the corresponding counts
based on the sample from
31.3.2019 with $N=220$.

For question~\QuestionRef{D1_appdom_past}, by practitioner, we mean
``practitioner in dependable or mission-critical software
engineering.''  To include respondents from all areas of the
population or at any study stage, we generalize ``practitioner'' by
the term ``user''.  Below, for the questions~\QuestionRef{P2_use_past}
and \QuestionRef{P3_use_past}, we then refer to ``formal method user''
and ``\ac{FM}-non-user''.

\begin{table}
  \caption{Overview of categories of respondents 
    for
    \QuestionRef{D1_appdom_past},
    \QuestionRef{D3_motiv},
    \QuestionRef{P1_role_past}, 
    \QuestionRef{P2_use_past}, and
    \QuestionRef{P3_use_past}
  \label{tab:respcat}}
  \footnotesize
  \begin{tabularx}{1.0\linewidth}{>{\hsize=.45\hsize}L>{\hsize=.55\hsize}Lrr}
    \toprule
    \textbf{Category of respondents to \dots}
    & \textbf{Description} 
    & \textbf{Count} & \textbf{Fraction}
    \\\midrule
    \multicolumn{4}{l}{\dots~\textbf{Question~\QuestionRef{P1_role_past}:}}
    \\

    \rowcolor{lightgray}
    \textbf{Respondents with academic educational background~(AEB)}
    & Researchers in academia; bachelor, master, or PhD students; 
      lecturers, teachers, trainers, coaches
    & 156 & 72\%
    \\
    Academics with pure transfer experience 
    & \acs{AEB} cut with researchers in industry, consulting and managing
      practitioners, and external consultants;
      \textbf{without} engineering practitioners in industry and without tool
      provider stakeholders
    & 35 & 16\%
    \\
    Academics with practical experience 
    & AEB cut with engineering practitioners in industry
    & 41 & 21\%
    \\
    Academics with experience in transfer and practice 
    & AEB cut with researchers in industry, consulting and managing
      practitioners, and external consultants;
      \textbf{cut with} engineering practitioners in industry and without tool
      provider stakeholders
    & 31 & 14\%
    \\ 
    Practitioners incl.~transfer practitioners and industrial consultants,
    all with academic background 
    & AEB cut with researchers in industry, consulting and managing
      practitioners, external consultants, and
      engineering practitioners in industry 
    & 86 & 40\%
    \\ 
    \rowcolor{lightgray}
    Pure academics 
    & AEB without respondents specifying additional roles
    & 66 & 31\%
    \\
    \midrule
    \rowcolor{lightgray}
    \textbf{Respondents not specifying an educational background~(NEB)}
    & The complement of AEB 
    & 60 & 27\%
    \\
    Respondents not specifying an educational background and being researchers
    in industry 
    & \acs{NEB} intersected with researchers
      in industry
    & 13 & 6\%
    \\
    Consultants 
    & NEB intersected with consulting or managing practitioners and external consultants
    & 23 & 11\%
    \\
    \rowcolor{lightgray}
    Pure practitioners 
    & NEB cut with engineering practitioners in industry
    & 23 & 11\%
    \\
    Tool provider stakeholders not specifying an educational background 
    & NEB cut with stakeholder of an \ac{FM} tool or service
      provider
    & 5 & 2.3\%
    \\
    Non-academic \ac{FM} non-users 
    & NEB cut with ``I have not used \acp{FM} in any specific role.''
    & 19 & 9\%
    \\\midrule
    \rowcolor{lightgray}
    Practitioners incl.~industrial consultants 
    & Consulting and managing
      practitioners, external consultants, and
      engineering practitioners in industry 
    & 108 & 50\%
    \\
    \ac{FM} users (all)
    & Respondents having used \acp{FM} in one or another way and context in the
      past
    & 212 & 98\%
    \\
    \rowcolor{lightgray}
    \ac{FM} users (beyond students)
    & Excl. ``only-in-course'' respondents
    & 202 & 93.5\%
    \\\midrule
    \multicolumn{4}{l}{\dots~\textbf{Question~\QuestionRef{D1_appdom_past}:}}
    \\
    \ac{FM} non-users
    & Respondents who chose ``I have not used \acp{FM} in any academic or
      industrial domain.''   
    & 36 & 17\%
    \\\midrule
    \multicolumn{4}{l}{\dots~\textbf{Question~\QuestionRef{D3_motiv}:}}
    \\
    Respondents with no motivation
    & Respondents who selected ``no'' for all
      motivating factors 
    & 9 & 4\%
    \\\midrule
    \multicolumn{4}{l}{\dots~\textbf{Questions~\QuestionRef{P2_use_past} and \QuestionRef{P3_use_past}:}}
    \\
    Non-practitioners including \ac{FM} non-users
    & Respondents who chose ``no experience or no knowledge'', ``studied
      in (university) course'' or ``applied in lab, experiments, case studies''
      for all \ac{FM} techniques. This group includes laypersons.  
    & 46 & 21\%                         
    \\\midrule
    Sample (N)
    & All valid responses
    & 216 & 100\%
    \\\bottomrule
  \end{tabularx}
\end{table}

\clearpage

\subsection{Geographical Analysis of the Sample}
\label{sec:data-geosample}

\begin{figure}
  \includegraphics[width=.9\linewidth]{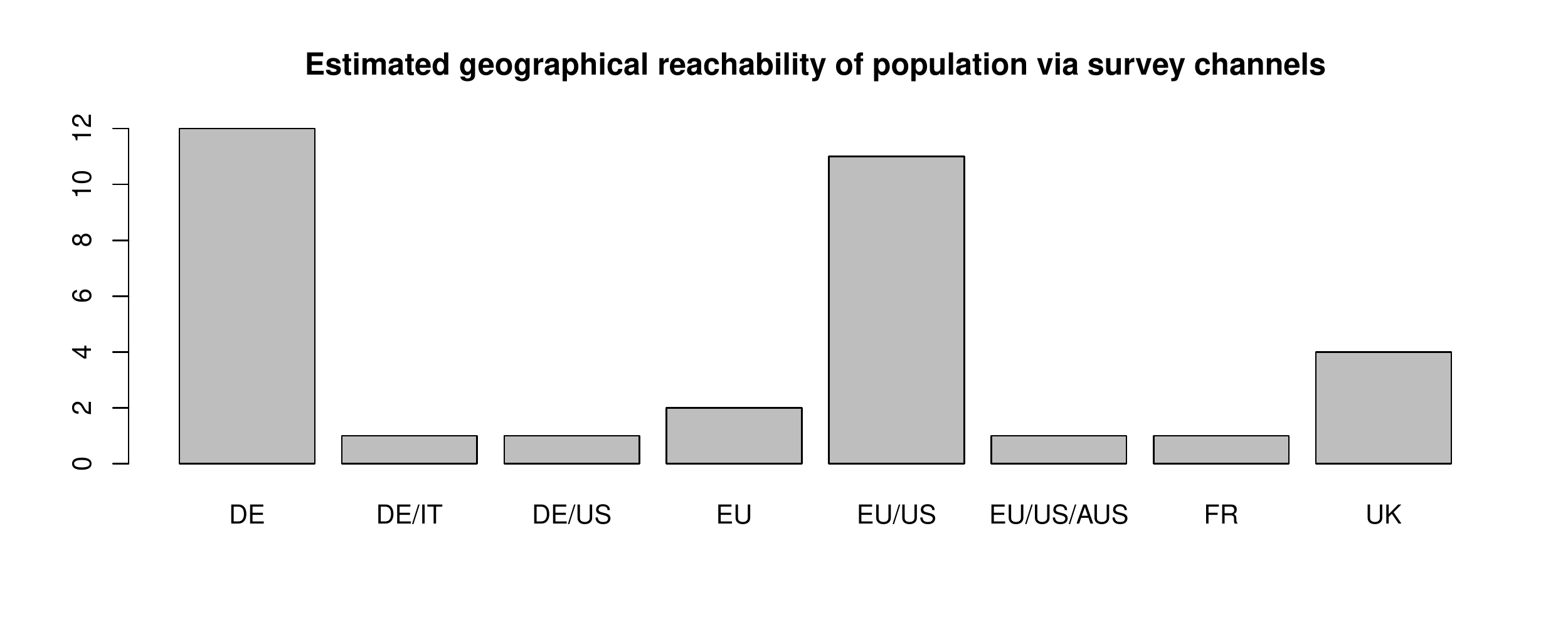}\newline
  \includegraphics[width=.9\linewidth]{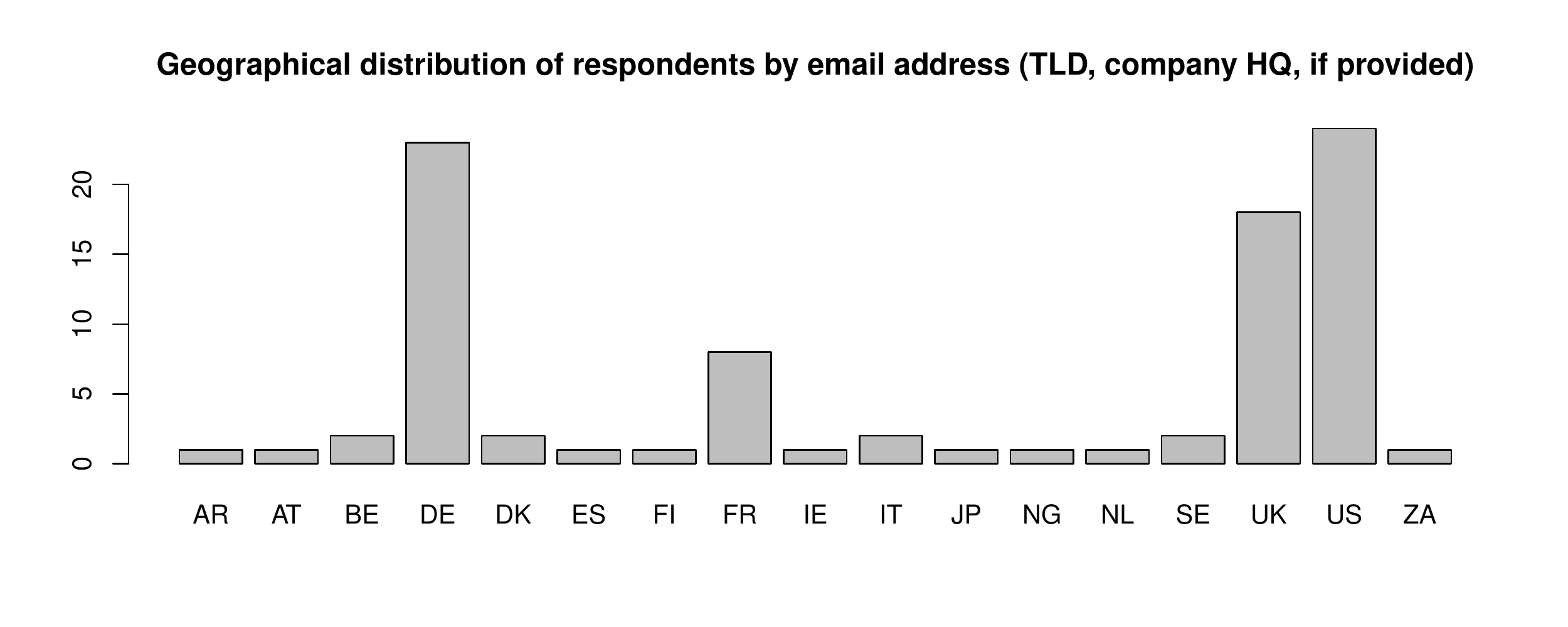}\newline
  \caption{Geographical analysis of the sample. \emph{Legend:}
    \ac{TLD}, \ac{HQ}
    \label{fig:data-geosample}} 
\end{figure}

\Cref{fig:data-geosample} shows geographical aspects of the sample for this
study.

\subsection{Usage Intent ($\mathit{UFM}_i$) by Purpose (for Analysis of \RQRef{2})}
\label{sec:data-intent-by-purpose}

\begin{figure}
  \centering
  \includegraphics[width=\linewidth]{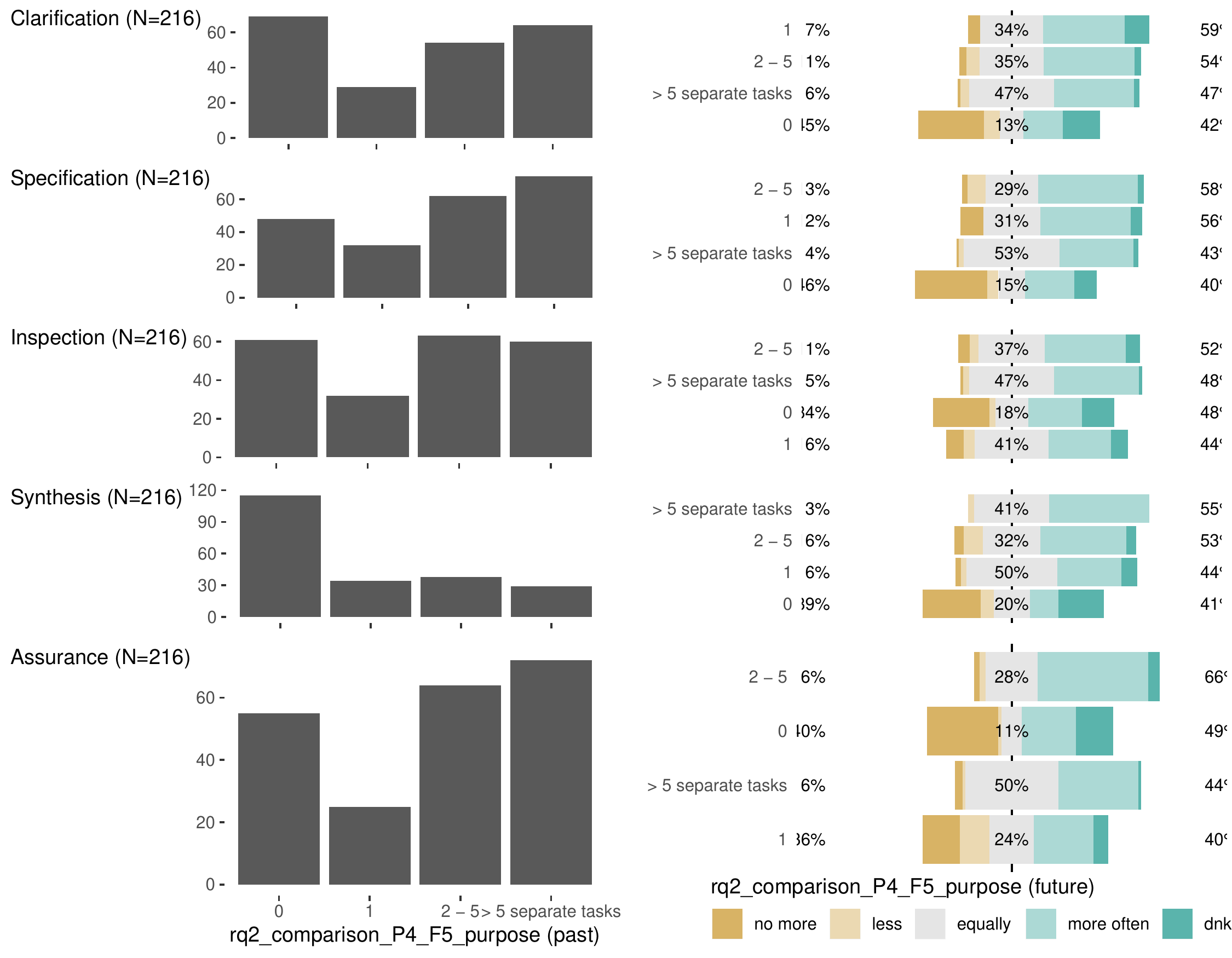}
  \caption{Comparison of past and future usage intent by purpose
    \label{fig:rq2_comparison_P4_F5_purpose}}
\end{figure}

The comparison in \Cref{fig:rq2_comparison_P4_F5_purpose} (and in the
figures of the \Cref{sec:data-intent-by-tech,sec:data-intent-cbfm-mbfm})
contains two columns.

The left column describes for each purpose~(\egs specification) how
often~(\egs in 2 to 5 separate tasks) respondents have used \acp{FM}
in the past~($\mathit{UFM}_p$).

The right column describes for each purpose the usage intent~($\mathit{UFM}_i$)
depending on how often respondents have used \acp{FM} in the past
($\mathit{UFM}_p$).  The horizontal bars representing the $\mathit{UFM}_p$ frequency
categories are listed in descending order by the overall size of both
$\mathit{UFM}_i$ groups ``more often'' and ``dnk''.  We chose to keep
\emph{dnk}-answers visible despite the readability inconvenience
caused by the \emph{dnk}s influencing the ordering.  However, in the
majority of cases the largest group of respondents intending to
increase \ac{FM} use in the future is visible first or near the top.

\subsection{Code-based \vs Model-based FMs for Assurance \vs Inspection}
\label{sec:data-intent-cbfm-mbfm}

\begin{figure}
  \centering
  \includegraphics[width=.8\textwidth]{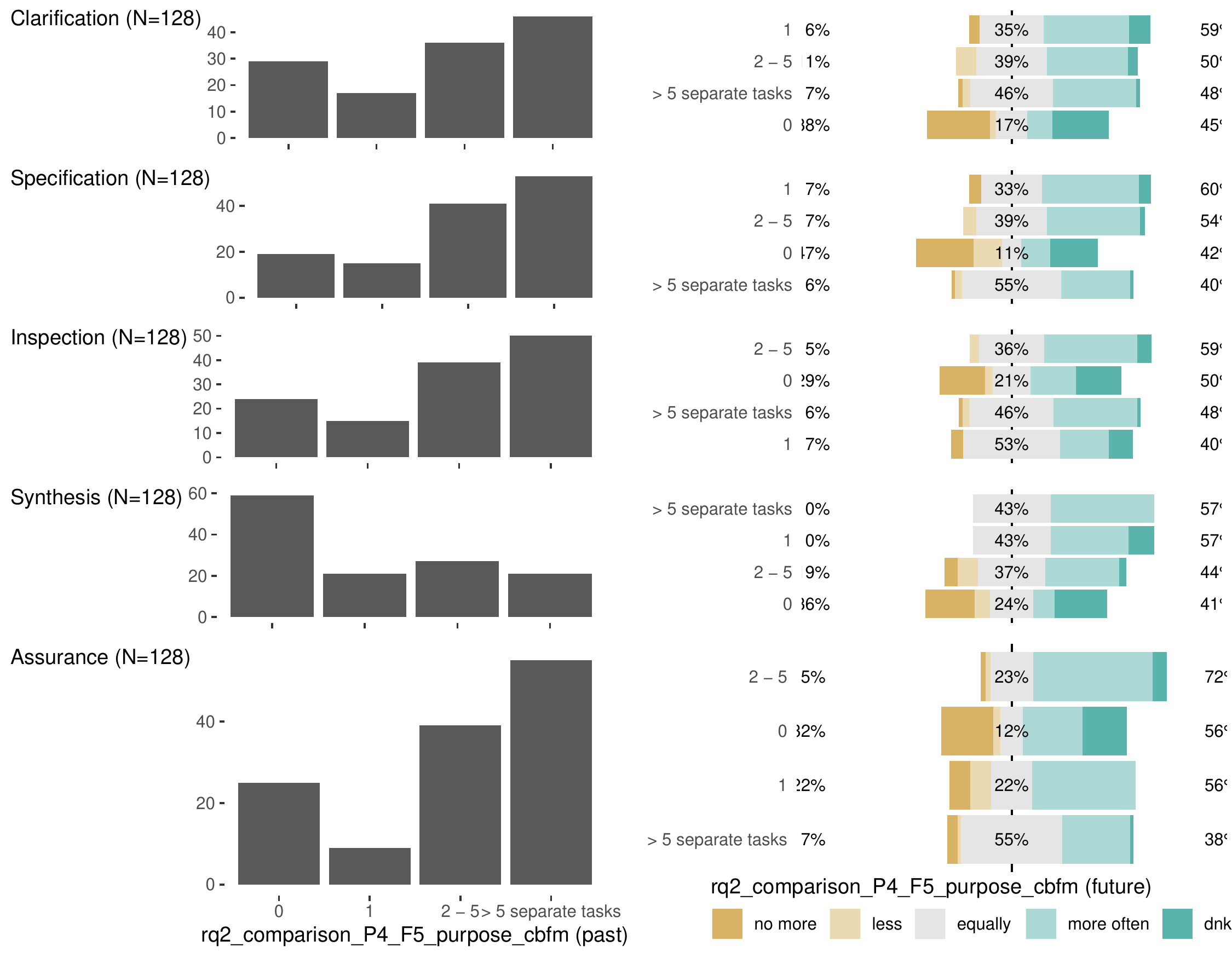}
  \includegraphics[width=.8\textwidth]{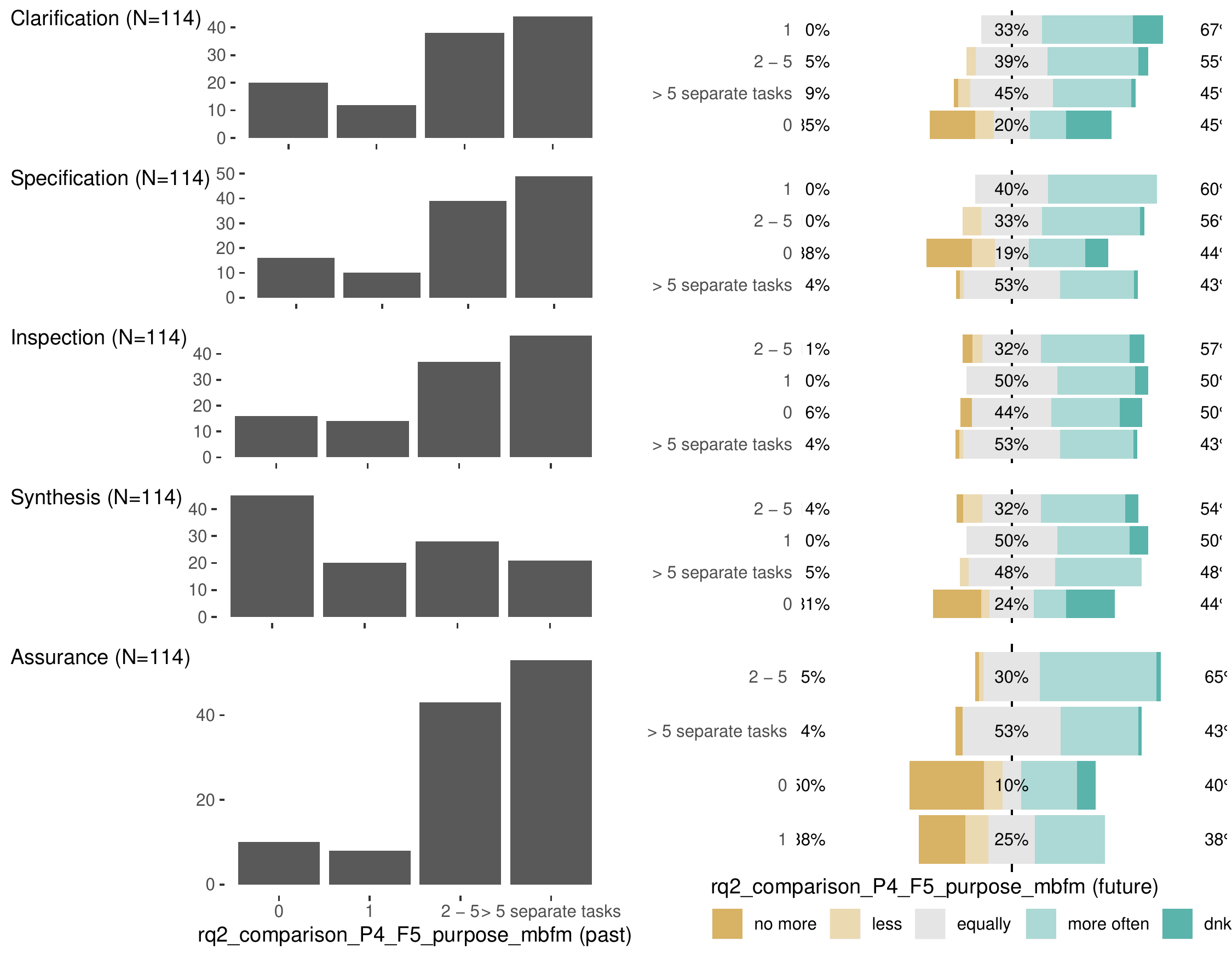}
  \caption{Comparison of past and future usage for code-based (top half) and model-based
    \acp{FM} (bottom half) by purpose
    \label{fig:rq2_comparison_cbfm_mbfm}}
\end{figure}

The data for this comparison is summarised in \Cref{fig:rq2_comparison_cbfm_mbfm}.

\clearpage

\subsection{Usage Intent ($\mathit{UFM}_i$) by FM Class (for Analysis of \RQRef{2})}
\label{sec:data-intent-by-tech}

\includegraphics[height=.9\textheight]{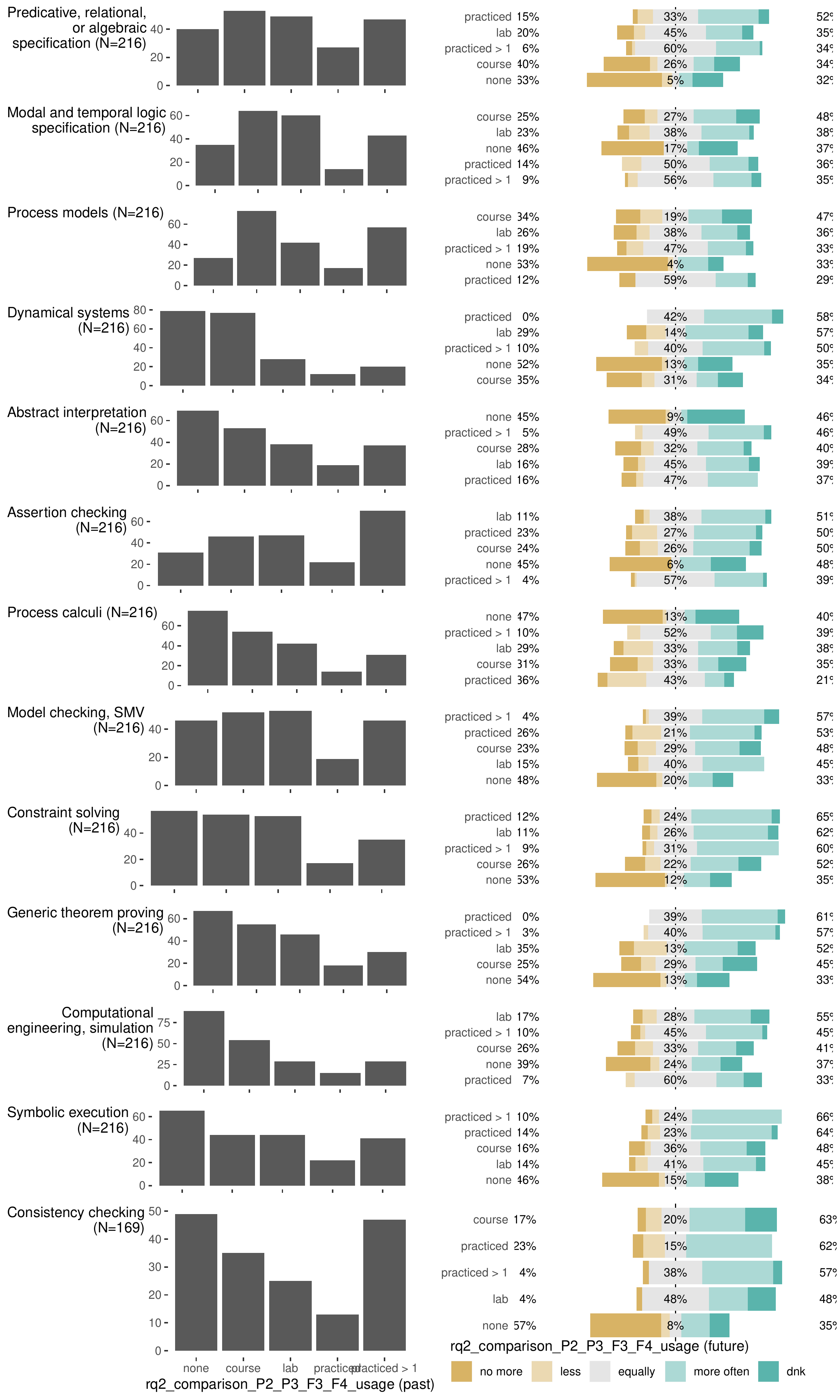}

\clearpage

\subsection{Data for the Analysis of \RQRef{3}}
\label{sec:data-rq3}

\Cref{fig:rq3_heatmapP4_O1} and the following figures in this section
show pairs of matrices, so-called ``heat\-maps'', useful for association
analysis between categorical and ordinal variables.  The cells in the
matrices represent combinations of the scales, each cell containing
data about the \emph{mode} and \emph{median} of ``degree of
difficulty'' ratings, their \emph{proportion} of \emph{tough} ratings,
and the \emph{actual numbers} of data points.  Both the colour
gradient~(red to white) and the solid vertical lines in the cells
represent the tough proportions~(left = 0 to right = 100\%), with the
dotted vertical line signifying the 50\% margin.

\begin{figure}[h]
  \includegraphics[width=.85\linewidth]{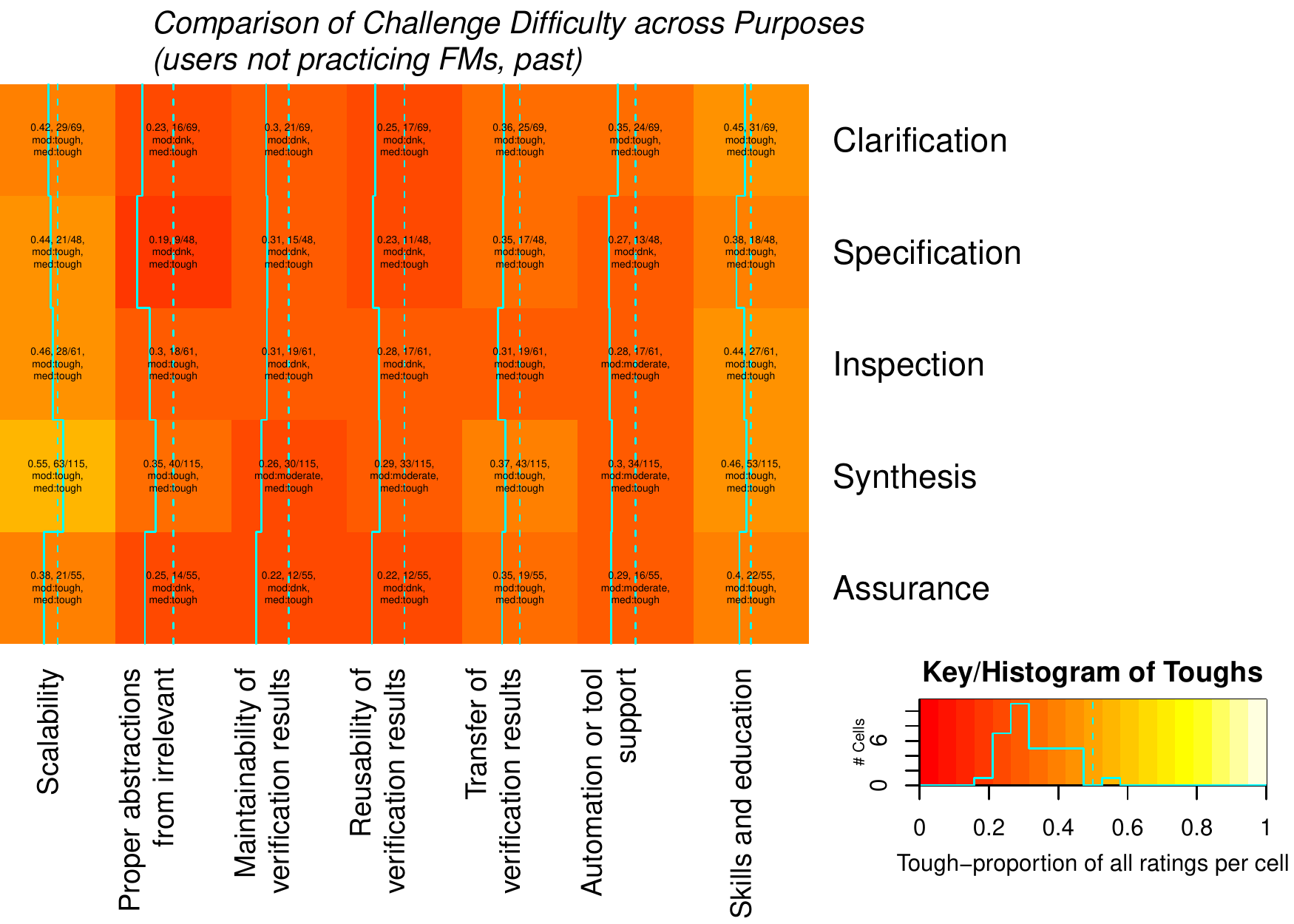}
  \includegraphics[width=.85\linewidth]{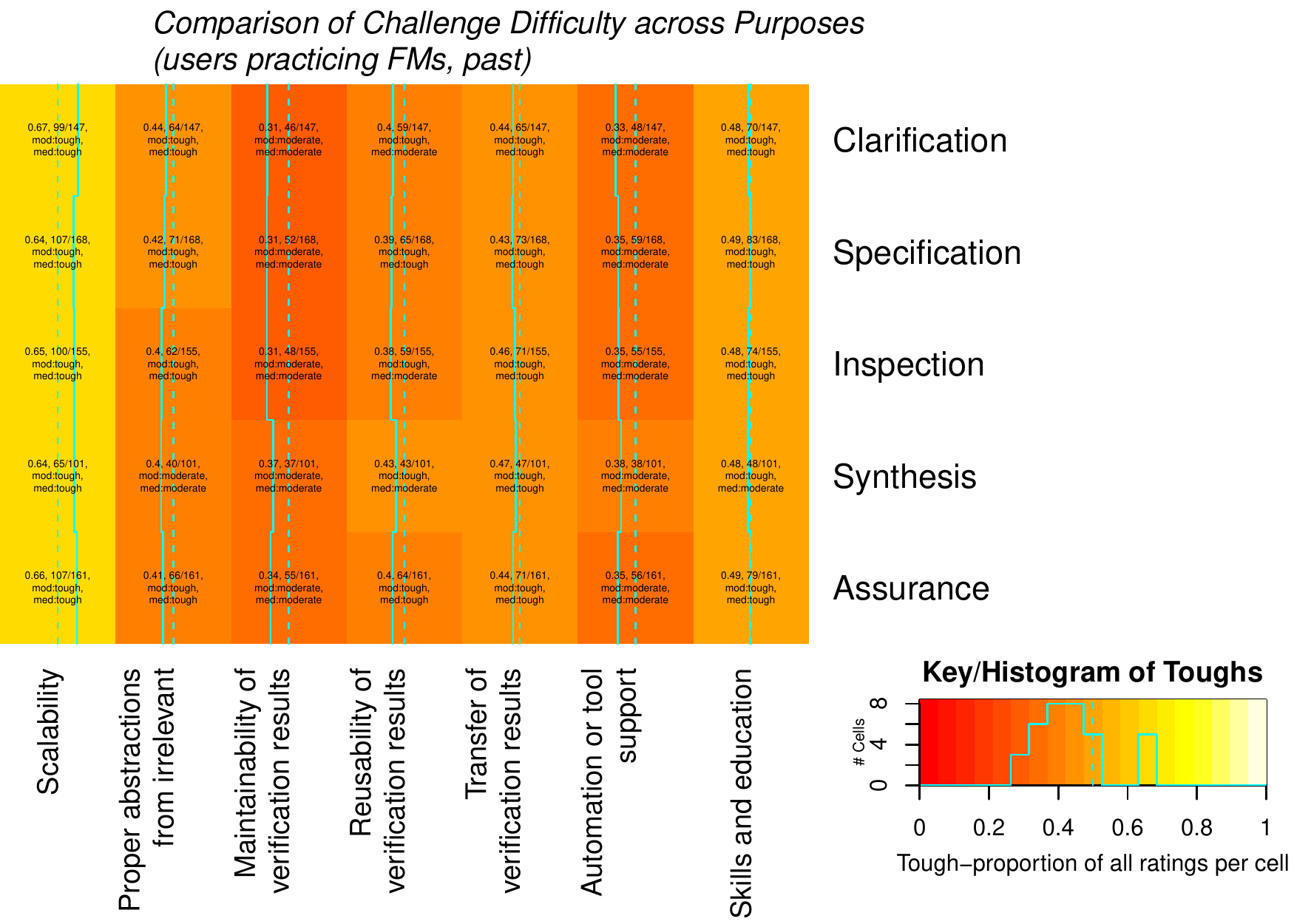}
  \caption{Comparison of challenge difficulty across purposes ($\mathit{UFM}_p$)
    \label{fig:rq3_heatmapP4_O1}}
\end{figure}

\clearpage

\begin{figure}
  \includegraphics[width=\linewidth]{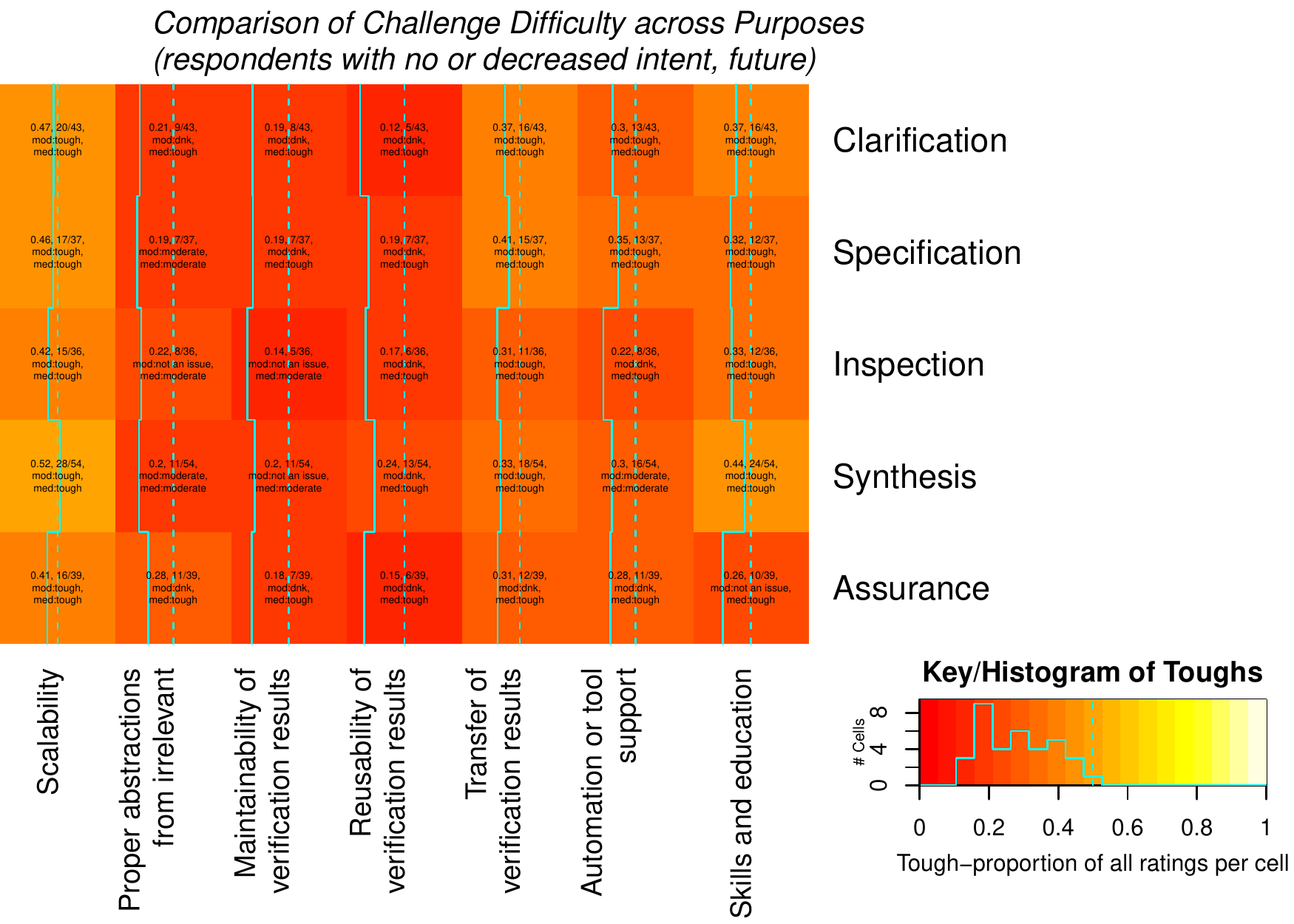}
  \includegraphics[width=\linewidth]{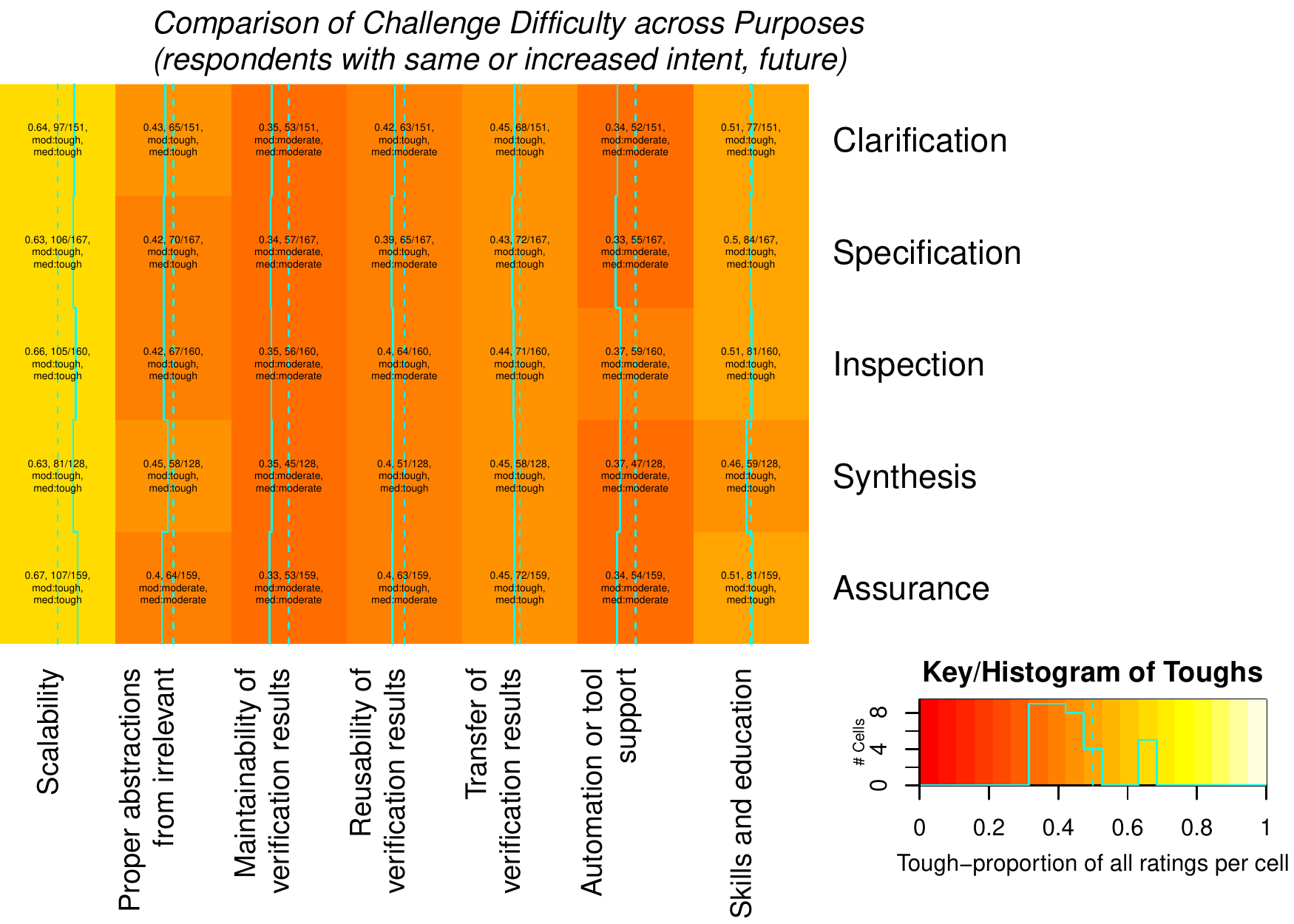}
  \caption{Comparison of challenge difficulty across purposes ($\mathit{UFM}_i$)
    \label{fig:rq3_heatmapF5_O1}}
\end{figure}

\begin{figure}
  \includegraphics[width=\linewidth]{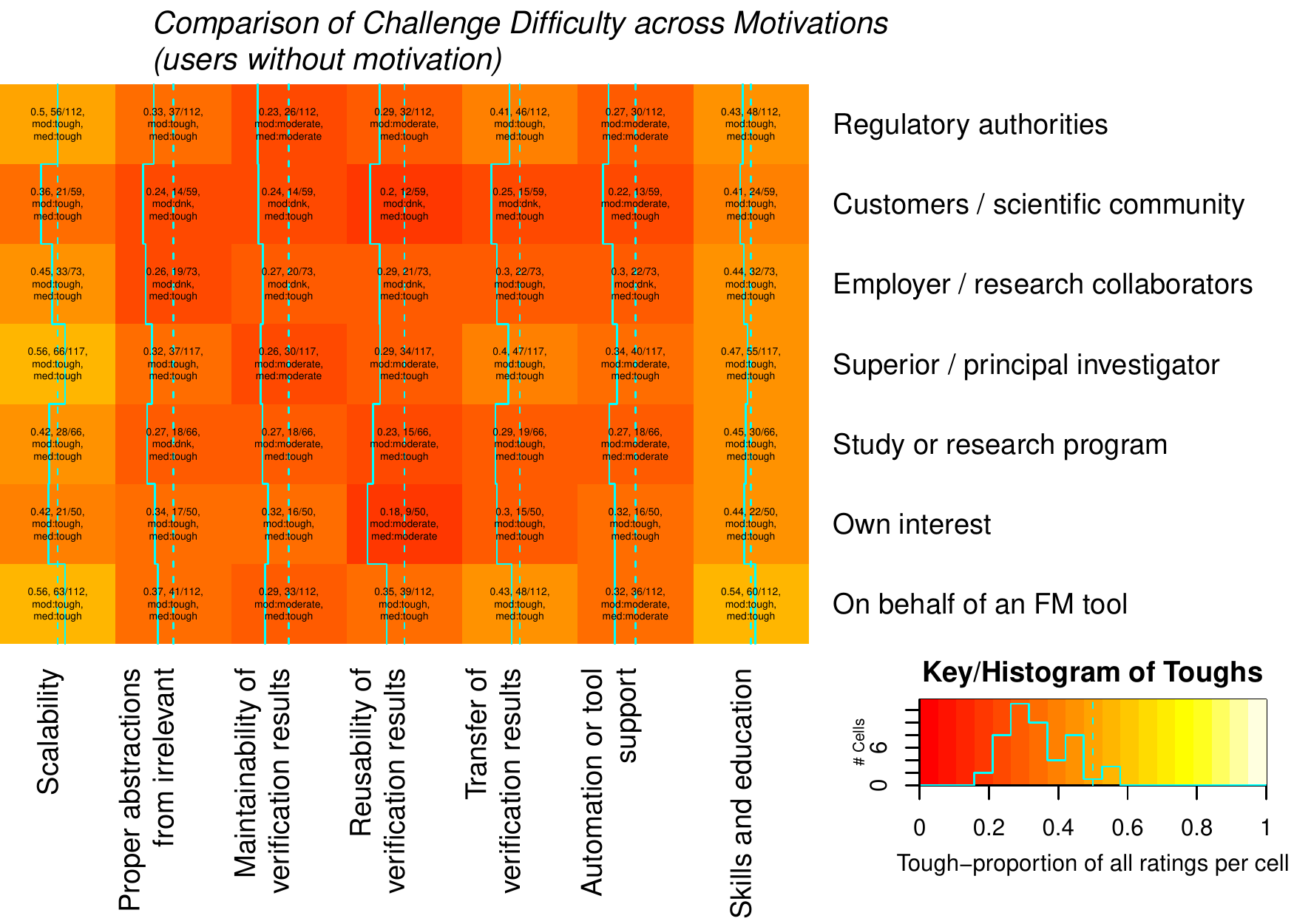}
  \includegraphics[width=\linewidth]{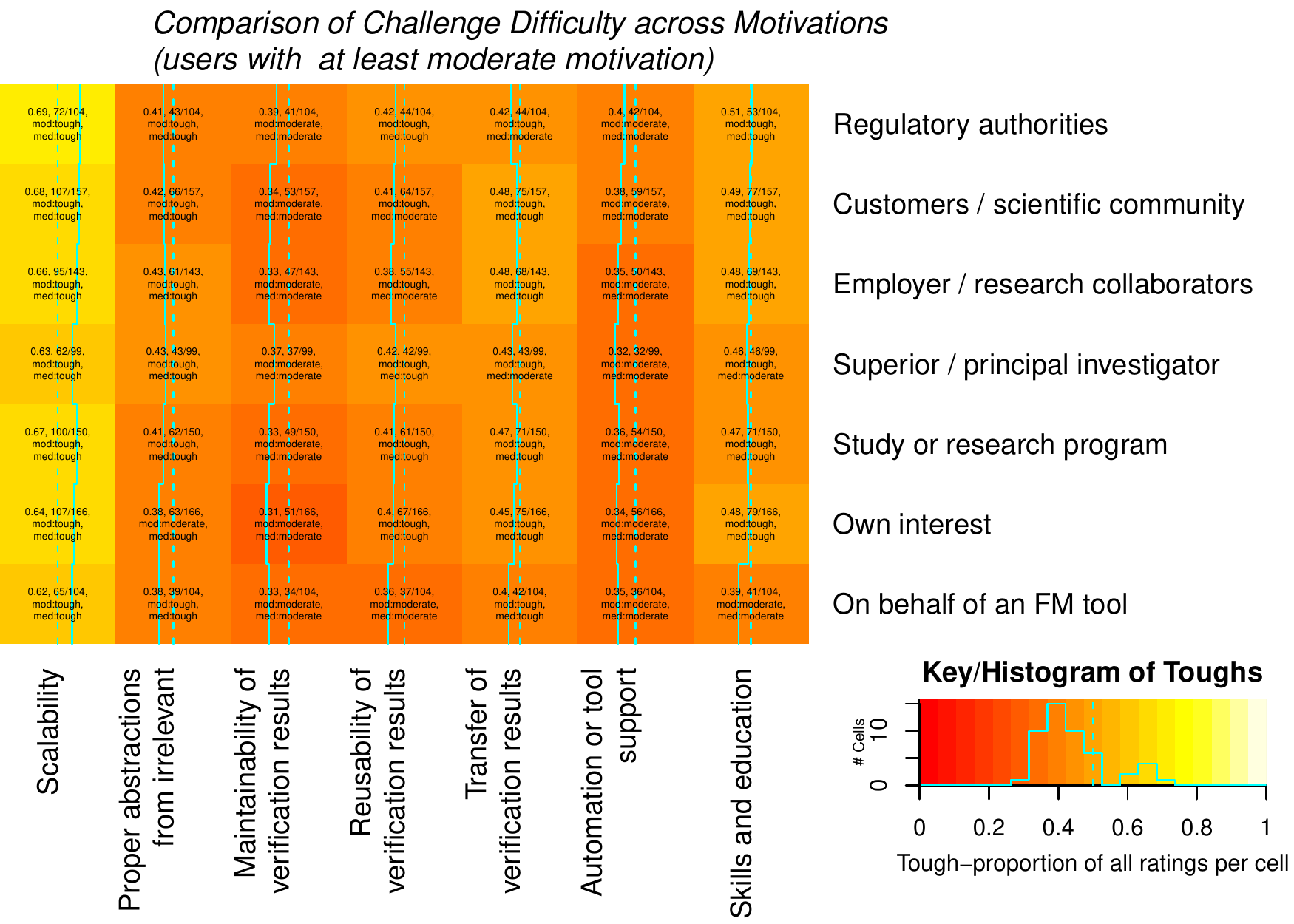}
  \caption{Comparison of challenge difficulty across motivations
    \label{fig:rq3_heatmapD3_O1}}
\end{figure}

\begin{figure}
  \includegraphics[width=\linewidth]{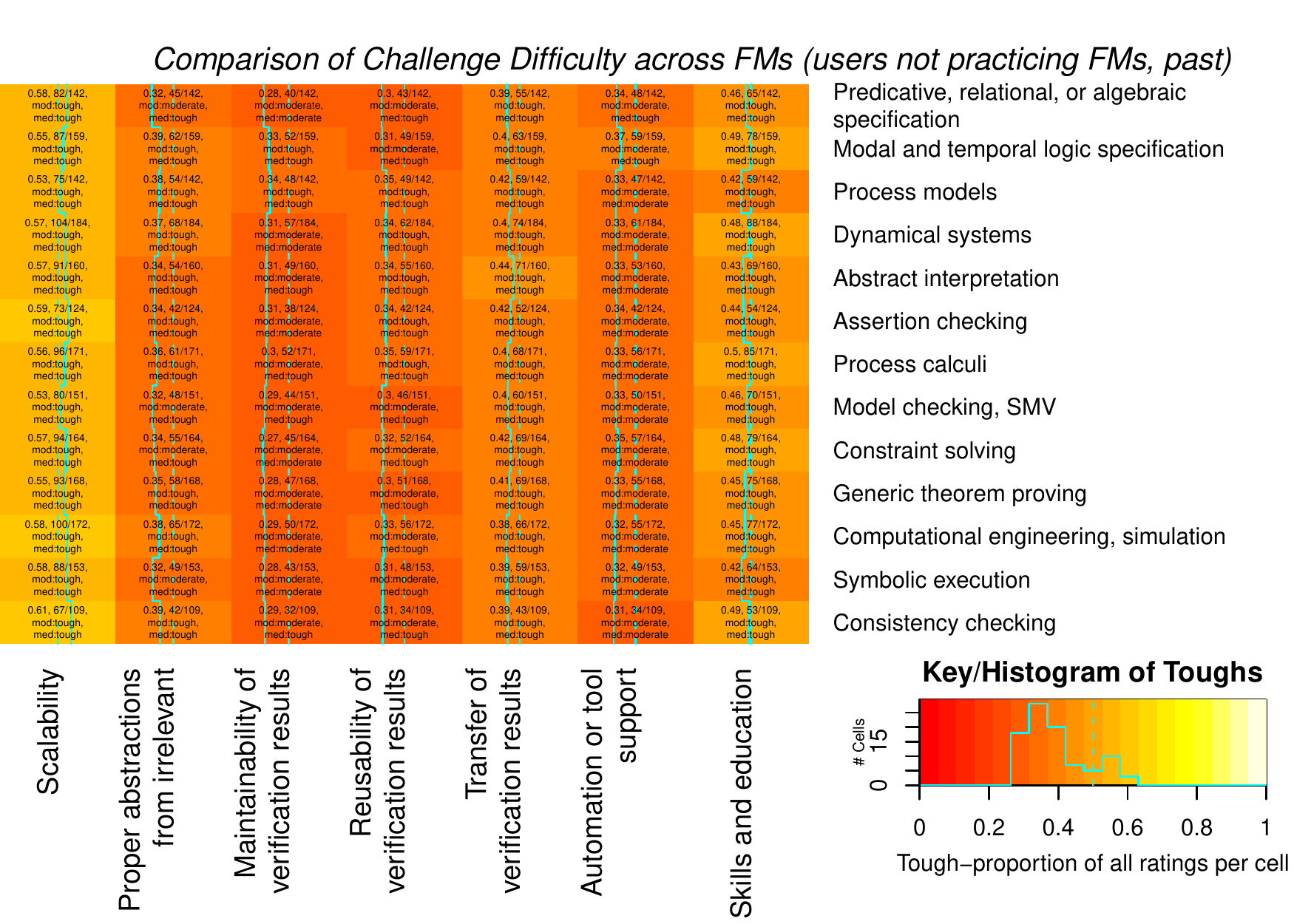}
  \includegraphics[width=\linewidth]{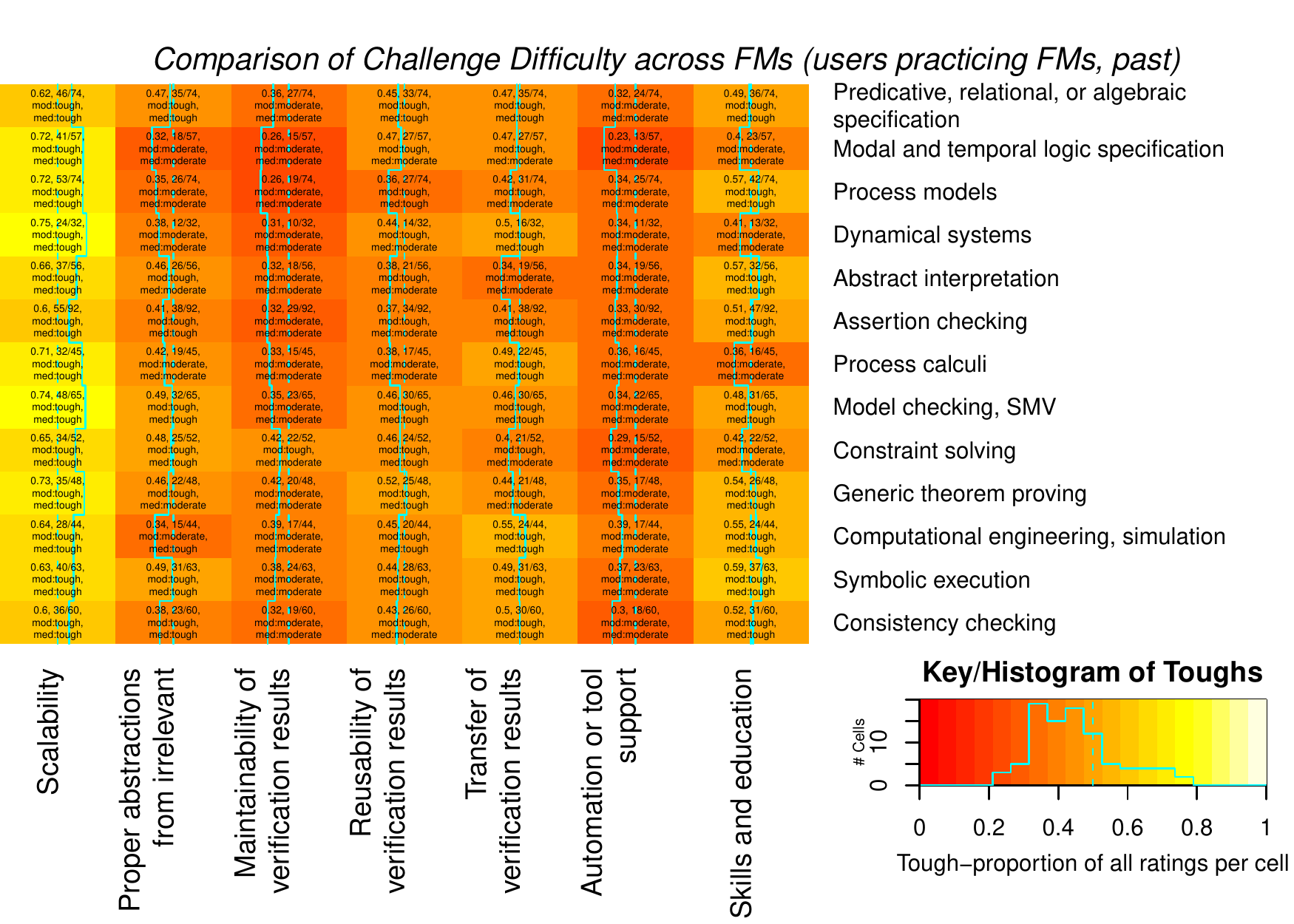}
  \caption{Comparison of challenge difficulty across \ac{FM} classes ($\mathit{UFM}_p$)
    \label{fig:rq3_heatmapP2_P3_O1}}
\end{figure}

\begin{figure}
  \includegraphics[width=\linewidth]{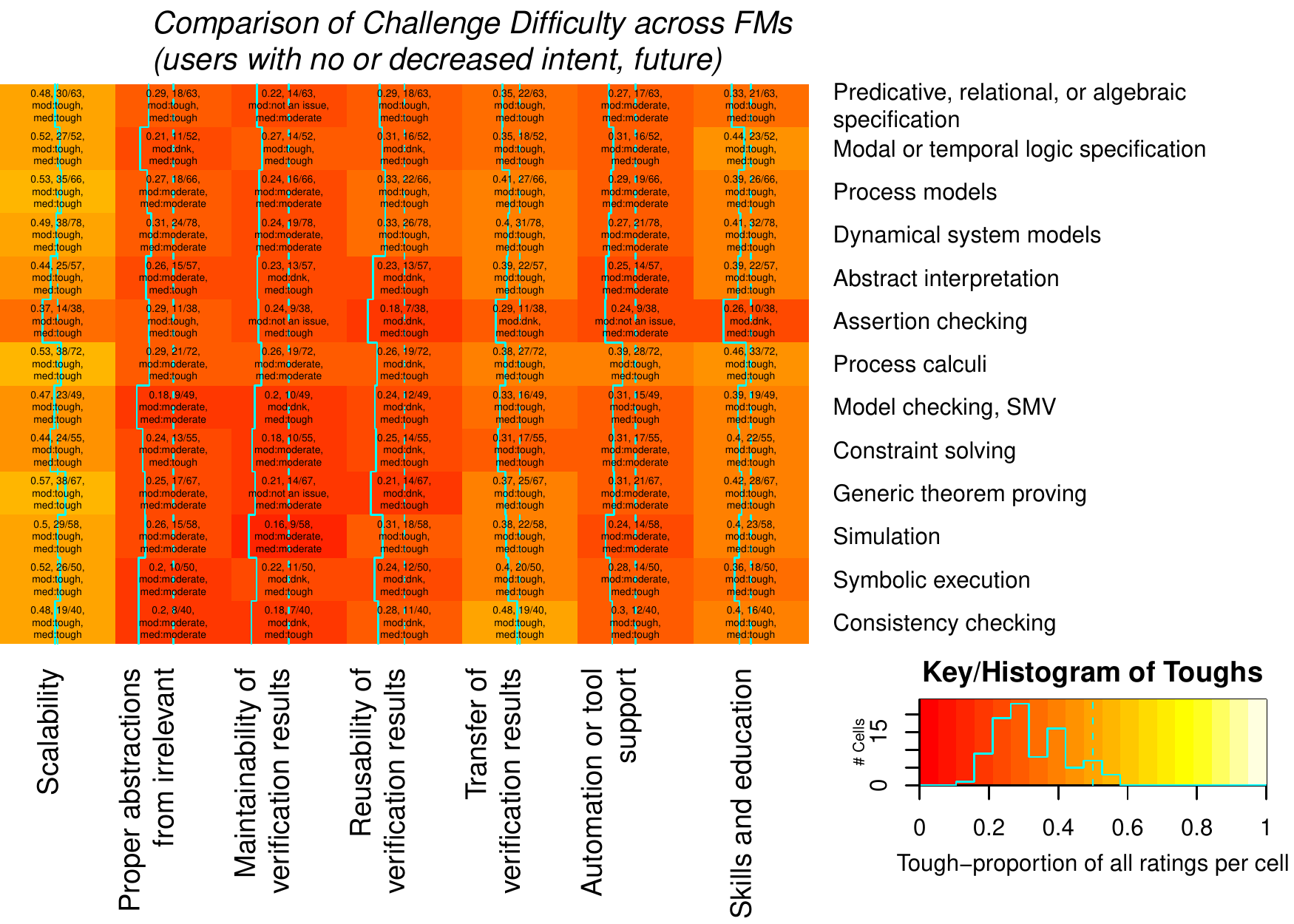}
  \includegraphics[width=\linewidth]{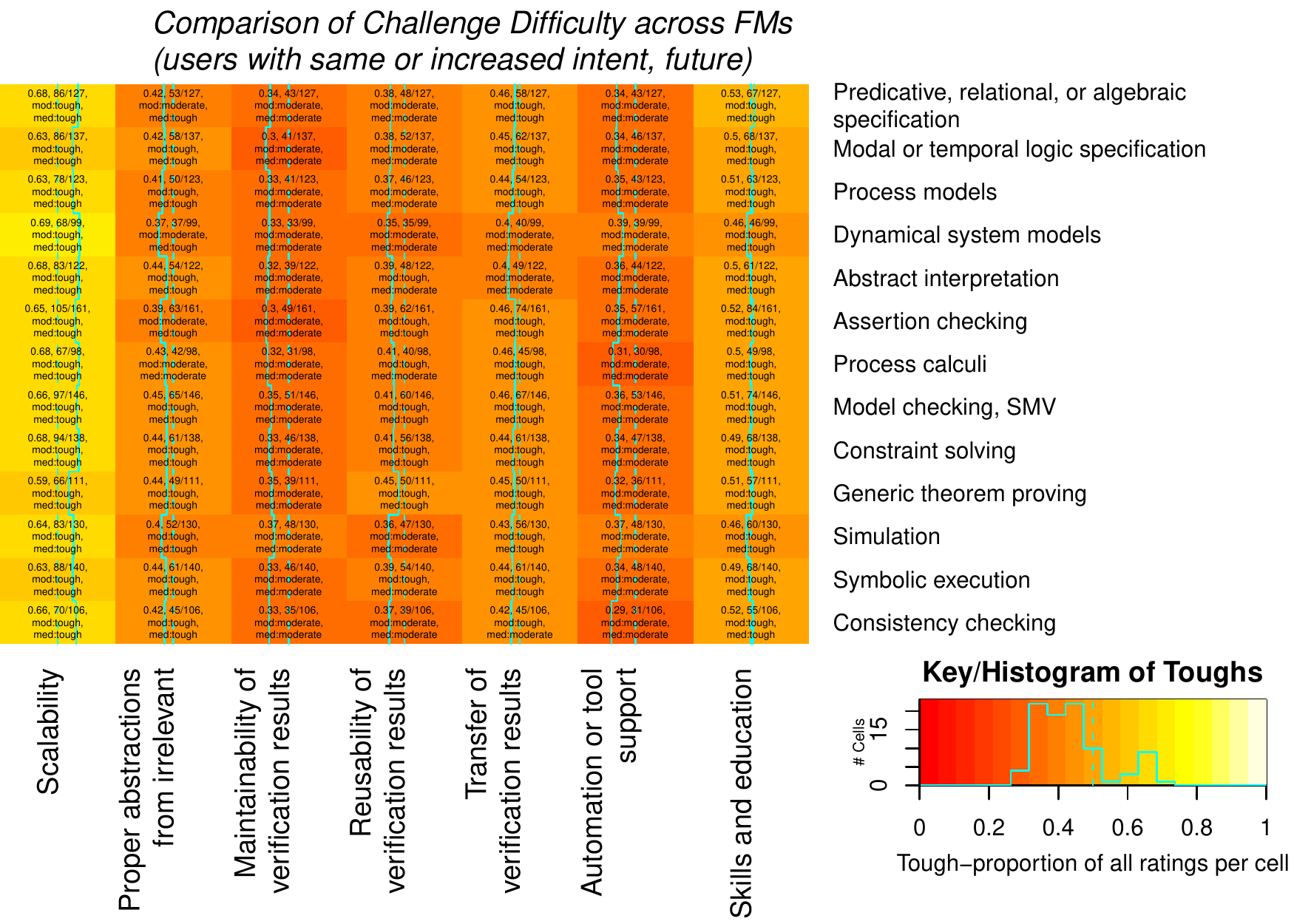}
  \caption{Comparison of challenge difficulty across \ac{FM} classes ($\mathit{UFM}_i$)
    \label{fig:rq3_heatmapF3_F4_O1}}
\end{figure}

\begin{figure}
  \includegraphics[width=\linewidth]{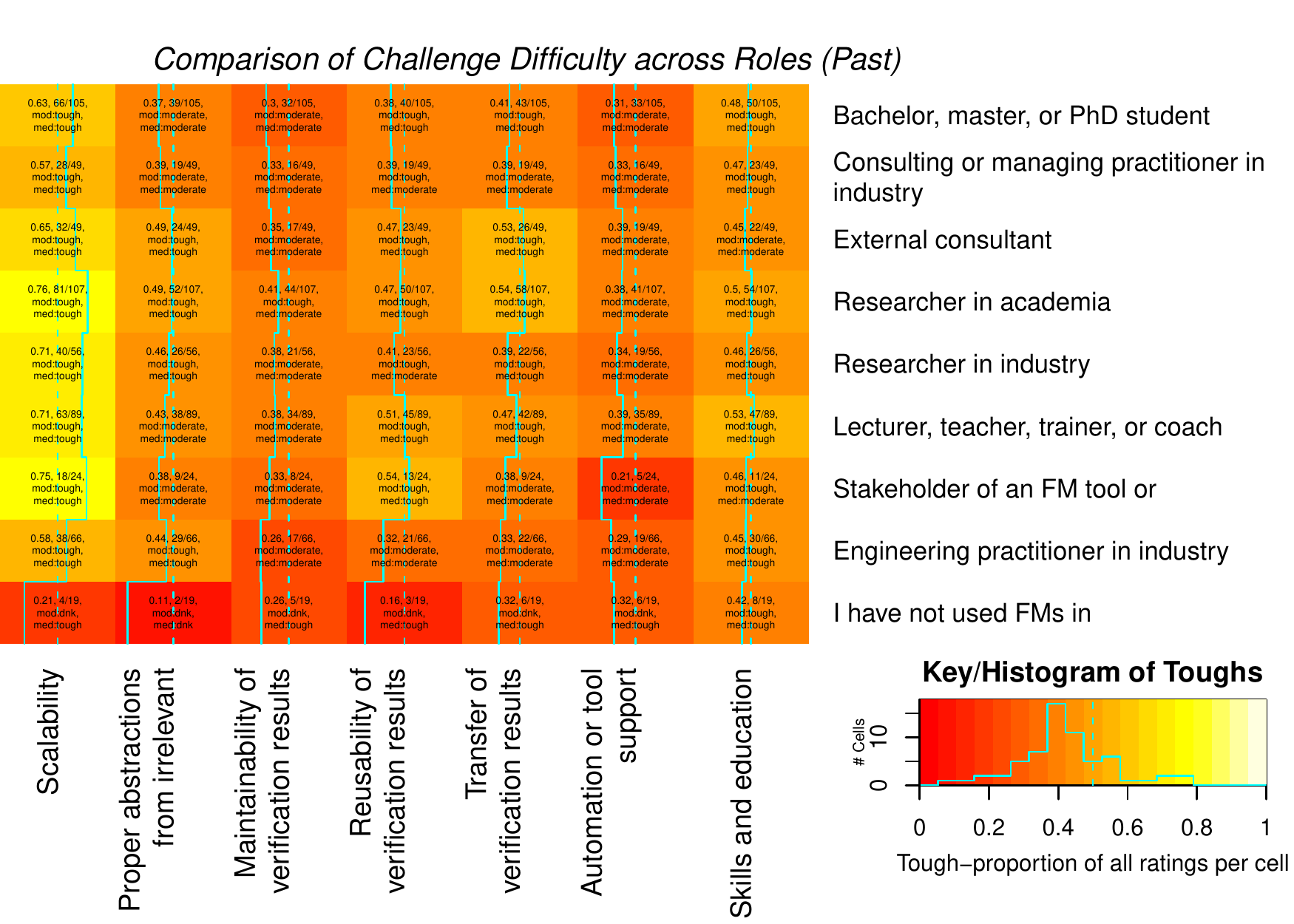}
  \includegraphics[width=\linewidth]{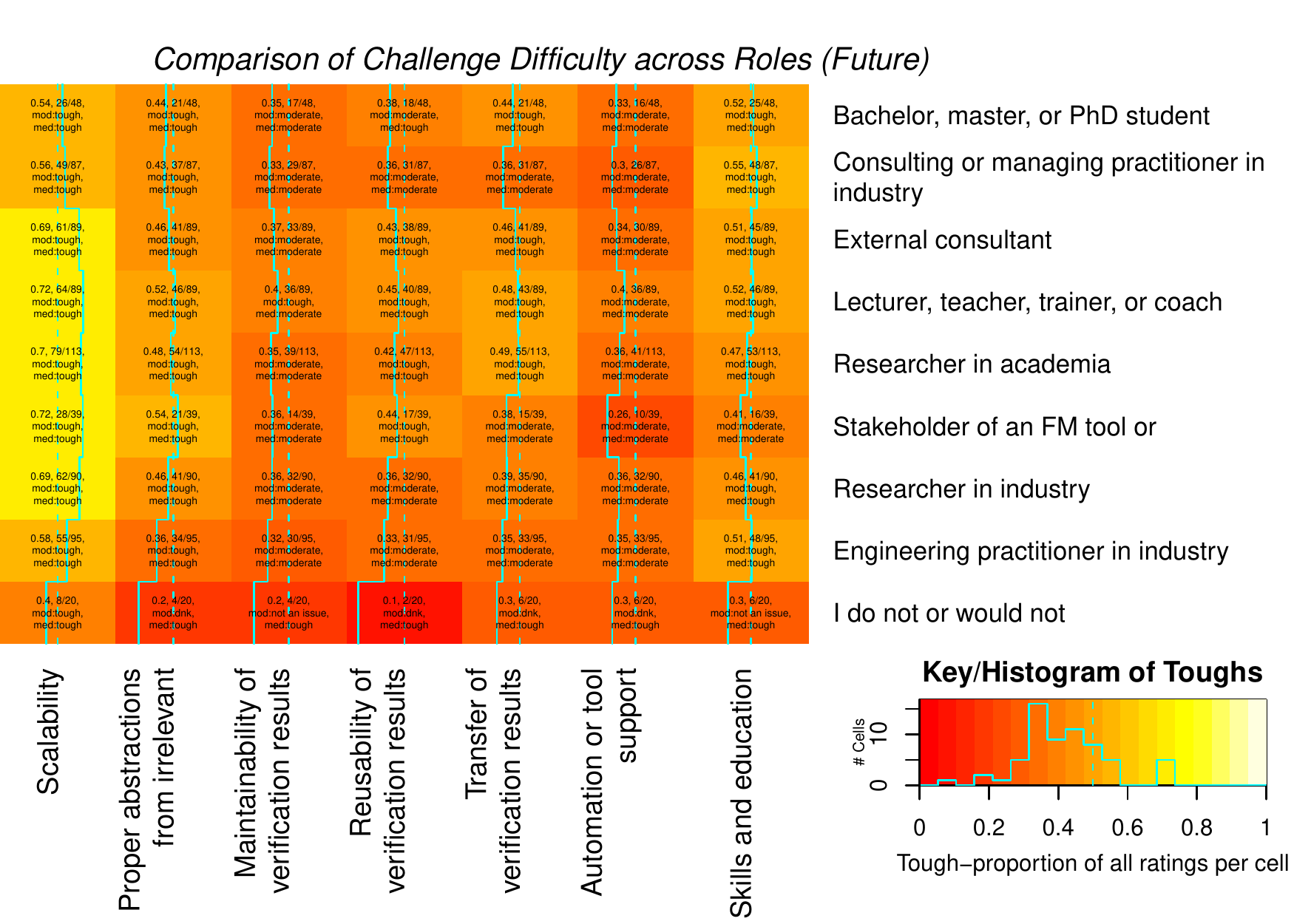}
  \caption{Comparison of challenge difficulty across roles
    \label{fig:rq3_heatmapP1F2}}
\end{figure}

\begin{figure}
  \includegraphics[width=\linewidth]{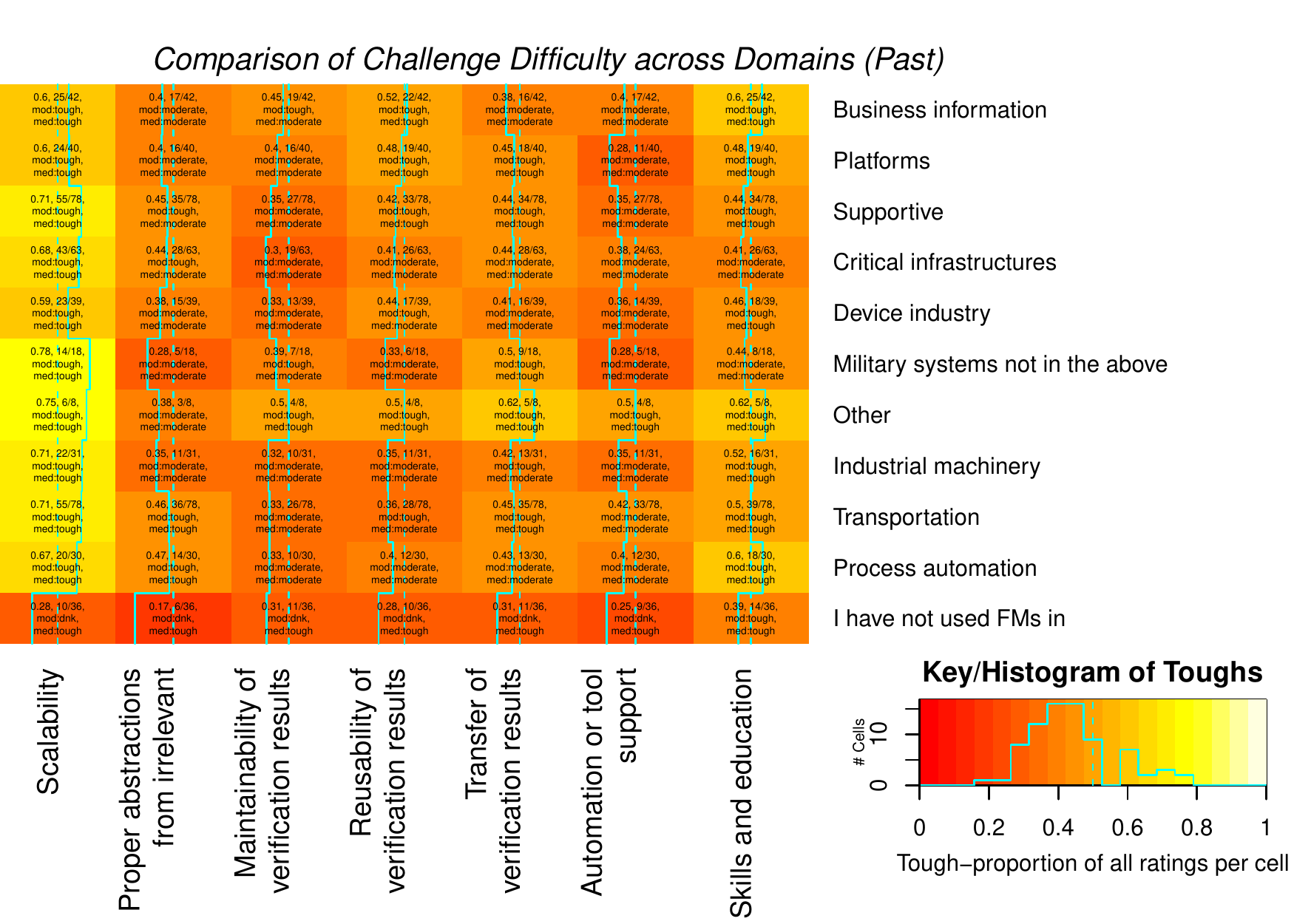}
  \includegraphics[width=\linewidth]{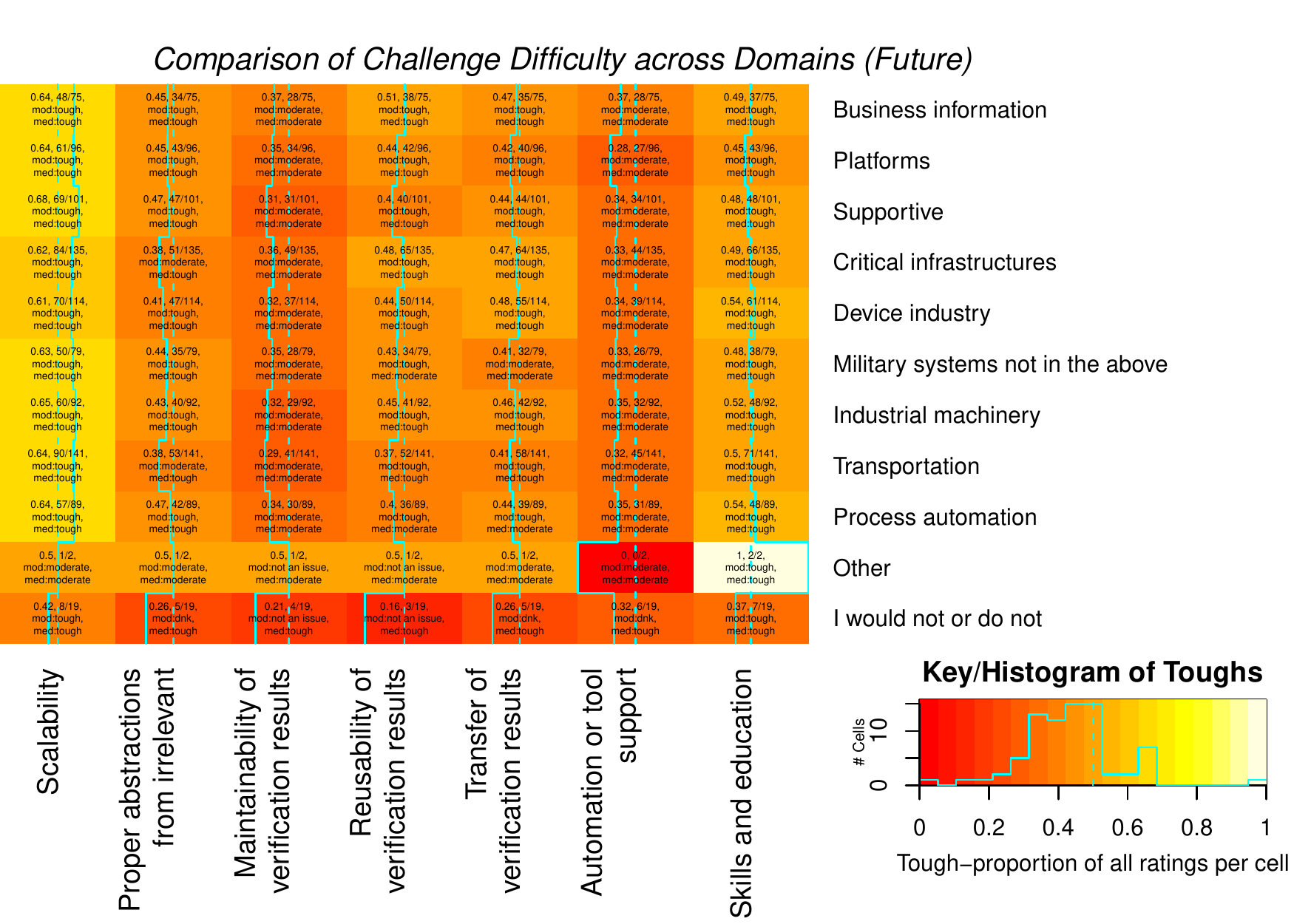}
  \caption{Comparison of challenge difficulty across domains
    \label{fig:rq3_heatmapD1F1}}
\end{figure}

\clearpage

\begin{figure}
  \centering
  \includegraphics[width=.8\linewidth]{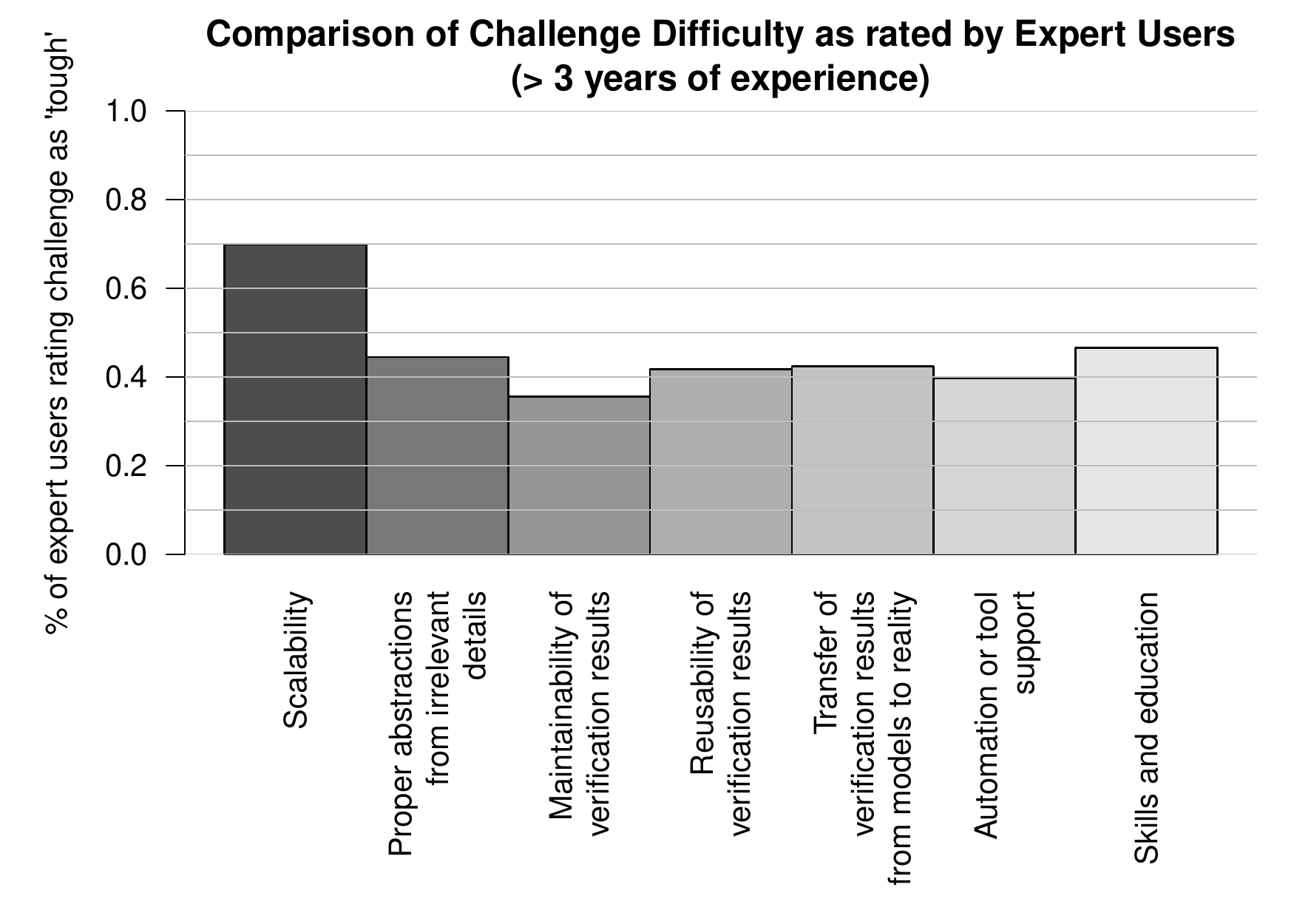}
  \includegraphics[width=.8\linewidth]{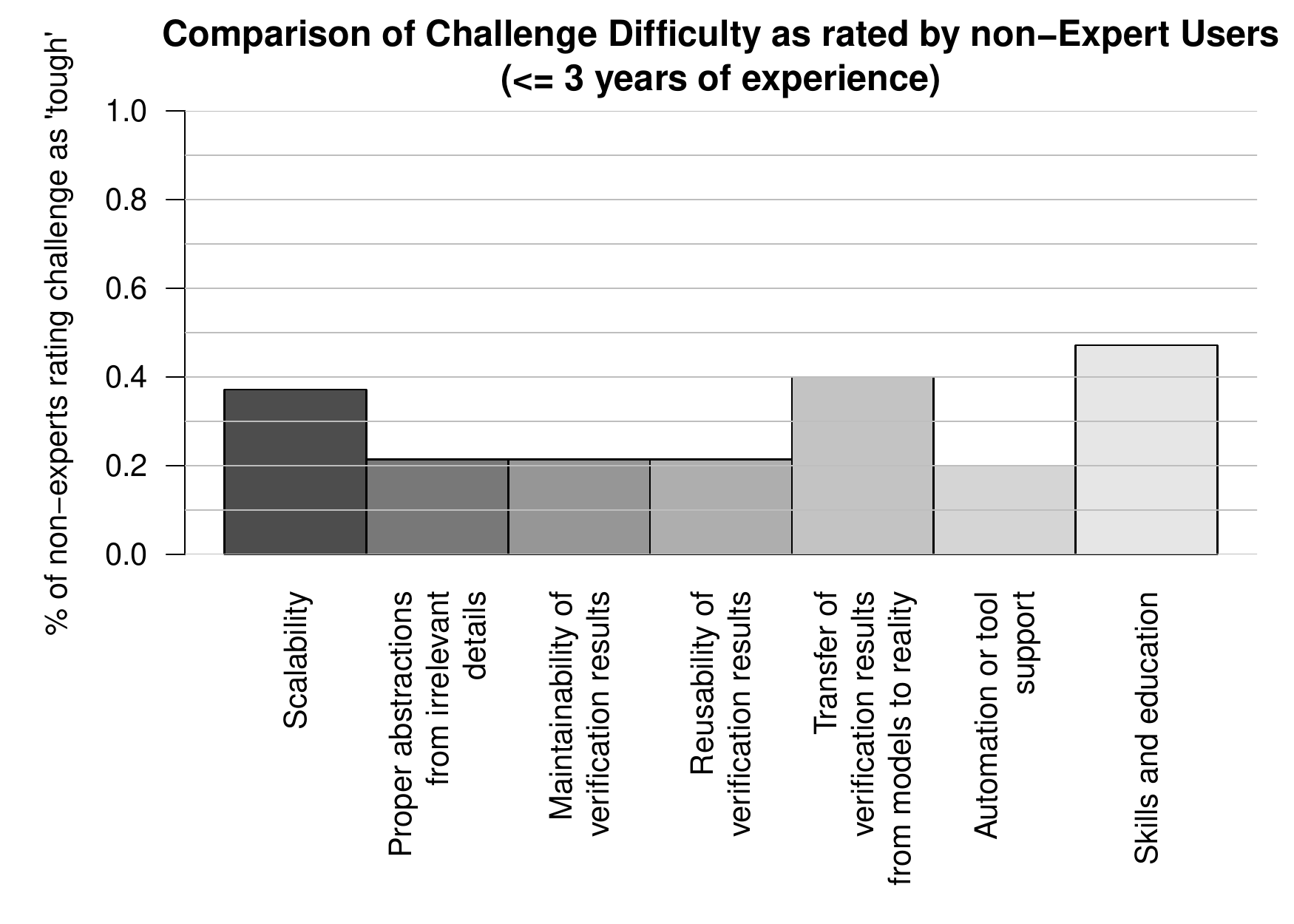}
  \caption{Comparison of expert and non-expert users by their
    perception of challenge difficulty
    \label{fig:rq3_heatmapD2O1}}
\end{figure}

\subsection{Details on the Systematic Map}
\label{sec:relat-work-deta}

\Cref{tab:relwork:details} contains the data we collected from the
literature for the systematic map.

\begin{landscape}
\scriptsize
\tablecaption{Details for the classification of related work
\label{tab:relwork:details}}
\tablefirsthead{%
  \toprule
  \textbf{Reference}
  & \textbf{Motivation}
  & \textbf{Approach}
  & \textbf{Result}
  & \textbf{Relation}
  & \textbf{List of Obstacles}
  \\\midrule
}
\tablehead{%
  \toprule
  \textbf{Reference}
  & \textbf{Motivation}
  & \textbf{Approach}
  & \textbf{Result}
  & \textbf{Relation}
  & \textbf{List of Obstacles}
  \\\midrule
}
\tabletail{%
\midrule \multicolumn{6}{r}{{Continued on next page}} \\ \midrule}

\begin{xtabular}{
    >{\raggedright\arraybackslash}p{1.5cm}
    >{\raggedright\arraybackslash}p{2cm}
    >{\raggedright\arraybackslash}p{3cm}
    >{\raggedright\arraybackslash}p{3cm}
    >{\raggedright\arraybackslash}p{2.5cm}
    >{\raggedright\arraybackslash}p{3cm}}
  \citet{Gerhart1976-ObservationsFallibilityApplications}
  & Pinpoint fallibility in the use of \acp{FM}
  & Evaluation of methodologies, error classification and analysis,
  using several examples of fundamental algorithms
  & Identification of three classes of errors: specification errors,
  systematic construction errors, proved program errors. Discussion of
  potential causes of errors of each class.
  & Observations contain obstacles, recommendations their alleviation
  & Formality gap, %
  appropriate abstraction and structuring, %
  lack of skills and education %
  \\
  \citet{Jackson1987-PowerLimitationsFormal}
  & Generic evaluation 
  & Expert opinion
  & \acp{FM} have inherent limitations
  & Aim of method transfer, design of applicable \acp{FM}
  & Inherent informality of formalisation, difficult to communicate to
  customers, lack of expressiveness/freedom of expression, lack of
  methodology
  \\
  \citet{Bjorner1987} %
  & Proposes a software Development method based on FM
  & Personal opinion, experience
  & Identify 3 main challenges: education, hiring, and tool support.
  & Challenges could be considered obstacles
  & Lack of skills and education, lack of tools, changeability/compatibility with
    existing process (method culture)
  \\
  \citet{Hall1990} %
  & Present and test \ac{FM} myths.
  & They evaluate \acp{FM} on one larger case study (~50.000 lines of
    Objective C Code) where they use Z (550 Z schemas to define 280
    operations) to develop a CASE tool.
  & Formal methods are powerful tools which must be better
    understood by developers at large.
  & Rejection of common Hypothesis about \acp{FM}. Evaluation of a single
    \ac{FM} by means of a case study.
  & Myth 1: improper abstraction; transfer of v.results;
    (myth 2-4: skills and directed education;)
    myth 5: time-budget restrictions;
    myth 6: improper abstraction, tool support (usability);
    myth 7: scalability
  \\
  \citet{Wing1990-specifiersintroductionformal}
  & \ac{FM} adoption
  & Literature review, summary, and analysis
  & Overview/taxonomy of \acp{FM}; analysis of limitations
  & Justifies the \ac{FM} classification used in our questionnaire
  & Proper abstraction (formality gap, neglected environmental
  assumptions)
  \\
  \citet{Bloomfield1991} %
  & Introduction to formal methods
  & Reference to technology transfer, existing case studies, and
    tool support
  & At the present time, formal methods are good for the description
    of sequential properties of systems, and for communication
    protocols, although they do not yet address temporal properties
    and concurrency particularly well.
  & Investigation of state of the art, no comparison with future.
  & Handling of incomplete specs, lack of verification tools,
    costly training, changing management style
  \\
  \citet{Austin1993-Formalmethodssurvey}
  & Lack of acceptance in industry
  & Literature survey and questionnaire
  & Identify obstacles and suggest to improve education and
  standardisation and to perform case studies and define metrics. 
  & Our questionnaire has less focus on representation
  and methodology and excludes questions on benefits, suggestions.
  Their sample mainly covers Z/VDM users in the UK.
  Our analysis of past use is more elaborate.
  & Math, tools, lack of cost/benefit evidence, change resistance
  \\
  \citet{Craigen1993} %
  & Determine state of the art about the use of \ac{FM} in practice
  & Analysis of $12$ case studies
  & \acp{FM} are beginning to be used seriously and successfully by
    industry
  & Study of \acp{FM} in industrial practice
  & Stated as recommendations: scalability, lack of tool support,
    lack of skills/education, transfer of verif. obl./results
    from/to code; resource constraints
  \\
  \citet{Fraser1994-Strategiesincorporatingformal}
  & Lack of \ac{FM} adoption, improvement of RE
  & Discuss benefits and problems of \ac{FM} adoption, literature study 
  & Present a two-dimensional framework for assessing strategies for
  incorporating formal specifications in software development 
  & Add suggestions on \ac{FM} introduction
  & %
  Lack of method and tool support,
  lack of skills/training, %
  not suitable for requirements prototyping, %
  lack of cost/benefit evidence
  \\
  \citet{Bowen1995} %
  & Re-examine Hall's myths, introduce 7 new Myths.
  & Argumentation and mentioning of case studies (although no
    reference to the studies are given).
  & More real links between industry and academia are required, and
    the successful use of formal methods must be better publicized.
  & Rejection of common Hypothesis about \acp{FM}. Evaluation with
    reference to case studies.
  & Time-budget-restrictions, lack of tool support, lack of
    integration in current process, scalability (multi-tech)
  \\
  \citet{Bowen1995a} %
  & Identify maxims that may help in the application of formal
    methods in an industrial setting.
  & Based on observations (by ourselves and others) on a number of
    recently completed and in-progress projects
  & 10 Hypotheses on how to improve the success of \ac{FM} usage
  & Investigation of \ac{FM} usage; Argumentation and Examples.
  & Stated as commandments: lack of tool support (documentation guidelines), proper
    abstraction, budget restrictions (bad cost-benefit ratio), lack of experts (skills and
    education); compatibility with current process (lack of quality
    culture); lack of reuseability
  \\
  \citet{Lai1995} %
  & Investigate reasons why academic methods are not adapted by
    industry
  & Personal experience
  & Provide $5$ reasons: practicality, too academic, Education,
    Resistance to change, difficulty to re-invest
  & The reasons can be considered as obstacles
  & Practicality (being too academic): scalability, skills and
    education, resistance to change (compat with existing process);
    budget restrictions
    \\
  \citet{Heisel1996} %
  & In practice \acp{FM} are not widely used
  & Personal Opinion/Observation
  & Proposes a pragmatic approach to FM
  & Method tries to overcome obstacles
  & Lack of skills/education, lack of experts, improper abstraction
    (low useability, low correctness); budget restrictions;
    compatibility with existing process
  \\
  \citet{Hinchey1996} %
  & Identify reasons for industry's reluctance to take formal
    methods to heart.
  & Experience in editing a collection of essays on the industrial
    application of FM.
  & Misconception of Myths, Standards, Tools, and Education are
    obstacles to use \ac{FM} in industry.
  & They identify obstacles. However, they to not provide evidence
    against or in favor of them.
  & Wrong skills and education,
    lack of clarification (bad
    reputation/misconception);
    lack of standards/regulation; lack
    of tools; 
  \\
  \citet{Holloway1996-Impedimentsindustrialuse}
  & \ac{FM} adoption
  & Position statement, experience report, expert opinion
  & List of impediments to \ac{FM} adoption
  & Do not measure usage intent but highlight lack of transfer efforts
  & Inadequate tools and examples, inadequate transfer
  \\
  \citet{Lai1996} %
  & Academic methods (FM) not used in communication industry
  & Personal opinions; critize research transfer and suggest
    improvements that might also be helpful for \ac{FM} transfer to practice
  & $8$ Reasons why academia needs to do industry research, $7$
    catalysts, $12$ industry relevant factors
  & Reasons can be considered obstacles
  & Lack of empirical evidence, lack of skills/education,
    scalability, proper abstractions; compatibility with existing
    process, lack of tool support
  \\
  \citet{Pfleeger1997-Investigatinginfluenceformal}
  & Investigate effectiveness of \acp{FM}
  & Case study and effect analysis: Comparison of change requests for
  FM-based and non-FM-based code fragments as a result of postdelivery
  problems caused by these fragments. 
  & \ac{FM} have a positive effect on code quality
  & Empirical evidence for \ac{FM} effectiveness shown for \acp{FM} in design
  phase but not in more general, however only one system
  & Cost-benefit (fault removal effectiveness)
  \\
  \citet{Knight1997} %
  & \acp{FM} are not accepted by industry
  & Elementary Field Experiment
  & Evaluation of Z, PVS, and state-charts
  & Evaluation of \acp{FM} in practice
  & Integration in existing process (tools, methods, environments,
    people); proper and useful abstractions (incl. usability/comprehensibility); tool 
    support (group development, collaborative eng.); evolution
    and spec/proof maintainability; budget constraints
  \\
  \citet{Galloway1998}
  & Lack of \ac{FM} adoption, improvement of requirements
  & Single case study in the lab
  & Applied discrete \acp{FM}~(\egs Z, PVS, and state-charts) for 
  specifying and aggregating requirements of aircraft engine control systems
  & Interesting discussion of \ac{FM} weaknesses in a very relevant
  application context
  & Inadequate abstraction (difficult to integrate discrete and continuous
  abstractions); lack of training; lack of interest from engineers;
  \acp{FM} are perceived as too expensive to apply
  \\
  \citet{Heitmeyer1998} %
  & \ac{FM} tools are useful but not used in industry since people are
    not skilled enough
  & Reference to a few case studies
  & Provide a set of guidelines how to make \ac{FM} more usable
  & Usability as obstacle
  & Tool support, proper abstraction, process changeability/compatibility
  \\
  \citet{Snook2001} %
  & Lack of empirical investigation on the use of FM
  & $5$ Structured Interviews
  & Improved quality of software with little or no additional
    lifecycle costs
  & Empirical investigation of the benefits of \ac{FM} in industry
  & Lack of skills/education, improper abstractions
  (understandability); transfer of verif.results (models to reality)
  \\
  \citet{Bowen2005} %
  & Re-examination of their 1995 commandments 10 Years later
  & Personal experience, reference to literature
  & empirically validate commandments with little conclusion
  & They investigate the use of \ac{FM} in industry
  & Tool support still an issue despite some case studies
  \\
  \citet{Bicarregui2009} %
  & No recent study on the industrial use of \acp{FM}.
  & Structured Questionnaire on $62$ industrial projects
  & $4$ challenges where identified
  & Empirical Study identifying obstacles
  & Lack of tool support, lack of empirical evidence; lack of
    experience (skills/education), budget restrictions
  \\
  \citet{Woodcock2009} %
  & State of the art of industrial use of \ac{FM} (extension of
    \cite{Bicarregui2009})
  & Structured Questionnaire on $62$ industrial projects
    (claimed to be most comprehensive review ever
    made of formal methods application in industry) 
  & Identify several challenges
  & Empirical Study identifying obstacles
  & Budget/resource constraints (high entry costs, cost-benefit),
    lack of tool support (automation) 
  \\
  \citet{Miller2010} %
  & \acp{FM} are not widely used in Industry
  & $3$ Case studies about the use of Model Checking in Industry
  & Model Checking can be effectively used to find errors early in
    the development process for many classes of models
  & They investigate the applicability of one \ac{FM} in industry
  & Lessons from case studies: scalability and proper abstraction
    (useability) 
  \\
  \citet{Parnas2010} %
  & \acp{FM} are not widely used in industry
  & Argumentation/Personal Experience
  & Provides reasons why \acp{FM} are not used and suggestions for
    improvement
  & Provides Obstacles
  & Proper abstraction, lack of empirical evidence, lack of tool
    support; maintainability/transfer (refinement,step-by-step)
  \\
  \citet{Mohagheghi2012-empiricalstudystate}
  & Adoption of \ac{MBE} in \ac{SE} 
  & Tool evaluations based on
  \ac{TAM}~\citep{Riemenschneider2002-Explainingsoftwaredeveloper},
  interviews, survey 
  & Evaluation of \ac{PEOU}, \ac{PU}, current, and future use
  & Also use IVs current and future use; little focus on FM; but MDE is
  sometimes based on FM, MDE adoption can improve \ac{FM} adoption; 
  & Lack of training, lack of maturity, broken tool chains, high cost
  of adoption (tool integration) 
  \\
  \citet{Davis2013-StudyBarriersIndustrial}
  & Identify barriers to \ac{FM} adoption and suggestions for
  barrier mitigation
  & Interviews with 31 practitioners from the US aerospace domain
  & Top barriers: education, tools, work environment;
  top mitigations: education, tool integration, evidence
  of \ac{FM} benefits; occasional non-barriers: evidence on savings, \ac{FM}
  complexity, training/skills
  & Similar research questions, open-ended interview questions;
  restricted to one domain and geography 
  & Education, tools, environment, engineering, certification,
  misconceptions, scalability, evidence of benefits, cost
  \\
  \citet{Liebel2016-Modelbasedengineering}
  & Adoption of \ac{MBE} (incl.~FM) in embedded \ac{SE}
  & Online survey on needs, positive/negative effects, and
  shortcomings of MDE adoption
  & SotA and challenge assessment: \acp{FM} not used widely; their data suggests a need of
  \ac{FM} adoption; 30\% of the responses from industry declare the need for \acp{FM} as a reason to adopt
  MDE; median of responses suggests that \ac{MBE} adoption has a positive effect on FM
  adoption; %
  & Few \ac{FM} users as participants
  & Lack of tool support, bad reputation, rigid development processes %
  \\
  \citet{Ferrari2019-SurveyFormalMethods}
  & Lack of \ac{FM} adoption in railway domain
  & Review of \ac{FM} literature, \ac{FM} projects, and \ac{FM} tools according to
  DESMET~\citep{Kitchenham1997-DESMETmethodologyevaluating}; survey
  among practitioners
  & UML dominates as the \ac{MBE} language, many \acp{FM} and FM-based tools are
  used, B dominates as the FM; tool ranking/selection matrix
  & Analyse maturity of \acp{FM} from literature review; %
  evaluate relevance/quality/maturity of \ac{FM} and \ac{FM} tool features from
  subjective assessment of survey respondents; %
  & Difficulty to learn, lack of tool qualifications, lack of expressiveness
  \\
  \citet{Klein2018-Formallyverifiedsoftware}
  & \ac{FM} adoption
  & Large case study, measurement of proof effort
  & \acp{FM} can scale to real systems, mixed assurance levels are possible
  & Evidence for scalability contradicting the belief of our responses
  & Incompleteness of theorems for abstracting from all hardware features
  \\
  \\\bottomrule	
\end{xtabular}
\end{landscape}

\subsection{Mapping of Studies to Challenges for \RQRef{3}}
\label{sec:map-lit-chall}

In addition to \Cref{tab:challenges} in \Cref{sec:analysis-rq3},
\Cref{tab:challenges:details} provides the complete lists of surveyed
studies mapped to the corresponding challenges.

\begin{table}[ht]
  \caption{Mapping of studies to challenge names (with the number of studies
    in parentheses)
    \label{tab:challenges:details}}
  \footnotesize
  \begin{tabularx}{\textwidth}
    {>{\bfseries\hsize=.2\hsize}L
    >{\hsize=.8\hsize}L}
    \toprule
    \textbf{Challenge Name}
    & \textbf{Supported by}
    \\\midrule
    Scalability (7)
    & \citet{Hall1990,Miller2010,Bowen1995,Lai1995,Lai1996,Craigen1993,Craigen1995a}
    \\\midrule
    Skills \& Education (13)
    & \citet{Bjorner1987,Bicarregui2009,Galloway1998,
      Hall1990,Barroca1992,Hinchey1996,Bowen1995a,Lai1995,Lai1996,Heisel1996,Craigen1993,Craigen1995a,Snook2001}
    \\\midrule
    Transfer of Proofs (8)
    & \citet{Jackson1987-PowerLimitationsFormal,Parnas2010,
      Hall1990,Craigen1993,Craigen1995a,Snook2001,Bloomfield1991,Barroca1992}
    \\\midrule
    Reusability (2)
    & \citet{Barroca1992,Bowen1995a}
    \\\midrule
    Abstraction (12)
    & \citet{Jackson1987-PowerLimitationsFormal,Galloway1998,Parnas2010,Miller2010,
      Hall1990,Barroca1992,Bowen1995a,Lai1996,Heitmeyer1998,Heisel1996,Knight1997,Snook2001}
    \\\midrule
    Tools \& Automation (16)
    & \citet{Bjorner1987,OHearn2018,
      Hall1990,Bloomfield1991,Bowen1995,Hinchey1996,Bowen1995a,Bowen2005,Bicarregui2009,Woodcock2009,Parnas2010,Lai1996,Heitmeyer1998,Craigen1993,Craigen1995a,Knight1997}
    \\\midrule
    Maintainability (3)
    & \citet{Barroca1992,Knight1997,Parnas2010}
    \\\midrule
    \cellcolor{lightgray}
    Resources (11)
    & \citet{Hall1990,Woodcock2009,
      Craigen1993,Craigen1995a,Bloomfield1991,Bowen1995,Bowen1995a,Lai1995,Heisel1996,Knight1997,Bicarregui2009}
    \\\cmidrule{1-2}
    \cellcolor{lightgray}
    Process Compatibility (12)
    & \citet{Bjorner1987,OHearn2018,
      Bloomfield1991,,Bowen1995,Bowen1995a,Lai1995,Hinchey1996,Lai1996,Heitmeyer1998,Heisel1996,Knight1997,Craigen1995a}
    \\\cmidrule{1-2}
    \cellcolor{lightgray}
    Practicality \& Reputation (6)
    & \citet{Lai1995,Parnas2010,Galloway1998,
      Lai1996,Glass2002,Bicarregui2009}
    \\\bottomrule	
  \end{tabularx}
\end{table}

\subsection{All Answers to Open Questions}
\label{sec:answ-open-quest}

\Cref{tab:openanswers} provides all answers to the open questions of
our questionnaire.  We fixed a small number of spelling mistakes in
the responses during the preparation of this table.

\begin{landscape}
\tiny
\tablecaption{Open answers to the questions \QuestionRef{D3_motiv},
  \QuestionRef{P3_use_past}, \QuestionRef{P4_purpose_past},
  \QuestionRef{F4_use_future}, \QuestionRef{F5_purpose_future}, and
  \QuestionRef{O1_obst}; R\dots respondent
\label{tab:openanswers}}
\tablefirsthead{%
  \toprule
  \textbf{R}
  & \textbf{Date}
  & \textbf{D3o} Further motivations to use FMs
  & \textbf{P3o} Other FMs used
  & \textbf{P4o} Used FMs for other purp.
  & \textbf{F4o} Future use of other FMs
  & \textbf{F5o} Future use for other purp.
  & \textbf{O1o} Further obstacles to FM use
  & \textbf{General Feedback} 
  \\\midrule
}
\tablehead{%
  \toprule
  \textbf{No.}
  & \textbf{Date}
  & \QuestionRef{D3_motiv} %
  Further motivations to use FMs
  & \QuestionRef{P3_use_past} %
  Other FMs used
  & \QuestionRef{P4_purpose_past} %
  Used FMs for other purp.
  & \QuestionRef{F4_use_future} %
  Future use of other FMs
  & \QuestionRef{F5_purpose_future} %
  Future use for other purp.
  & \QuestionRef{O1_obst} %
  Further obstacles to FM use
  & \textbf{General Feedback} 
  \\\midrule
}
\tabletail{%
\midrule \multicolumn{9}{r}{{Continued on next page}}\\}
\tablelasttail{\\\bottomrule}

\begin{xtabular}{
    ll
    >{\raggedright\arraybackslash}
    p{2.2cm}
    >{\raggedright\arraybackslash}
    p{1.7cm}
    p{1.7cm}
    p{1.7cm}
    p{1.7cm}
    p{2.7cm}
    p{2.9cm}}

2 & 2017/08/14 &  & none & none & none & none & none & cool \\ 
4 & 2017/08/14 & Science shall be best and great. & No, that appear approximately complete to me. & That is pretty all. & No & No & In real operative situations are always hidden obstacles as well as those listed already. Imagine real and nominal definitions were really first order logic. Add operational definition and stay sound. & Abstraction, reduction, exemplification, representation, knowledge, experience, skill and the art of analysing the taxonomies under selection are necessary or in need. \\ 
8 & 2017/08/18 & Scientific curiosity &  &  &  &  &  &  \\ 
9 & 2017/08/18 & Test-Generation for small problems, like MCDC coverage &  &  &  &  & high license costs &  \\ 
19 & 2017/08/25 & I just liked logic. Actually never cared about applying it, but research positions are in FM so well :-) &  &  &  &  & goes completely against any existing software development process: FM require that you write specifications before implementing. I don't believe anymore that anybody would do this any time. &  \\ 
21 & 2017/08/28 & Error elimination &  &  &  &  &  &  \\ 
22 & 2017/08/28 & Completeness, accuracy &  &  &  &  &  & Survey not suitable for beginners (like me) with the FMs. \\ 
28 & 2017/09/12 &  &  &  &  &  & Budgetary restrictions &  \\ 
32 & 2017/10/16 &  &  &  &  &  & environment &  \\ 
34 & 2017/10/20 &  &  &  &  &  & The cost-benefit relationship is not slanted in FM's favour: substantial improvements in software engineering will not come from application of FM but through process improvement, better training and education, and better control of requirements. Formal methods have uses for verifying critical components but they will not solve the big problems in software engineering. &  \\ 
35 & 2017/10/25 & Supporting the design and construction of reliable and dependable systems & Many dialects of Petri nets an Automata &  &  &  &  &  \\ 
36 & 2017/10/25 & Improved level of confidence versus more traditional means &  & reverse engineering &  &  & The main obstacle to adoption by industry is the "use" of engineers not able to handle abstraction, leading to poor results. &  \\ 
37 & 2017/10/25 & The need to achieve and demonstrate the highest possible integrity of systems  & None & None & None & None & Perceived cost and difficulty of use requiring specialist knowledge &  \\ 
38 & 2017/10/25 & No. &  & None & None &  & There are no specific barriers (apart from not having needed practice with them) apart from I find my Formalised Methods (to strict procedures) gets the task done well enough. & I make a clear distinction between the mathematically Formal Methods and the procedure based Formalised Methods. \\ 
39 & 2017/10/25 &  & B method &  &  &  &  &  \\ 
40 & 2017/10/25 & Cost and quality of finished product & SPARK Ada programming language, Z, CSP &  &  &  & Broken market/market-for-lemons in software quality &  \\ 
43 & 2017/10/25 & They are the only way to guarantee certain properties of software and its documentation & Specification: TLA+, Z & Fault and failure analysis &  &  &  &  \\ 
44 & 2017/10/25 & none & none & no &  & No &  &  \\ 
45 & 2017/10/25 & Clearly thinking about the problem and correctness &  &  &  &  &  &  \\ 
47 & 2017/10/25 &  &  &  & I work with 'time triggered' systems.  These can be modelled effectively (semi-formally) without a complete mathematical model.  Many of our customers think this is far more advanced than they require ... &  & No customer demand.   &  \\ 
48 & 2017/10/25 & "Bugs", software defects, are semantic inconsistencies in code. Formal methods acknowledge the necessary semantics and help prevent defects. & Currently integrating a degree of formal methods into software requirements and design modelling method & Drive construction of good code. Properties of a good proof are also properties of good code: minimal steps, logical progression of steps, ... Writing code to be prove-able (even if never explicitly proven) yields better code. & I don't understand this question. & None that I can think of now & The software industry's cultural predisposition to informal approaches ("But that's not the way we've always done it!") &  \\ 
49 & 2017/10/26 & Combining traditional and Formal methods based Software V \& V &  &  &  &  &  &  \\ 
50 & 2017/10/26 & I love logic and maths. & my experience is mostly as a researcher and teacher, something for which the above does not allow me to tick a box & to make exam questions :-) &  &  &  &  \\ 
54 & 2017/11/04 & Belief it helps construct systems with much less errors &  &  &  &  & Most of practitioners don't and don't want to know formal methods. So there is a strong need for "hidden" use of formal methods, like compilers. & FM have strong potentials but are also difficult to use in industry practices. Transferring very good academia results into industry practices is challenging. \\ 
56 & 2017/11/06 & no &  &  &  &  &  &  \\ 
58 & 2017/11/23 & To reduce costs in system development. &  & Reduce costs. &  &  &  &  \\ 
59 & 2017/11/23 & simplifies sometimes things, because of enforcement of a systematic approach &  &  &  &  &  &  \\ 
60 & 2017/11/23 & The beauty of mathematics &  &  &  &  &  &  \\ 
61 & 2017/11/23 & research program &  &  &  &  & ignorant persons &  \\ 
64 & 2017/11/23 & Elegance and precision & VDM, Z, Event-B &  &  &  &  & Interesting set of questions \\ 
65 & 2017/11/23 & Maintenance cost and reliability &  &  &  &  &  &  \\ 
66 & 2017/11/23 & I consider it the foundation of principled software engineering & Unifying Theories of Programming &  & rapid prototyping calculation tools &  & lack of time &  \\ 
67 & 2017/11/23 & Higher level of trust &  &  & pseudo-occam, SPARK-ada &  & Not very user friendly, too abstract and diverse syntax between different FMs, productivity &  \\ 
68 & 2017/11/23 & Increasing engineering reliability & The TRIO specification language &  & UML with formal application-dependent tailoring. &  &  &  \\ 
71 & 2017/11/23 & Teaching how to build high-quality software  &  &  &  &  &  &  \\ 
72 & 2017/11/23 &  &  &  &  &  & Engineers will not be able to apply formal methods. &  \\ 
73 & 2017/11/23 & Achieving correct, fault-free software &  &  &  & In lectures and teaching FMs. &  &  \\ 
76 & 2017/11/25 & no &  &  &  &  &  &  \\ 
77 & 2017/11/26 & Thesis - tool verification &  &  &  &  &  &  \\ 
79 & 2017/11/29 & productivity of the *whole* process &  &  &  &  &  &  \\ 
80 & 2017/11/30 & Soundness checking of automation &  &  &  &  &  &  \\ 
83 & 2017/12/06 &  &  &  &  &  &  & Well chosen questions which do not leave me guessing. Relevant to future FM research and practice. \\ 
84 & 2017/12/06 & None - impractical for most systems &  &  &  &  &  &  \\ 
85 & 2017/12/06 & Identified as industrial best practice &  &  &  &  &  &  \\ 
87 & 2017/12/06 & Desire to develop systems that I have solid evidence perform correctly under all scenarios. &  &  &  &  & Aligning academic research with industrial need. Getting tool developers to invest in FM. &  \\ 
90 & 2017/12/19 & business and research differentiation &  &  &  &  &  &  \\ 
91 & 2018/01/02 &  &  &  &  &  & Lack of tools and wide range of methods in existence. &  \\ 
93 & 2018/03/16 & I don't know & I haven't & I haven't & I don't know & Academia & I don't know &  \\ 
94 & 2018/07/30 & Niche market, outperform companies using "traditional" means  &  &  &  &  &  &  \\ 
96 & 2018/07/31 & verify protocols &  &  &  &  &  &  \\ 
97 & 2018/08/01 & I believe that software engineering should have the same mathematical underpinning as regular engineering (I majored in engineering) &  &  &  &  &  &  \\ 
98 & 2018/08/01 & To increase quality and reliability &  &  &  &  &  &  \\ 
100 & 2018/08/01 & needed & - & - & - & - & - & - \\ 
102 & 2018/08/02 & no & none &  & no & no &  &  \\ 
105 & 2018/08/02 & See Andy Galloway, Trevor Cockram, and John McDermid. Experiences with the application of discrete formal methods to the development of engine control ... & Independent reviewer of FM specifications & None &  &  &  &  \\ 
107 & 2018/08/02 & Pushing back the boundaries of knowledge. Making programs which are correct by construction. & FermaT Program Transformations &  &  &  &  &  \\ 
108 & 2018/08/03 & critical infrastructures and technology requirements &  & specification of hybrid communicating systems &  & future technological design methodologies &  &  \\ 
109 & 2018/08/03 & No further motivations to use FMs  &  &  &  &  &  &  \\ 
111 & 2018/08/03 & A candidate supplier offered FMs in a tender response to address an assurance requirement & reviewing consistent set of requirements expressed using Z, OBJ, etc &  &  &  & I need to provide assurance that is understandable to those who need to be assured & Formal Methods are perceived to be expensive to use, and they can be, but this can be offset by the benefits; there does not seem to be much work published on this that could be used to persuade the Customer and the budget holder... \\ 
113 & 2018/08/03 & Formal is the only way to obtain rigorous verification. &  &  &  &  &  &  \\ 
114 & 2018/08/04 & Strategic facilitation in large enterprises, where formal modelling can reveal gaps in the clients’ understanding of their own ecosystems. & Z, VDM, AXES & Requirements engineering & Protective Analysis (PAN) &  &  &  \\ 
115 & 2018/08/04 & As per my interpretation, FM is a consistent a logical way to describe a System/Product &  &  &  &  & Current corporations do not consider valuable FMs during product development, as a consequence, Engineers will not even try to use it & Looks really interesting and I will do some research from my side to learn a little bit more \\ 
117 & 2018/08/05 & To change the way the world produces software  &  & Security and safety analysis  & No & Safety and security analysis of Systems  &  Awareness of commercial potential  & Closed questionnaire is just a start \\ 
118 & 2018/08/05 & Improve general system reliability & TRIO , timed Automata, B & Requirement engineering &  & Research on innovative formalisms, foundational research &  &  \\ 
122 & 2018/08/06 & I have developed various safety-related gas sensors and other measurement devices. I was invited to become an assessor of these systems by a certifying body. I was shocked by what I discovered when I scrutinised the methods and approaches used by the embedded software developers whose work was being certified. I resolved to improve my own knowledge and practice so that I could deal authoritatively with some of the unsatisfactory situations that arose during various certifications. &  &  &  & I have great difficulty in persuading anyone to take an interest in FM. I represented the UK  on a European committee concerned with safety related gas systems. The ignorance and lack of interest in FM was startling - the German representatives were particularly hostile. So I only use FM in special cases with specific customers. There is much prejudice out there against FM. & See answer above. We have to publicise some  successes in order to lift the veil of ignorance about FM. &  \\ 
123 & 2018/08/06 & We provide customers an FM tool &  &  &  &  & Most tools have many limitations and do not support wide domains by themselves &  \\ 
126 & 2018/08/07 & Quality requirements &  &  &  &  & Complexity &  \\ 
129 & 2018/08/07 & Scientific curiosity. &  & Formalising international standards &  &  &  &  \\ 
131 & 2018/08/07 & It is a hard problem. &  &  &  &  &  &  \\ 
133 & 2018/08/07 & Improving confidence in software &  &  &  &  &  &  \\ 
134 & 2018/08/07 & Formal methods are the backbone of the science in "Computing Science". &  &  &  &  &  &  \\ 
136 & 2018/08/07 & To model complex phenomena in social sciences &  &  &  &  &  &  \\ 
137 & 2018/08/07 & No &  &  &  &  &  &  \\ 
138 & 2018/08/07 &  & Protocol analysis &  & Protocol analysis  &  &  &  \\ 
141 & 2018/08/08 &  & The lists essentially contains formal verification approaches. Formal methods are mathematically grounded approaches to design/analyse artefacts, e.g. the B method is a formal method to devise software components. I use the B method in my current job. & System-level analysis. Software design. &  & System analysis. Software design. &  &  \\ 
145 & 2018/08/08 & Curiosity &  &  & None &  &  &  \\ 
146 & 2018/08/08 & Interest / Research &  &  & As testing tools. &  & Industry mind set  &  \\ 
147 & 2018/08/08 & gain confidence in the algorithms we were reasoning about. &  &  &  &  & None. I am convinced that only FMs can help us from shitty and buggy software.  &  \\ 
150 & 2018/08/09 & The robustness and preciseness that the formal methods discipline can provide via its specifications and verifications techniques &  &  &  &  &  & Thank you very much for this survey. It is very constructive and important. It handles most of the issues encountered by any practitioner and user of formal methods.  \\ 
151 & 2018/08/09 &  &  &  &  &  &  & I only encountered FMs as a topic in my studies once. They do not play any role in my profession at the moment. \\ 
153 & 2018/08/09 & Quality &  &  &  &  &  &  \\ 
154 & 2018/08/09 & Main part of business model of our startup &  &  &  & Sorry for not disclosing our Future Extent of Formal Methods Use & Hypes (eg AI) that direct decision makers in other directions & Sorry for not disclosing our Future Extent of Formal Methods Use \\ 
157 & 2018/08/10 &  &  &  &  &  &  & I think FM is misplaced in our study program SE, especially it being mandatory. It is very rarely used in industry (for a reason) and there are so many more important things to learn in order to become the Software Engineers of tomorrow (eg. Entrepreneurship). \\ 
160 & 2018/08/10 &  &  &  &  & I am a PhD student using FM for formal verification. I plan on continuing to do this... &  &  \\ 
161 & 2018/08/11 &  &  &  &  &  & Time constraints, financial constraints, refactoring efforts &  \\ 
167 & 2018/08/13 & They are fun and solve a "real" problem &  &  &  &  &  &  \\ 
168 & 2018/08/13 &  & Abstract state Machines, Petri Nets, LTL, CTL, SAT and SMT solvers &  &  &  &  &  \\ 
169 & 2018/08/13 & Strong (potential) guarantees unobtainable by other methods &  &  & Requirement falsification (through simulation), dynamic/online assurance (monitoring and reasoning) &  & Lack of appreciation and understanding from other (CS and engineering) communities & Nice questions overall. In those with relative answers (e.g., more/less frequent than before), it would be nice to have my previous answers on the same page (to check what I answered exactly).  \\ 
171 & 2018/08/13 & I have developed both VDM and Rely/Guarantee concepts & VDM, Rely/Guarantee, Separation Logic &  & Rely/Guarantee &  & education &  \\ 
174 & 2018/08/13 &  & model-based testing  &  &  &  &  &  \\ 
176 & 2018/08/14 &  & Petri Nets &  &  &  &  & very strange you put Petri nets to Section P2 as a description technique, and process calculi to Section P3 as a reasoning technique. In fact, Petri nets offer much more reasoning techniques than process calculi do \\ 
178 & 2018/08/15 &  &  &  &  &  & Software Projects often have a large existing code base. The assumptions that hold in that code base need to be thoroughly analyzed first. Moreover, I am missing automatic mapping from source code to formal logic models / descriptions. & I've missed questions towards why people do not use formal methods in a given domain. What are the obstacles that they are currently facing? \\ 
179 & 2018/08/15 & Interest in reliable critical systems &  &  &  &  & Changed focus of interest at my work &  \\ 
182 & 2018/08/16 & I believe in doing things right or doing it not at all. & code generation based on allegories &  &  &  & The way academia communicates FMs to engineers in practice. &  \\ 
183 & 2018/08/16 & Interest &  &  &  &  &  &  \\ 
186 & 2018/08/17 & Teaching FM &  &  & Combinations of formal and semi-formal methods &  &  &  \\ 
187 & 2018/08/19 & FM = only method able to solve the problem  &  &  &  &  &  &  \\ 
193 & 2018/08/26 &  & Z & Karnaugh Maps ? for circuit design &  &  &  &  \\ 
195 & 2018/09/04 & business in FM & mcrl2 &  &  &  &  &  \\ 
197 & 2018/09/24 & FMs have been in existence since 1967. they have never made it into the mainstream. & CSP & Finding synchronization bugs &  &  &  &  \\ 
198 & 2018/09/24 & need for rigorous specification and verification &  &  &  &  &  &  \\ 
199 & 2018/09/24 & State of the art in control engineering and automation &  &  &  &  &  &  \\ 
204 & 2018/09/26 &  &  &  &  &  & Move to management level &  \\ 
207 & 2018/10/05 & Only reasonable way to get my results &  &  &  &  &  &  \\ 
208 & 2018/10/05 & I never felt tempted to use formal methods &  & To clearly define concepts &  &  &  &  \\ 
213 & 2018/12/16 & Regulatory authority &  &  &  &  &  & ok \\ 
215 & 2019/02/04 & Functional safety standards and their application & Modelling languages &  & rather semi-FMs, are more practical and easy to apply & support test case generation and validation & customers ability to make use of them as well in the specification and development process of critical systems &  \\ 
217 & 2019/03/20 &  &  &  & UML &  &  &  \\ 
218 & 2019/03/22 & increase confidence in system security &  &  &  &  &  &  \\ 
221 & 2020/04/20 &  &  &  & Static analyzer e.g. PVS Studio &  & 1. The tools don't even work. 2. The tools aren't cost effective. 3. We would have to train all the programmers, the QA team, the systems engineering team, and (importantly) the company executives.  &  
\end{xtabular}
\end{landscape}

\subsection{Copy of the Advertisement Flyer}
\label{sec:questionnaire-flyer}

\includegraphics[width=.7\textwidth]{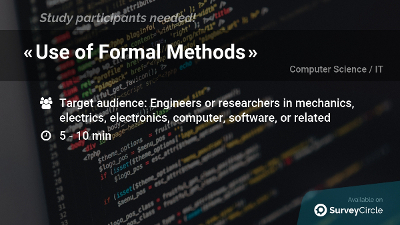}

\subsection{Screenshot of the Twitter Poll}
\label{sec:questionnaire-twitter}

\includegraphics[width=.7\textwidth]{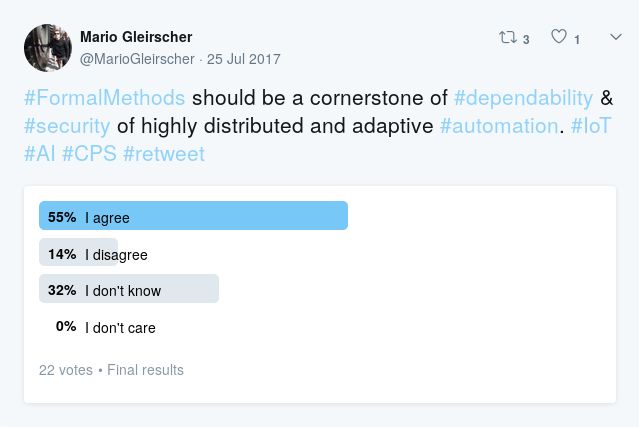}

\subsection{Copy of the Questionnaire}
\label{sec:questionnaire-complete}

\normalsize

The PDF export of our on-line questionnaire \textbf{on the next page}
corresponds to the questionnaire we used for the sample taken until
31.3.2019 with $N=220$.
Since 26.5.2019, an extended version of the questionnaire had been
available online at
\begin{quotation}
  \url{https://goo.gl/forms/FnKNQtTmI3A6BekM2}.
\end{quotation}
We crafted this questionnaire using Google Forms~\citep{Google2018}.
We use numbered identifiers for each question category, demographic
questions are prefixed with a ``D'', questions about past \ac{FM}
use~($\mathit{UFM}_p$) with a ``P'', about future or intended \ac{FM}
use~($\mathit{UFM}_i$) with an ``F'', questions about obstacles with an ``O''.
Open questions are suffixed by an ``o''.

\raggedright

\includegraphics[page=1,width=.8\textwidth]{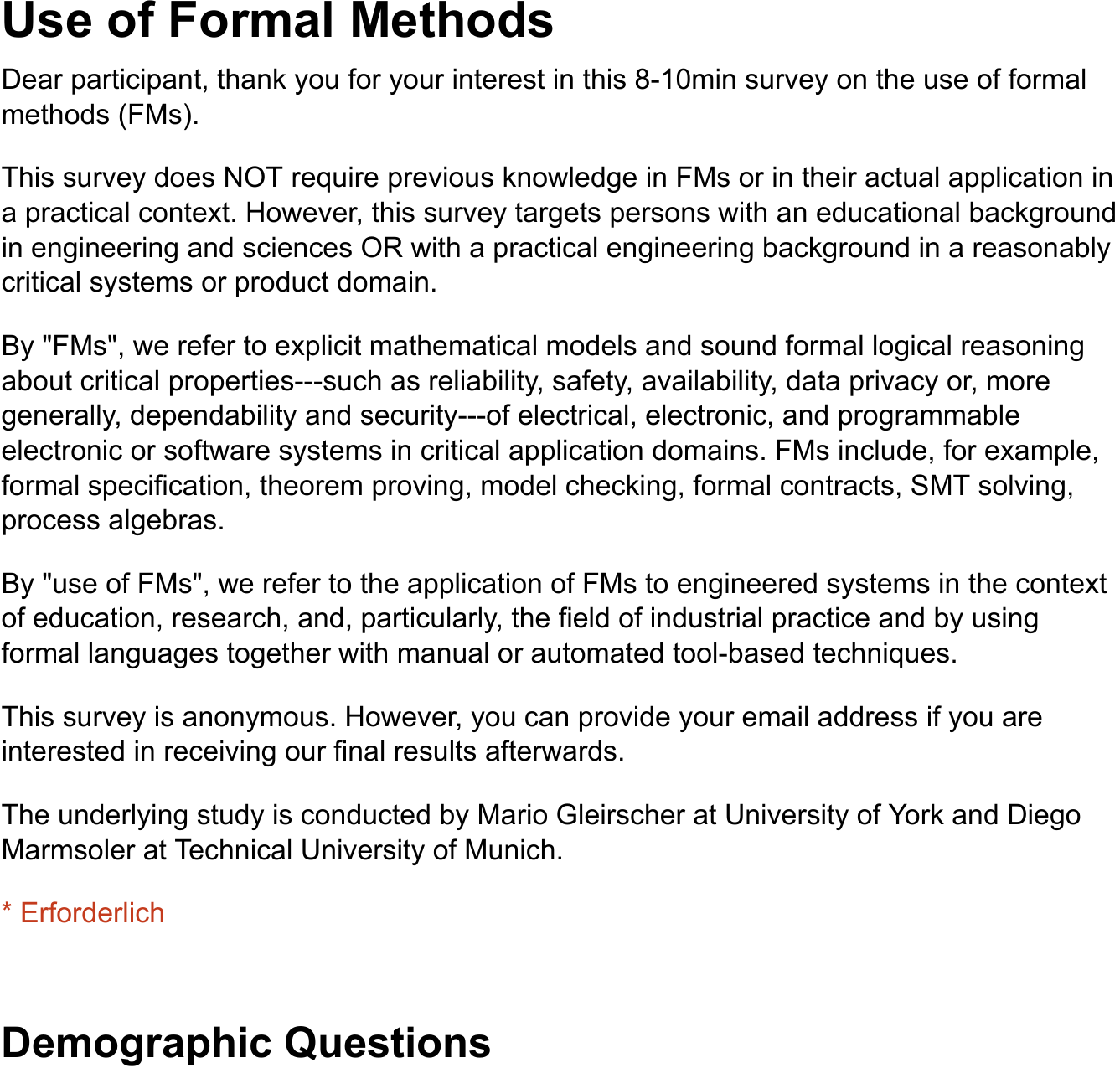}\\
\includegraphics[page=2,width=.8\textwidth]{questionnaire-2018-07-31-cropped}\\
\includegraphics[page=3,width=.8\textwidth]{questionnaire-2018-07-31-cropped}\\
\includegraphics[page=4,width=.8\textwidth]{questionnaire-2018-07-31-cropped}\\
\includegraphics[page=5,width=.8\textwidth]{questionnaire-2018-07-31-cropped}\\
\includegraphics[page=6,width=.8\textwidth]{questionnaire-2018-07-31-cropped}\\
\includegraphics[page=7,width=.8\textwidth]{questionnaire-2018-07-31-cropped}\\
\includegraphics[page=8,width=.8\textwidth]{questionnaire-2018-07-31-cropped}\\
\includegraphics[page=9,width=.8\textwidth]{questionnaire-2018-07-31-cropped}\\
\includegraphics[page=10,width=.8\textwidth]{questionnaire-2018-07-31-cropped}\\
\includegraphics[page=11,width=.8\textwidth]{questionnaire-2018-07-31-cropped}\\
\includegraphics[page=12,width=.8\textwidth]{questionnaire-2018-07-31-cropped}

\end{document}